\documentclass[]{article}

\usepackage{fullpage}
\usepackage{wrapfig}
\usepackage{graphicx}
\usepackage[backend=biber, style=phys, articletitle=false, eprint=true]{biblatex}
\usepackage{fancyhdr}
\usepackage{hyperref}
\usepackage{amsmath}
\usepackage{mathrsfs}
\usepackage{bbold}
\usepackage{color}

\graphicspath{{./figures/}}

\fancypagestyle{firstpage}{%
	\rhead{NT@UW-22-10}
	\cfoot{}

}

\title{Entanglement between Valence and Sea Quarks in Hadrons of 1+1 Dimensional QCD}
\author{Peter J. Ehlers \\ \it Department of Physics, University of Washington, Seattle, WA 98195-1560, USA}

\begin{document}
\maketitle

\thispagestyle{firstpage}
\pagenumbering{roman}

\renewcommand{\abstractname}{}

\begin{abstract}
	The conceptual interpretation of valence- and sea-quark separation, which is a key aspect of the parton model and of an intuitive picture of hadron structure, becomes obscured by quantum effects in QCD. This suggests that there may be measures of entanglement between quark degrees of freedom that are present in QCD, but absent in the intuitive picture with a clear valence-sea (VS) separation. In this paper, we define the first rigorous measure of VS entanglement in QCD in an attempt to bring conceptual clarity to this issue, and, potentially, to find a measure of the applicability of the parton model to QCD bound states. This VS entanglement vanishes in the large-$N_c$ limit, and it remains low when finite-$N_c$ states resemble their large-$N_c$ counterparts. We perform a numerical study of VS entanglement in 1+1 dimensional discrete light-cone quantized QCD, and in the process develop a method for building the color-singlet basis of 1+1d QCD that is manifestly complete and orthogonal by construction. We calculate this VS entanglement entropy for the first time and find that it is relatively low for the first few excited states of both mesons and baryons compared to all other states in the spectrum, with the VS entropy of ground state hadrons providing a minimum. We also see that for ground state mesons the entropy is well described in the $1/N_c$ approximation. These results suggest that low energy hadrons may be the only QCD bound states for which the large-$N_c$ expansion, and perhaps the parton model, provide an accurate description. This work also provides the first evidence that the VS entanglement entropy of QCD in 3+1d, which would likely serve as an order parameter for the transition between quark and hadron degrees of freedom, may be perturbatively accessible through a large-$N_c$ expansion.
\end{abstract}

\pagebreak

\thispagestyle{empty}
\tableofcontents

\pagebreak

\pagenumbering{arabic}

\section{Introduction}
\label{sec:intro}

It has been well established for decades now that QCD is the underlying theory of the strong interactions. Though it is described simply in quantum field theory with an $SU(3)$ gauge field coupled to fermionic matter, it exhibits the rich tapestry of phenomena that we observe in nuclear processes. In particular, quarks in QCD behave as nearly free particles at large momentum scales beyond 1 GeV (dubbed "asymptotic freedom") \cite{Gross:1973id,Politzer:1973fx}, but for lower energies the interaction is strong enough that quarks and gluons become confined and are only observed within bound states known as hadrons \cite{Wilson:1974sk}. Despite having the theory of the strong interactions in hand, QCD has never been solved analytically, even when no matter fields are present. A multitude of techniques have been developed over the decades to model the strong interaction phenomena generated by QCD. 

Given that there are still difficulties in computing rigorous results in hadronic physics, it may prove useful to approach problems in nuclear physics from a different perspective. There has been a great deal of recent interest in quantum information theory and its application to other areas of physics \cite{Bauer:2022hpo}, instigated by a surge of interest and developments in quantum computing \cite{Byrnes:2005qx,Jordan:2011ci,Jordan:2012xnu,Jordan:2014tma,Zohar:2012ay,Zohar:2012xf,Banerjee:2012pg,Banerjee:2012xg,Tagliacozzo:2012df,Stannigel:2013zka,Zohar:2013zla,Zohar:2014qma,Marcos:2014lda,Wiese:2014rla,Garcia-Alvarez:2014uda,Kasper:2015cca,Mezzacapo:2015bra,Bazavov:2015kka,Martinez:2016yna,Muschik:2016tws,Banuls:2017ena,Yeter-Aydeniz:2017ubh,Raychowdhury:2018osk,Kaplan:2018vnj,Klco:2018zqz,Macridin:2018oli,Yeter-Aydeniz:2018mix,Alexandru:2019nsa,Luo:2019vmi,Surace:2019dtp,Klco:2019evd,Mil:2019pbt,Haase:2020kaj,Shaw:2020udc,Surace:2019dtp,Raychowdhury:2018osk,Banuls:2019bmf,Kharzeev:2020kgc,Kreshchuk:2020dla,Kreshchuk:2020aiq,Davoudi:2020yln,Mueller:2020vha,Ciavarella:2021nmj,Barata:2020jtq,Kurkcuoglu:2021dnw,Bauer:2021gup,Davoudi:2021ney,Atas:2021ext,ARahman:2021ktn}. A key concept in quantum information science is the notion of quantum entanglement and the corresponding numerical quantity known as entanglement entropy. Entanglement is very important to quantum computing as it in principle allows for quantum computers to exceed the capabilities of classical computers, and the entanglement entropy serves as a resource that is exploited by the algorithms. Mapping with the entanglement structure of physical theories can lead to more efficient simulations of these theories with quantum computers. There have been a number of important results regarding entanglement entropy in field theory \cite{Bombelli:1986rw,Srednicki:1993im,Callan:1994py,Holzhey:1994we,Calabrese:2004eu,Calabrese:2009qy,Ryu:2006bv,Ryu:2006ef}, but these have been primarily focused on  the subjects of black holes, conformal field theories and holography.

In nuclear physics, a quantum information approach to phenomena has seen relatively few applications, but this is rapidly changing. A paper by Kharzeev and Levin \cite{Kharzeev:2017qzs} obtained some interesting results for the entanglement entropy of deep inelastic scattering for the purpose of better understanding the parton model description of the proton. Deep inelastic scattering only probes part of the internal structure of the proton, so the unknown remainder of the proton structure which goes into the ensuing hadron shower that is not measured in these experiments is handled statistically, resulting in a determinable entropy. Prior to this paper, the investigation of entanglement in nuclear physics had been rather sparse \cite{Simak:1988qp,Bialas:1999wi,Reznik:2002fz,EHSNA22:2005ygs,Kutak:2011rb,Muller:2011ra,Balasubramanian:2011wt,Peschanski:2012cw,Hsu:2012gk,Lello:2013bva,Seki:2014cgq,Wang:2014lua,Kovner:2015hga,Ho:2015rga,Peschanski:2016hgk}, but many papers have been written on the subject since then \cite{Beane:2019loz,Berges:2017hne,Shuryak:2017phz,Cervera-Lierta:2017tdt,Baker:2017wtt,Hagiwara:2017uaz,Neill:2018uqw,Beane:2018oxh,Kovner:2018rbf,Berges:2018cny,Feal:2018ptp,Tu:2019ouv,Peschanski:2019yah,Armesto:2019mna,Castorina:2020cro,Ramos:2020kaj,Gotsman:2020bjc,Robin:2020aeh,Duan:2020jkz,Ramos:2020kyc,Iskander:2020rkb,H1:2020zpd,Afik:2020onf,Germano:2020nbh,Germano:2021brq,Li:2020bys,Feal:2020myr,Kharzeev:2021yyf,Klco:2021biu,Florio:2021xvj,Duan:2021kmf,Kharzeev:2021nzh,Beane:2021zvo,Skokov:2021yez,Zhang:2021hra,Hentschinski:2021aux,Duan:2021clk,Klco:2021cxq,Low:2021ufv,Gong:2021bcp,Dvali:2021ooc,Ramos:2022gia,Dumitru:2022tud,Liu:2022ohy,Liu:2022hto,Afik:2022kwm,Liu:2022qqf,Hentschinski:2022rsa,Kou:2022dkw,Wang:2022noa,Fedida:2022izl}. Many of these papers focus on entanglement in field theories between two non-overlapping regions of space, with only a few looking into entanglement between other degrees of freedom \cite{Balasubramanian:2011wt}. Part of the goal of this project is to go beyond the usual focus on bipartite spatial entanglement and introduce a more general measure of entanglement. We will use this new measure to define and investigate the quantum entanglement between the quarks that compose the hadrons in QCD. An example of an earlier attempt to generalize entanglement is the entwinement \cite{Balasubramanian:2016xho,Balasubramanian:2018ajb,Erdmenger:2019lzr}.

The parton model, a major predecessor of QCD \cite{Bjorken:1968dy,Feynman:1969ej,Bjorken:1969ja}, was a major step toward developing the theory of strong interactions and in many ways still informs our intuitive understanding of it. In the parton model, a hadron is composed of a small handful of "valence" quarks, corresponding to the flavor quantum numbers of the hadron, that are embedded inside of a bubbling soup of "sea" quarks and gluons. However, a more accurate picture originating from QCD more directly would suggest that the hadronic state is an ambiguous superposition of states with many possible numbers of partons, with the only clear pattern being that the total flavor numbers for these quark ensemble states are fixed. Thus the notions of valence and sea quarks seem to have limited utility in the more accurate theory, despite providing a useful picture in the past. What this may be indicating is that there can be a high degree of entanglement between the valence and sea quarks that is being captured by QCD but is largely omitted by the parton model. QCD is in some sense more quantum mechanical than the parton model, so it stands to reason that one of the major differences between them would be rooted in a purely quantum mechanical concept such as entanglement. Through some notion of valence-sea (VS) entanglement, one may be able to quantify the degree to which the parton model picture makes rigorous sense for a given hadronic state in QCD.

By "parton model," we are referring to a model where hadrons are approximated as pointlike constituent particles, with valence partons embedded in a bath of sea partons. For example, in deep inelastic scattering of an electron off of a proton, the process can be approximated by the scattering of the electron off of an individual parton inside the proton that is then integrated over what is known as a parton distribution function (PDF) \cite{Ellis:1978ty, Collins:1989gx} or its generalizations \cite{Diehl:2003ny,Ji:2003ak,Ji:2004wu}. The PDF essentially tells us the probability of finding a specific parton within the proton that has a fraction $x$ of its total momentum. This turns out to be the first order approximation in what is known as the twist expansion \cite{Collins:1989gx}. In addition, the model assumes that the distribution of quarks can be separated into valence and sea contributions, and that for each sea quark there is an antiquark with identical properties. For example, the up quark PDF $f_u(x)$ in a proton should be able to split into valence and sea contributions $f_u(x)=f_{u,V}(x)+f_{u,S}(x)$ with $f_{u,S}(x)=f_{\bar{u}}(x)$ in the parton model. The quark-antiquark symmetry of the parton sea assumes that it can be divided neatly into quark-antiquark pairs, which can presumably be used to identify which quarks are part of the sea and which ones are valence quarks. This all implies that the wavefunction for the proton roughly factorizes into valence and sea parts $\psi(x)\sim\psi_V(x)\psi_S(x)$. If this was strictly accurate, then the entanglement between valence and sea partons would be zero. This is the sense in which valence-sea entanglement can measure the applicability of the parton model, as a large degree of entanglement would indicate a strong violation of the assumptions inherent to the parton model as described here.

Additionally, there is no reason to assume that entanglement is accessible through perturbative methods, so an entanglement-based investigative approach to QCD could provide new insights into hadron structure. For VS entanglement in particular, it is known that in large-$N_c$ QCD where diagrams with quark-antiquark pairs are suppressed \cite{tHooft:1973alw,tHooft:1974pnl,Witten:1979kh,Witten:1980sp,Dashen:1993jt,Dashen:1994qi,Kaplan:1995yg,Kaplan:1996rk}, the hadrons in the spectrum have only gluons in their parton seas. Thus there can be no VS entanglement between quarks in the large-$N_c$ limit, which shows that a next to leading order calculation in the large-$N_c$ expansion is necessary to capture such a property of the hadrons. This further motivates an investigation into VS entanglement as a means of gaining more nonperturbative information about the makeup of hadrons.

In previous work, we designed a model for helicity entanglement in nucleons \cite{Beane:2019loz}. In this model, the null-plane formalism was used to separate the dynamical, spin-based generators of Poincar\'{e} symmetry from the kinematical, momentum-based generators. It was argued that it may be sensible to represent a nucleon and its first few excitations through a handful of chiral basis states defined by the helicity and isospin representations of the valence quarks. Fitting experimental data for static baryon properties such as the axial coupling $g_A$ and masses to their results in the chiral basis, one could find a change of basis matrix between baryon and chiral states, and from there the entanglement entropy between the different valence representations could be calculated for the nucleon. It was found that this entropy quickly grew to near its maximum value after just a few chiral states were introduced. This paper can be viewed as somewhat of a continuation of our previous work in that it seeks to learn more about the nucleon through the entanglement of its constituent quarks with the remaining quarks in the nucleon, but in this case making fewer assumptions about the nucleon state.

The results of this work are obtained using QCD in 1 space and 1 time dimension, which was extensively analyzed in \cite{Pauli:1985ps,Brodsky:1997de,Hornbostel:1988fb,Hornbostel:1988ne}. While QCD in lower dimensions lacks some of the features of full QCD, such as fermion spin and dynamical gauge field components, confinement still occurs here, and its spectrum has been essentially solved in the large-$N_c$ limit \cite{tHooft:1974pnl}. The relative simplicity of the lower dimensional theory combined with the presence of confinement makes it a useful model for the low energy spectrum of QCD. From our work, we find that the VS entanglement that we have defined converges to a finite value fairly quickly for the lowest energy hadrons, and the values they approach tend to be rather small. For mesons, this is a result of their wavefunctions strongly resembling their large-$N_c$ counterparts, with this connection diminishing as we move further along the excited meson spectrum. For baryons, we cannot draw a resemblance to the large-$N_c$ wavefunction specifically, but a similar connection to large-$N_c$ physics more broadly is likely underlying the small values in that case as well. For ground state mesons, the $N_c$ dependence of the VS entropy also fits nicely to a $1/N_c$ curve, which suggests that a large-$N_c$ expansion of the VS entropy should converge rapidly. If this convergence holds in real QCD, it would open the possibility of calculating it perturbatively with the large-$N_c$ expansion. The VS entanglement of real QCD would almost certainly be scale dependent, and so it may serve as an order parameter for the transition between quark and hadron degrees of freedom, which would show that entanglement between quarks is an important ingredient in the process of hadronization.

This paper will proceed as follows. In Section \ref{sec:eoe}, we will define two notions of quark entanglement: single quark (SQ) entanglement, which we will show is related to parton distribution functions, and valence-sea (VS) entanglement, which will be the primary interest of the rest of the paper. Section \ref{sec:tHooft} will define 1+1d QCD in the Hamiltonian formalism as well as the 't Hooft model, which is the large-$N_c$ limit of this theory. Section \ref{sec:DLCQ} will discuss discrete light-cone quantization (DLCQ) as it applies to this work, which is the main tool used to get our entanglement results from 1+1d QCD. More details about our calculational methods can be found in the appendices. Section \ref{sec:VSE} will show how the VS entanglement entropy relates to matrix elements that can be calculated in DLCQ. Section \ref{sec:plots} will contain all of our numerical data and calculational results, which is followed by our conclusions about the work in Section \ref{sec:conclu}.

Appendix \ref{Algorithm} discusses the procedure we use to define a set of color singlet basis states in DLCQ using Young's orthogonal form for the permutations of quarks in a Fock state. The advantage of using this basis is that it is manifestly orthonormal and complete from the beginning. Other papers that have done similar calculations \cite{Hornbostel:1988fb,Hornbostel:1988ne,Burkardt:1989wy} mention rendering the Hamiltonian in an incomplete basis as an intermediate step, so to our knowledge the method we use to construct the 1+1d QCD Hamiltonian in DLCQ has not yet been described in the literature. Appendix \ref{PrincipalValue} details our choice of principal value in DLCQ and why this eliminates the two-quark potential term given in Eq.~\eqref{V2} from the 1+1d QCD Hamiltonian. This choice is needed to prevent the formation of hadrons with unphysical masses when the coupling strength is much larger than the quark masses.

\section{Extended Definition of Entanglement}
\label{sec:eoe}

\subsection{Definition}
We can quantify valence-sea entanglement by defining an associated entropy stemming from a bipartition of the valence and sea quark degrees of freedom. However, in QCD, there is no way to make a distinction between the valence and the sea degrees of freedom outside of the specific flavors of the valence quarks. A general Fock state that is a component of a hadronic state can have multiple quarks that have the same flavor as a valence quark in the hadron, so we will need a prescription to determine which quarks count as valence quarks and which ones are part of the sea. Our approach is to duplicate the Hilbert space, apply some operators to the density matrix that pull quarks of the proper flavors that match the valence flavor structure, and then trace out the original Hilbert space to obtain a density matrix that acts in the valence quark Fock state basis. This effectively adds a contribution to the valence density matrix for every possible way of separating valence quarks from the sea for a given Fock state. This means that our definition of the valence density matrix is agnostic regarding the properties of the valence quarks aside from flavor, and that Fock states with a higher number of quarks will be weighted more heavily than those with fewer quarks because our procedure sums over possible valence quark assignments rather than averaging over them.

To illustrate how this formalism works, we will begin with a simple example. Consider a quantum system where particles can occupy two sites labeled $A$ and $B$. Each site can either be in an occupied state $\left|1\right>$ or in an unoccupied state $\left|0\right>$. The most general state one can make in this system is
\begin{equation}
\left|\psi\right>=\alpha\left|00\right>+\beta\left|01\right>+\gamma\left|10\right>+\delta\left|11\right>, \quad \alpha^*\alpha+\beta^*\beta+\gamma^*\gamma+\delta^*\delta=1,
\end{equation}
with amplitudes $\alpha, \beta, \gamma, \delta$, where the left digit in the basis states correspond to site $A$ while the right digit corresponds to site $B$. The general reduced density matrix for site $A$ is given by
\begin{equation}
\rho_A = \mathrm{Tr}_B[\left|\psi\right>\left<\psi\right|] = (\alpha\left|0\right>+\gamma\left|1\right>)(\alpha^*\left<0\right|+\gamma^*\left<1\right|)+(\beta\left|0\right>+\delta\left|1\right>)(\beta^*\left<0\right|+\delta^*\left<1\right|).
\end{equation}
As we will see later, the usual notion of bipartite entanglement can always be obtained using the extended formalism. We can define another reduced density matrix given by
\begin{equation}
\rho_{2A} = \mathrm{Tr}_1[X_A\left|\psi\right>_1\left|00\right>_2\left<00\right|_2\left<\psi\right|_1X_A],
\end{equation}
where we have duplicated the system with our general state $\left|\psi\right>$ in the original space $1$ and we have chosen our starting vector in the auxiliary space to be the state $\left|00\right>$. The operator $X_A$ swaps the state in site $A1$ with the state in site $A2$, so that
\begin{equation}
X_A\left|\psi\right>_1\left|00\right>_2 = \alpha\left|00\right>_1\left|00\right>_2+\beta\left|01\right>_1\left|00\right>_2+\gamma\left|00\right>_1\left|10\right>_2+\delta\left|01\right>_1\left|10\right>_2.
\end{equation}
Putting this into the expression for $\rho_{2A}$ and tracing out region $1$ (and dropping the $2$ subscripts on the remaining state vectors), we get
\begin{equation}
\rho_{2A} = (\alpha\left|00\right>+\gamma\left|10\right>)(\alpha^*\left<00\right|+\gamma^*\left<10\right|)+(\beta\left|00\right>+\delta\left|10\right>)(\beta^*\left<00\right|+\delta^*\left<10\right|) = \rho_A\otimes(\left|0\right>\left<0\right|)_B,
\end{equation}
which will give the same results for any measure of entanglement as $\rho_A$ does by itself. Note that instead of choosing $\left|00\right>$ for our initial auxiliary vector we could have chosen any separable state $\left|\phi_A\right>\left|\phi_B\right>$ and achieved the same result with $\rho_{2A} = \rho_A\otimes\left|\phi_B\right>\left<\phi_B\right|$.

\begin{figure}
	\centering
	\includegraphics[width=0.7\textwidth]{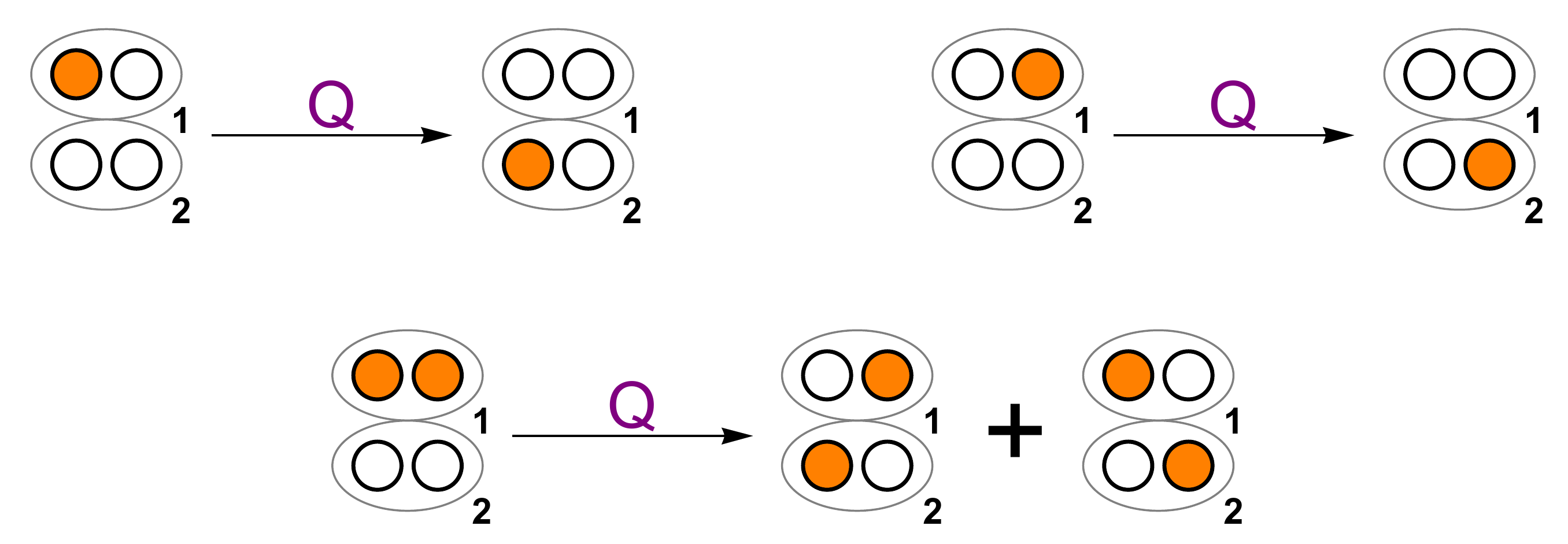}
	\caption{Illustration of how the operator $Q$ acts upon states in the two-site example. When only one site is occupied $Q$ moves it into the duplicate Hilbert space. When both are occupied, $Q$ adds two contributions, one where the first particle is treated as the valence particle along with another where the second is treated as the valence particle.}
	\label{Qaction}
\end{figure}

This system can also help illustrate what is meant by quark entanglement in this paper. Instead of the operator $X_A$, which probes entanglement between sites $A$ and $B$, we use the Hermitian operator $Q$ given by 
\begin{align}
\nonumber Q =&~ \sigma_{A,1}^+\otimes\mathbb{1}_{B,1}\otimes\sigma_{A,2}^-\otimes\mathbb{1}_{B,2} + \mathbb{1}_{A,1}\otimes\sigma_{B,1}^+\otimes\mathbb{1}_{A,2}\otimes\sigma_{B,2}^- \\
& + \sigma_{A,1}^-\otimes\mathbb{1}_{B,1}\otimes\sigma_{A,2}^+\otimes\mathbb{1}_{B,2}+ \mathbb{1}_{A,1}\otimes\sigma_{B,1}^-\otimes\mathbb{1}_{A,2}\otimes\sigma_{B,2}^+,
\end{align}
where $\sigma^-=\left(\begin{matrix} 0 & 1 \\ 0 & 0 \end{matrix}\right)$ removes a particle from an occupied state and $\sigma^+=\left(\begin{matrix} 0 & 0 \\ 1 & 0 \end{matrix}\right)$ adds a particle to an unoccupied state. We strategically choose $\left|00\right>$ for our initial auxiliary vector here because it is fully unoccupied, so that
\begin{align}
\nonumber Q\left|\psi\right>_1\left|00\right>_2 &= (\sigma_{A,1}^-\otimes\mathbb{1}_{B,1}\otimes\sigma_{A,2}^+\otimes\mathbb{1}_{B,2}+ \mathbb{1}_{A,1}\otimes\sigma_{B,1}^-\otimes\mathbb{1}_{A,2}\otimes\sigma_{B,2}^+)\left|\psi\right>_1\left|00\right>_2 \\
&= \beta\left|00\right>_1\left|01\right>_2+\gamma\left|00\right>_1\left|10\right>_2+\delta(\left|01\right>_1\left|10\right>_2+\left|10\right>_1\left|01\right>_2).
\end{align}
The action of $Q$ on the components of $\left|\psi\right>_1\left|00\right>_2$ is illustrated in Fig.~\ref{Qaction}. This makes it so that our procedure is not generating completely artificial entanglement, guaranteeing that the entanglement we are measuring strictly originates from the original state and not the auxiliary one. The reduced density matrix for quark entanglement can then be written as 
\begin{align}
\nonumber \rho_Q &= \mathcal{N}\mathrm{Tr}_1[Q\left|\psi\right>_1\left|00\right>_2\left<00\right|_2\left<\psi\right|_1Q] \\
&= \mathcal{N}\left((\beta\left|01\right>+\gamma\left|10\right>)(\beta^*\left<01\right|+\gamma^*\left<10\right|)+\delta^*\delta(\left|01\right>\left<01\right|+\left|10\right>\left<10\right|)\right), \\
\mathcal{N}^{-1} &= \beta^*\beta+\gamma^*\gamma+2\delta^*\delta=1+\delta^*\delta-\alpha^*\alpha.
\end{align}
Note that the one particle states in $\left|\psi\right>$ show up as a pure state contribution to the density matrix and are only counted once, while the two particle state $\left|11\right>$ contributes twice in the form of an identity matrix. This is a general pattern that will hold for valence-sea entanglement as well; Fock states that only contain valence quarks do not contribute to the entropy, and Fock states with many sea quarks generally contribute more to the entropy than those with fewer sea quarks. This is because every possible way of separating valence quarks from the sea is summed over in the density matrix, so our measure of valence-sea entanglement is very sensitive to the presence of many sea quarks.

\begin{figure}
	\centering
	\includegraphics[width=0.5\textwidth]{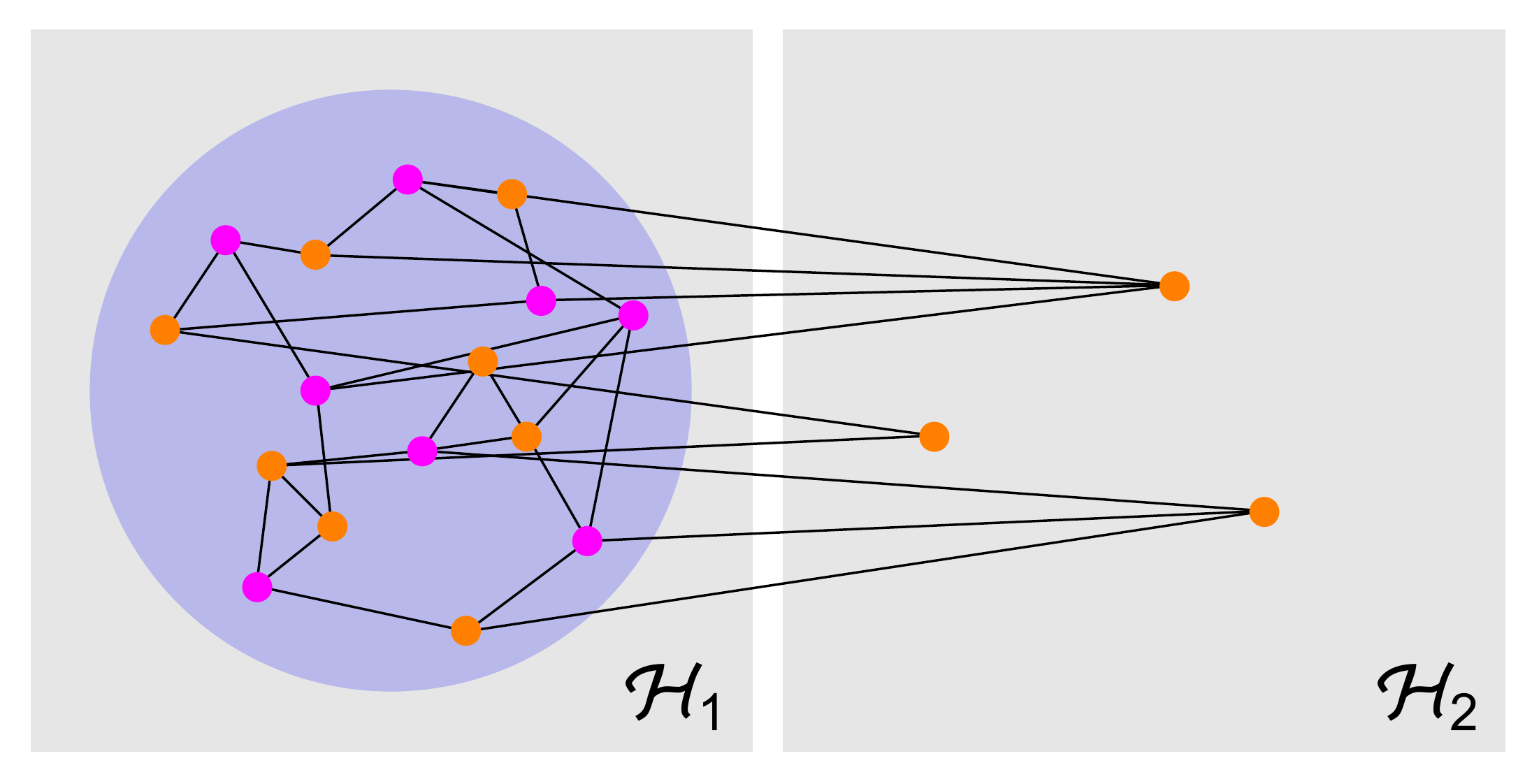}
	\caption{Illustration of the procedure for obtaining the valence-sea entanglement for a baryon in QCD. Our procedure pulls three quarks with appropriate flavors out of the baryon state in $\mathcal{H}_1$ and moves them into the vacuum state in $\mathcal{H}_2$. The entanglement between the valence quarks and the parton sea becomes the entanglement between $\mathcal{H}_1$ and $\mathcal{H}_2$, which can then be analyzed using standard methods for bipartite entanglement.}
	\label{VSBaryon}
\end{figure}

The procedure for obtaining the reduced density matrix in a general setting is as follows. Given an initial pure state $\left|\psi\right>$, the definition of the reduced density matrix is given by
\begin{equation}
  \rho = \mathcal{N}~\mathrm{Tr}_1\left[\mathcal{O}_{12}\left|\psi\right>_1\left|0\right>_2\left<0\right|_2\left<\psi\right|_1\mathcal{O}^\dagger_{12}\right].
\end{equation}
This requires the duplication of the Hilbert space so that $\mathcal{H}\rightarrow\mathcal{H}_1\otimes\mathcal{H}_2$. This duplication serves the purpose of creating an artificial bipartition of the Hilbert space so that notions of entanglement can be rigorously defined even when no bipartition exists in the original Hilbert space. The operator $\mathcal{O}_{12}$ acts on both Hilbert spaces so that the initial states $\left|\psi\right>$ and $\left|0\right>$ are transformed into an entangled state and the density matrix is non-trivial. This is illustrated for the example of a baryon in Fig.~\ref{VSBaryon}, where the operator $\mathcal{O}_{12}$ moves three valences quarks out of the baryon in $\mathcal{H}_1$ and into the vacuum in $\mathcal{H}_2$. The specific choice of operator and initial state $\left|0\right>_2$ defines the type of entanglement being represented by $\rho$. Additionally, there may be a normalization factor $\mathcal{N}$ necessary to ensure that the trace of $\rho$ is fixed appropriately.

There are two main reasons why this definition is appropriate. First, this formalism can be used to replicate the entanglement measures of any system that is already bipartite, including spatial entanglement. To see this, consider a state $\left|\psi_{AB}\right>$ in a Hilbert space $\mathcal{H}=\mathcal{H}_A\otimes\mathcal{H}_B$, such as the two-state example we walked through previously. We can define the density matrix in this formalism by choosing an operator $X_{12}(A)$ that swaps components of the state within $A$ so that
\begin{equation}
  X_{12}(A)\left|\psi_A\psi'_B\right>_1\left|\phi_A\phi'_B\right>_2 = \left|\phi_A\psi'_B\right>_1\left|\psi_A\phi'_B\right>_2,
\end{equation}
along with a separable initial state $\left|0\right>=\left|0_A0'_B\right>$. In the example, this is just the operator that swaps sites $A1$ and $A2$. It is straightforward to show that $X_{12}(A)$ is both a Hermitian and unitary operator. Using a set of separable basis states $\left|\phi_{i,A}\phi'_{j,B}\right>$, we can write $\left|\psi\right>=\sum_{i,j}\alpha_{ij}\left|\phi_{i,A}\phi'_{j,B}\right>$. The density matrix is then defined by
\begin{align}
\nonumber \rho &= \mathcal{N}~\mathrm{Tr}_1\left[X_{12}(A)\left|\psi_{AB}\right>_1\left|0\right>_2\left<0\right|_2\left<\psi_{AB}\right|_1X_{12}^\dagger(A)\right] \\
\nonumber &= \sum_{i,j}\alpha_{ij}\sum_{k,l}\alpha^*_{kl}\mathrm{Tr}_1\left[\left|0_A\phi'_{j,B}\right>_1\left|\phi_{i,A}0'_B\right>_2\left<\phi_{k,A}0'_B\right|_2\left<0_A\phi'_{l,B}\right|_1\right] \\
\nonumber &= \sum_{i,k}\left(\sum_{j}\alpha_{ij}\alpha^*_{kj}\right)\left|\phi_{i,A}0'_B\right>_2\left<\phi_{k,A}0'_B\right|_2 \\
  &= \rho_{AB}\otimes\left|0'_B\right>\left<0'_B\right|,
\end{align}
where $\rho_{AB}=\mathrm{Tr}_B\left[\left|\psi_{AB}\right>\left<\psi_{AB}\right|\right]$ is the usual notion of bipartite entanglement between subspaces $A$ and $B$. Since $\rho_{AB}$ is equal to a tensor product of $\rho_{AB}$ with a pure state density matrix $\left|0'_B\right>\left<0'_B\right|$, the entanglement entropy of $\rho$ is equal to the entanglement entropy of $\rho_{AB}$, so our formalism reduces to the usual notion of entanglement when the original Hilbert space is already bipartite.

Second, this construction allows for the VS entanglement of mesons to be zero in the large-$N_c$ limit. The operator corresponding to VS entanglement will take a quark operator, similar to the operator $Q$ from the example, for each valence quark from $\mathcal{H}_1$ and move them into $\mathcal{H}_2$ before tracing out $\mathcal{H}_1$. The initial state $\left|0\right>$ is taken to be the free quark vacuum state so that there are no particles in the duplicate space to move into the original space and create entirely artificial entanglement, which was also a consideration in the example. For mesons  in the large-$N_c$ limit, which have no sea quarks, the operator effectively dumps the entire meson state into $\mathcal{H}_2$, making $\rho$ a pure state density matrix with zero entropy, which must be the case if this is to be a sensible measure of entanglement between valence and sea quarks. This also applies to the parton model state outlined in the introduction where the state $\psi(x)\sim\psi_V(x)\psi_S(x)$ would have $\psi_V(x)$ moved to $\mathcal{H}_2$ while $\psi_S(x)$ is left in $\mathcal{H}_1$ and traced out, leaving a pure state density matrix defined only by $\psi_V(x)$. Details about how VS entanglement is defined using this formalism will be given in Section \ref{sec:VSE}.

\subsection{Single-Quark Entanglement}
As an aside, we can also look at how an individual quark is entangled with the rest of a hadron. A single quark operator of flavor $i$ in the light-cone formalism is given by
\begin{align}
\label{Qdef}
 Q_i &= \int dx^-dx^{d-2}_\perp~\bar{q}_{i,2}(x^-,\vec{x}_\perp)\gamma^+q_{i,1}(x^-,\vec{x}_\perp), \\
\label{Qdef2}
  \bar{Q}_i &= -\int dx^-dx^{d-2}_\perp~\bar{q}_{i,1}(x^-,\vec{x}_\perp)\gamma^+q_{i,2}(x^-,\vec{x}_\perp),
\end{align}
where $d=2$ or $4$ is the spacetime dimension. In these equations and the work going forward, we use the Kogut-Soper convention \cite{Kogut:1969xa,Brodsky:1997de} $x^\pm=(x^0\pm x^{d-1})/\sqrt{2}$ for the light-cone coordinates and $\gamma^\pm=(\gamma^0\pm \gamma^{d-1})/\sqrt{2}$. The quark fields in these expressions are full, unprojected Dirac spinors \cite{Beane:2013oia, Belitsky:2005qn}, though the $\gamma^+$ in between them will eliminate parts of these spinors in any practical calculation. Using this as our entanglement operator and applying several conservation laws, it can be shown that the resulting density matrix is given by (in light-cone gauge)
\begin{align}
 \rho_i &= \mathcal{N}\sum_\lambda\int^1_0 dx\int^\infty_{-\infty} dk^{d-2}_\perp~f_{i,\lambda}(x,\vec{k}_\perp)\left|q_{i,\lambda}(x P^+,\vec{k}_\perp)\right>\left<q_{i,\lambda}(xP^+ ,\vec{k}_\perp)\right|, \\
  f_{i,\lambda}(x,\vec{k}_\perp) &= \int \frac{dz^-dz^{d-2}_\perp}{2(2\pi)^{d-1}}e^{-i x P^+ z^-}e^{i \vec{k}_\perp.\vec{z}_\perp}\left<\Psi(P^+ ,\vec{P}_\perp)\right|\bar{q}_i(z)\Pi_\lambda\gamma^+q_i(0)\left|\Psi(P^+ ,\vec{P}_\perp)\right>,
\end{align}
where $\left|q_{i,\lambda}(k^+,\vec{k}_\perp)\right>=b^\dagger_{i,\lambda}(k^+,\vec{k}_\perp)\left|0\right>$ represents a single quark state in $d=2$ or $4$ spacetime dimensions with momentum $k$, flavor $i$, and helicity $\lambda$, while $\left|\Psi(P^+ ,\vec{P}_\perp)\right>$ represents a hadron with total momentum $P$, $x$ is the fractional momentum $k^+/P^+$, and $\Pi_\lambda$ is a helicity projection operator, given by $\Pi_\lambda=\frac{1}{2}(\mathbf{I}+\lambda\gamma^5)$. As with our position coordinates, the light-cone momenta are defined with the convention $p^\pm=(p^0\pm p^{d-1})/\sqrt{2}$. With $d=4$, the expression for $f_{i,\lambda}(x,\vec{k}_\perp)$ turns out to be the expression for a parton distribution function (PDF) in QCD, including the dependence on helicity and transverse momentum. This provides an interpretation of PDFs as a measure of quark entanglement within a hadron.

The expression for the SQ density matrix above was specific to light-cone gauge. In general, a Wilson line is necessary to maintain gauge invariance in both the PDF and another line for the SQ state vectors in the duplicate space. This work is entirely in light-cone gauge as it is the natural choice for both PDFs and 1+1d QCD.

\subsection{Other Definitions of Reduced Density Matrices}
So far, we have defined our reduced density matrix by using a partial trace over the original Hilbert space. In the literature, given a Hilbert space that can be partitioned into two regions $\mathcal{H}=\mathcal{H}_A\otimes\mathcal{H}_B$, there are three equivalent ways to define reduced density matrices in a field theory \cite{Balasubramanian:2016xho}:
\begin{itemize}
	\item[1] The standard method of doing a partial trace over the pure state density matrix $\rho=\left|\psi\right>\left<\psi\right|$ to get a reduced density matrix $\rho_A=\mathrm{Tr}_B[\rho]$. In theories with gauge symmetry, this may become more complicated as extra physical degrees of freedom may need to be temporarily added to the Hilbert space to perform the partial trace, as in \cite{Donnelly:2014gva}.
	\item[2] One can also find the density matrix by searching for the matrix $\rho_A\in\mathcal{H}_A$ that has expectation values with other operators in $\mathcal{H}_A$ equal to those of the original pure state $\left|\psi\right>$. In other words, $\rho_A$ must satisfy
	\begin{equation}
	\mathrm{Tr}_A[\rho_A\mathcal{O}_A]=\left<\psi\right|(\mathcal{O}_A\otimes\mathbf{I}_B)\left|\psi\right>, \quad \forall\mathcal{O}_A\in \mathcal{H}_A.
	\end{equation}
	The space of operators in $\mathcal{H}_A$ can be defined as the subset of operators in $\mathcal{H}$ that have the form $\mathcal{O}_A\otimes\mathbf{I}_B$.
	\item[3] The last definition is to write the density matrix as a path integral, and is primarily used when applying the replica trick to calculate the entropy of a state \cite{Callan:1994py,Headrick:2012fk,Donnelly:2014gva}. One starts with the pure state density matrix
	\begin{equation}
	\rho=\int\mathcal{D}\phi\mathcal{D}\phi' Z_\psi(\phi,\phi')\left|\phi\right>\left<\phi'\right|,
	\end{equation}
	where $\left|\phi\right>$ is some state that diagonalizes the constituent field operators that make up $\psi$, and $Z_\psi(\phi,\phi')=\left<\phi'|\psi\right>\left<\psi|\phi\right>$ will be our path integral. Assuming that $\left|\phi\right>$ represents a minimum energy state for a given set of conserved quantum numbers, $Z_\psi(\phi,\phi')$ can be written as a Euclidean path integral with an insertion of some simple operators with the appropriate quantum numbers, whose upper and lower functional bounds are given by $\phi'$ and $\phi$, respectively. Assuming that this functional basis can be partitioned so that $\left|\phi\right>=\left|\phi_A\right>\left|\phi_B\right>$, the reduced density matrix is obtained by setting $\phi'_B=\phi_B$ and integrating, so that
	\begin{equation}
	\rho_A=\int\mathcal{D}\phi_A\mathcal{D}\phi'_A \left(\int\mathcal{D}\phi_B Z_\psi(\phi_A,\phi_B;\phi'_A,\phi_B)\right)\left|\phi_A\right>\left<\phi'_A\right|.
	\end{equation}
\end{itemize}
Our discussion of a broadened definition of entanglement has been limited to definition 1 above. For SQ entanglement, the density matrix can be obtained from definition 2 by matching to matrix elements of the form $\left<\psi\right|\bar{q}_i(x)\Gamma\gamma^+q_i(y)\left|\psi\right>$ for some arbitrary spinor matrix $\Gamma$. A similar operator definition likely exists for VS entanglement, but some subtleties discussed below may require a slight modification to either this approach or the partial trace approach to get a more sensible definition of VS entanglement. Since definition 3 makes some assumptions about the hadron, it cannot be discussed in general terms here.

\section{1+1d QCD and the 't Hooft model}
\label{sec:tHooft}

This section will give details on the 1+1d QCD null plane Hamiltonian that will serve as the mathematical foundation for everything that follows. We will also reproduce the 't Hooft model in this formalism, which serves as a useful starting point for defining valence-sea (VS) entanglement as the mesons in this model have no sea and therefore no VS entanglement. It was first worked out in a paper by 't Hooft in 1974 \cite{tHooft:1974pnl} (see also section 3.2 of \cite{Manohar:1998xv} for a discussion of these results using this paper's conventions), and the 't Hooft model Hamiltonian has been known since the '90s \cite{Lenz:1991sa,Engelhardt:1994tk}. One appealing aspect of the 't Hooft paper is that it gives us a simple model that exhibits a Regge trajectory in its meson spectrum. The paper's method is to find the meson's wavefunction using the Bethe-Salpeter equation for the large-$N_c$ interactions in $d=1+1$ QCD. The large-$N_c$ limit makes it tractable to sum up all of the relevant diagrams and get a non-perturbative result for the wavefunction. It is also an example where light-cone coordinates are of more use than conventional space-time coordinates. 

\subsection{QCD in 1+1 Dimensions}
\subsubsection{Null Plane Hamiltonian}
The QCD action in $d=2$ dimensions reads 
\begin{equation}
  S=\int d^2x\left(\bar{q}(i\gamma^\mu D_\mu-m_q)q-\frac{1}{2}\mathrm{Tr}[F^{\mu\nu}F_{\mu\nu}]\right),
\end{equation}
where $D_\mu=\partial_\mu-igA_\mu$ is the covariant derivative and $m_q$ can in principle be a diagonal matrix in flavor space for $N_f$ flavors of fermions. To quantize on the null plane, we will need to separate the fermion fields into two parts using the projection operators $\Pi^\pm=\frac{1}{2}\gamma^\mp\gamma^\pm$ as in \cite{Beane:2013oia, Belitsky:2005qn}. With $q_\pm=\Pi^\pm q$ and choosing light-cone gauge $A^+=0$, the action becomes
\begin{equation}
  S=\int d^2x\left(\sqrt{2}iq_+^\dagger\partial^-q_++\sqrt{2}iq_-^\dagger\partial^+q_--\frac{1}{\sqrt{2}}(q_+^\dagger m_q\gamma^-q_-+q_-^\dagger m_q\gamma^+q_+)+\sqrt{2}gq_+^\dagger A^-q_++\frac{1}{2}(\partial^+A^{-,a})^2\right),
\end{equation}
where the index $a$ is the adjoint index for the $SU(N_c)$ gauge field. From this expression, we can see that both $q_-$ and $A^-$ are non-dynamical and should be integrated out. The resulting terms from the integration will be nonlocal in $x^-$, and at this stage we will take care to define our quantities rigorously. 

Define the inverse derivative to be 
\begin{equation}
  \frac{1}{\partial^+}\phi(x^+,x^-) = \frac{1}{2}\int_{-\infty}^{\infty}dy^-\epsilon(x^--y^-)\phi(x^+,y^-),
\end{equation}
where $\epsilon(x)$ is the sign function, whose derivative is twice the delta function. The advantage of this definition is that the inverse derivative obeys integration by parts; that is,
\begin{equation}
  \int_{-\infty}^{\infty}dx^-\left(\frac{1}{\partial^+}\phi(x)\right)\pi(x) = -\int_{-\infty}^{\infty}dx^-\phi(x)\left(\frac{1}{\partial^+}\pi(x)\right).
\end{equation}
Now we can define the fields to be
\begin{equation}
  \phi(x^+,x^-) = \frac{1}{\partial^+}(\partial^+\phi(x^+,x^-))+C_\phi(x^+).
\end{equation}
The function $C_\phi$ is necessary to allow $\phi$ to be a completely arbitrary field under a path integral.

Returning to the QCD action, we can use the above conventions to integrate out the non-dynamical fields and obtain
\begin{equation}
  S=\int dx^-dx^+\left((\sqrt{2}iq_+^\dagger)\partial^-q_+-q_+^\dagger \frac{m_q^2}{\sqrt{2}i\partial^+}q_+-\frac{1}{2}\left(\sqrt{2}g\frac{1}{\partial^+}(q_+^\dagger T^aq_+)\right)^2\right).
\end{equation}
We also get a couple of constraints
\begin{equation}
  \int_{-\infty}^\infty dx^-q_+^\dagger(x^+,x^-)=0, \quad \quad \quad  \int_{-\infty}^\infty dx^-\left(q_+^\dagger T^aq_+\right)(x^+,x^-)=0
\end{equation}
from integrating out $C_{q_-}(x^+)$ and $C_A^{-,a}(x^+)$, respectively. Note that through integration by parts we could have had to integrate out $C_{q_+}(x^+)$ as well to get the complex conjugate of the first constraint. This seems to imply that states with $p^+=0$ are not allowed in the Hilbert space. The second constraint implies that states in the Hilbert space must be invariant under {\it global} $SU(N_c)$ transformations in the null plane.

From the above expression for the Lagrangian, we can identify that the fundamental commutation relation and the null plane Hamiltonian are
\begin{align}
  \{(q_+)_{a,m}(x^-),(q_+^\dagger)_{b,n}(y^-)\} = \frac{1}{\sqrt{2}}\delta_{ab}\delta_{mn}\delta(x^--y^-), \\
  P^- = \int_{-\infty}^\infty dx^-\left(q_+^\dagger \frac{m_q^2}{\sqrt{2}i\partial^+}q_++g^2\left(\frac{1}{\partial^+}(q_+^\dagger T^aq_+)\right)^2\right),
\end{align}
where the subscripts $a,b$ represent fundamental $SU(N_c)$ color indices and $m,n$ represent flavor indices for $N_f$ flavors. It is clear from this expression that the Hamiltonian is composed of a kinetic term $P_0^-$ and a potential term $V$ proportional to $g^2$. If we write out the inverse derivatives in the potential, we get 
\begin{align}
\nonumber V &= \frac{g^2}{4}\int_{-\infty}^\infty dx^-dy^-dz^-\epsilon(x^--y^-)\epsilon(x^--z^-)(q_+^\dagger T^aq_+)(y^-)(q_+^\dagger T^aq_+)(z^-) \\
  &= \frac{g^2}{2}\int_{-\infty}^\infty dy^-dz^-(\pi\delta(0)-|y^--z^-|)(q_+^\dagger T^aq_+)(y^-)(q_+^\dagger T^aq_+)(z^-).
\end{align}
The divergent $\delta(0)$ term appears problematic at first glance, but upon closer inspection it is proportional to two factors of the global $SU(N_c)$ generator $\int dx^-q_+^\dagger T^aq_+$, which by the second constraint above must vanish for any state in our Hilbert space. In fact, this is the same infrared divergence that 't Hooft encountered in his derivation of the self-energy that eventually disappeared in the final result. Thus we can neglect the divergent term to write the potential as
\begin{equation}
  V = -\frac{g^2}{2}\int_{-\infty}^\infty dx^-dy^-|x^--y^-|(q_+^\dagger T^aq_+)(x^-)(q_+^\dagger T^aq_+)(y^-).
\end{equation}
This explicitly shows the linear confining potential of $d=2$ QCD. One would get a similar result for $d=2$ QED in the null plane.

\subsubsection{Momentum Space Potentials}
While the form of the Hamiltonian found above looks fairly simple, it is not particularly useful for analyzing the lowest lying energy states. To do this, we need to convert the Hamiltonian into a momentum space representation. The quark fields can be written as
\begin{equation}
  q_+(x) = \int_0^\infty \frac{dk^+}{4\pi k^+}\sqrt{\sqrt{2}k^+}\left(b(k)e^{-ik^+x^-}+d^\dagger(k)e^{ik^+x^-}\right),
\end{equation}
where $b(k)$ and $d(k)$ are the annihilation operators for the quark and antiquark, respectively. Note that these operators do not have a spin index because in $d=2$ spacetime dimensions there are only 2 components to a Dirac spinor, and one of them was integrated out in the process of getting to the null plane Hamiltonian. The factor of $\sqrt{\sqrt{2}k^+}$ is a remnant of the plane wave spinors that act as solutions to the Dirac equation. The fundamental anticommutation relations between the creation and annihilation operators are given by 
\begin{equation}
  \{b_{a,m}(k),b_{b,n}^\dagger(q)\} = \{d_{a,m}(k),d_{b,n}^\dagger(q)\} = 4\pi k^+\delta_{ab}\delta_{mn}\delta(k^+-q^+),
\end{equation}
where $a,b$ and $m,n$ continue to represent color and flavor indices, respectively.

We will not go through the derivation of how to get the expression for $P^-$ in terms of the $b(k)$ and $d(k)$ operators. Rather, we will categorize the different terms that appear in this expression and briefly discuss each one. In momentum space, we organize the Hamiltonian into the following terms:
\begin{equation}
  P^- = K+V_M+V_B+V_g+V_s+V_2-\Omega.
  \label{Pminus}
\end{equation}
$K$ is the free particle kinetic term, given by
\begin{equation}
  K = \int_0^\infty \frac{dk}{4\pi k}\left(b^\dagger(k)\frac{m_q^2}{2k}b(k)+d^\dagger(k)\frac{m_q^2}{2k}d(k)\right).
\end{equation}
Note that in this expression and in all of the ones to follow we are suppressing the superscript $+$ on the momenta. Unless otherwise stated, all momenta without a superscript now refers to the $+$ component. Also recall that $m_q$ can in principle be a real, diagonal matrix in flavor space. This term simply adds a factor of $\frac{m_q^2}{2k}$ to the energy for each particle and antiparticle in a given state vector.

The constant $\Omega$ is the formally divergent vacuum state energy, with contributions from both the free particle Hamiltonian and the potential. Since it is a constant and only contributes to the overall phase of a given state, we will neglect it going forward.

The first term in the potential, $V_M$, is given by 
\begin{align}
\nonumber V_M =& -\frac{g^2}{2}\left(\prod_{i=1}^{4}\int_0^\infty \frac{dk_i}{2\pi\sqrt{ 2k_i}}\right)\mathscr{P}\left[\frac{1}{(k_2-k_3)^2}\right](2\pi)\delta(k_1+k_2-k_3-k_4) \\
  & *\left(b_{a,m}^\dagger(k_1)d_{a,n}^\dagger(k_2)d_{b,n}(k_3)b_{b,m}(k_4)-\frac{1}{N_c}b_{a,m}^\dagger(k_1)d_{b,n}^\dagger(k_2)d_{b,n}(k_3)b_{a,m}(k_4)\right),
\end{align}
where $\mathscr{P}$ refers to the principle value of quantity in brackets. This term is labeled $V_M$ because, as we will discuss below, in the large-$N_c$ limit it is the only part of the potential (besides $V_2$) that contributes in the meson sector. We can see that the leading piece corresponds to a quark-antiquark pair in a color singlet being annihilated and recreated. If we suppose that this term represents some intermediate state that forms and decays during this process, then it would be a color singlet that has two flavor indices, just like a meson. This is a further indication that this term is specifically important for the description of mesons.

The term $V_B$ is given by 
\begin{align}
\nonumber V_B =&~ \frac{g^2}{4}\left(\prod_{i=1}^{4}\int_0^\infty \frac{dk_i}{2\pi\sqrt{ 2k_i}}\right)\mathscr{P}\left[\frac{1}{(k_2-k_3)^2}\right](2\pi)\delta(k_1+k_2-k_3-k_4) \\
  & *\left(b_{a,m}^\dagger(k_1)b_{b,n}^\dagger(k_2)b_{a,n}(k_3)b_{b,m}(k_4)-\frac{1}{N_c}b_{a,m}^\dagger(k_1)b_{b,n}^\dagger(k_2)b_{b,n}(k_3)b_{a,m}(k_4)+(b\rightarrow d)\right).
\end{align}
This term is labeled $V_B$ because in the large-$N_c$ limit it is the only part of the potential (besides $V_2$) that contributes in the baryon (or antibaryon) sector. This is also the only term that counts either quarks or antiquarks exclusively; that is, these are the only 4-quark terms that do not have both $b$'s and $d$'s in them. For mesons, however, this term is subleading in the $\frac{1}{N_c}$ expansion. In terms of diagrams, this seems to correspond to two (anti)quarks exchanging a gluon and swapping color charge in the process.

The term $V_g$ is given by 
\begin{align}
\nonumber V_g =&~ \frac{g^2}{2}\left(\prod_{i=1}^{4}\int_0^\infty \frac{dk_i}{2\pi\sqrt{ 2k_i}}\right)\frac{1}{(k_1+k_2)^2}(2\pi)\delta(k_1+k_2-k_3-k_4) \\
  & *\left(b_{a,m}^\dagger(k_1)d_{b,m}^\dagger(k_2)d_{b,n}(k_3)b_{a,n}(k_4)-\frac{1}{N_c}b_{a,m}^\dagger(k_1)d_{a,m}^\dagger(k_2)d_{b,n}(k_3)b_{b,n}(k_4)\right).
\end{align}
This term is labeled $V_g$ because it corresponds to a quark-antiquark pair in a {\it flavor} singlet but a color adjoint being annihilated and recreated. This suggests that the term represents some intermediate state with the same quantum numbers as a gluon, and the kinematical factor of $\frac{1}{(k_1+k_2)^2}$ appears to be the gluon propagator in light-cone gauge. It is subleading in the $\frac{1}{N_c}$ expansion.

The term $V_s$ is given by 
\begin{align}
\nonumber V_s =&~ \frac{g^2}{2}\left(\prod_{i=1}^{4}\int_0^\infty \frac{dk_i}{2\pi\sqrt{ 2k_i}}\right)\frac{1}{(k_1+k_2)^2}(2\pi)\delta(k_1+k_2+k_3-k_4)\left(\delta_{ad}\delta_{bc}-\frac{1}{N_c}\delta_{ab}\delta_{cd}\right) \\
  & *\left(b_{a,n}^\dagger(k_1)d_{b,n}^\dagger(k_2)(b_{c,m}^\dagger(k_3)b_{d,m}(k_4)-d_{d,m}^\dagger(k_3)d_{c,m}(k_4))+(h.c.)\right).
\end{align}
This term is labeled $V_s$ because it is the only term that breaks quark+antiquark number symmetry, allowing for the existence of sea quarks distinct from valence quarks. This part of the potential contributes to diagrams with virtual quark-antiquark loops in them. It is subleading in the $\frac{1}{N_c}$ expansion, as is known from the counting of $N_c$ factors in the diagrammatic expansion.

Finally, the term $V_2$ is given by
\begin{equation}
\label{V2}
  V_2 = -g^2\frac{N_c^2-1}{2\pi N_c}\int_0^\infty \frac{dk}{4\pi k}\left(b^\dagger(k)\frac{1}{2k}b(k)+d^\dagger(k)\frac{1}{2k}d(k)\right).
\end{equation}
This term is labeled $V_2$ since it has the only 2-quark terms in the potential. It corresponds to the self-energy of a quark, and its only effect is to shift the squared quark masses by a factor of $-g^2\frac{N_c^2-1}{2\pi N_c}$. It survives in the large-$N_c$ limit.

\subsection{Large-$N_c$ Limit}

Taking the large-$N_c$ limit in the Hamiltonian formalism corresponds to the vanishing of some terms in the (anti)commutators between field operators. This is because the anticommutator between two field operators includes a Kronecker delta for the color indices. The sum over a Kronecker delta is given by $\sum_a\delta_{ab}=1$, while the sum itself goes as $N_c$, so the delta function can be thought of as going as 1/${N_c}$. Thus the fundamental anticommutator between quark fields would seem to vanish in the large-$N_c$ limit, but this is not strictly true, such as in cases where the quark fields are guaranteed to have the same color index. To clearly see which commutators survive the large-$N_c$ limit, we need to work strictly with color singlet operators.

For the meson sector, the most basic color singlet operators we will need are of the form $b^\dagger b$, $d^\dagger d$, $b^\dagger d^\dagger$, and $d b$. For the first two operators, the commutators with a single field operator all have the form
\begin{equation}
  [b_{b,\ell}^\dagger(k) b_{b,m}(q), b_{a,n}^\dagger(p)] = 4\pi p~\delta_{mn}\delta(q-p)b_{a,\ell}^\dagger(k),
\end{equation}
which is order $1$ in the large-$N_c$ expansion. Thus all of their commutators will be $\mathscr{O}(1)$ and we can consider these operators to be of $\mathscr{O}(1)$ as well. The latter two operators, $b^\dagger d^\dagger$ and $d b$, are a bit more complicated. Between the two of them, we have 
\begin{align}
\nonumber [d_{b,m_2}(q_2)b_{b,m_1}(q_1), b_{a,n_1}^\dagger(k_1) d_{a,n_2}^\dagger(k_2)] =&~ N_c(4\pi k_1\delta_{m_1n_1}\delta(k_1-q_1))(4\pi k_2\delta_{m_2n_2}\delta(k_2-q_2)) \\
\nonumber & -(4\pi k_2\delta_{m_2n_2}\delta(k_2-q_2))b_{a,m_1}^\dagger(k_1)b_{a,n_1}(q_1) \\
  & -(4\pi k_1\delta_{m_1n_1}\delta(k_1-q_1))d_{a,m_2}^\dagger(k_2)d_{a,n_2}(q_2),
\end{align}
whose leading term is of $\mathscr{O}(N_c)$. However, the commutators with the former two operators is of $\mathscr{O}(1)$. For what follows, we will define the operator
\begin{align}
 L_{mn}(p,x) &= \frac{1}{\sqrt{4\pi N_cx(1-x)}}d_{a,n}(p(1-x))b_{a,m}(p x), \\
 [L_{m_1m_2}(q,y), L_{n_1n_2}^\dagger(k,x)] &= 4\pi k\delta_{m_1n_1}\delta_{m_2n_2}\delta(k-q)\delta(x-y)-\mathscr{O}(\frac{1}{N_c}).
\end{align}
The letter $L$ is chosen because this operator can be thought of as a generator (or annihilator) of a quark-antiquark loop in a color singlet configuration. With this definition, we now have all of the color singlet field operators we need for the meson sector ($b^\dagger b$, $d^\dagger d$, $L$, and $L^\dagger$) which scale as $\mathscr{O}(1)$ operators to leading order in the large-$N_c$ expansion.

The 't Hooft model is $d=2$ QCD in the large-$N_c$ limit such that the coupling constant $g \sim \frac{1}{\sqrt{N_c}}$. The potential is proportional to $g^2$, so the operators must go as $N_c$ or better to survive the large-$N_c$ limit. The only way to achieve this with a four-operator color singlet term is with something like $(b^\dagger d^\dagger)(d b)\sim N_cL^\dagger L$, which exists only in the $V_M$ term. The $V_2$ is also proportional to $g^2N_c$ as is evident from its definition above. Thus the null plane Hamiltonian for the 't Hooft model can be written as 
\begin{align}
 \lim\limits_{N_c\rightarrow\infty}P^- =& \lim\limits_{N_c\rightarrow\infty}(K^*+V_M), \\
 \lim\limits_{N_c\rightarrow\infty}K^* =&~ K+V_2 = \int_0^\infty \frac{dk}{4\pi k}\left(b^\dagger(k)\frac{M_q^2}{2k}b(k)+d^\dagger(k)\frac{M_q^2}{2k}d(k)\right), \\
\nonumber \lim\limits_{N_c\rightarrow\infty}V_M =& -\xi^2\frac{\pi}{N_c}\left(\prod_{i=1}^{4}\int_0^\infty \frac{dk_i}{2\pi\sqrt{ 2k_i}}\right)\mathscr{P}\left[\frac{1}{(k_2-k_3)^2}\right](2\pi)\delta(k_1+k_2-k_3-k_4) \\
\nonumber & *\left(b_{a,m}^\dagger(k_1)d_{a,n}^\dagger(k_2)d_{b,n}(k_3)b_{b,m}(k_4)\right) \\
  =& -\xi^2\int_0^\infty \frac{dk}{4\pi k}\int_0^1 dx\int_0^1 dy~\mathscr{P}\left[\frac{1}{(x-y)^2}\right]\frac{1}{2k}L_{mn}^\dagger(k,x)L_{mn}(k,y),
\end{align}
where $\xi^2=\frac{g^2N_c}{2\pi}$, which remains finite in the large-$N_c$ limit, and $M_q^2=m_q^2-\xi^2$.

\subsection{The Meson Wavefunction}
The meson state vector is given by 
\begin{align}
\nonumber \left|\psi, p\right> &= \int_0^\infty \frac{dk_1}{4\pi k_1}\int_0^\infty \frac{dk_2}{4\pi k_2}\Psi_{mn}(k_1,k_2;p)\frac{1}{\sqrt{N_c}}b_{a,m}^\dagger(k_1)d_{a,n}^\dagger(k_2)\left|0\right> \\
  &= \int_0^1 dx~\psi_{mn}(x)L_{mn}^\dagger(p,x)\left|0\right>, \\
  \Psi_{mn}(k_1,k_2;p) &= 4\pi k\delta(k-p)\sqrt{4\pi x(1-x)}\psi_{mn}(x),
\end{align}
where $\Psi_{mn}(k_1,k_2;p)$ is the total meson wavefunction, $k=k_1+k_2$ is the total meson momentum, $x=k_1/k$ is the fractional momentum of the quark ($1-x$ for the antiquark), and $\left|0\right>$ is the vacuum state. Since the Hamiltonian commutes with the momentum operator, this state is an eigenstate of $P^+$ with eigenvalue $p$. The state is normalized to $\left<\psi, q|\psi, p\right>=4\pi p\delta(p-q)$, so the wavefunction must have a normalization of
\begin{equation}
  \int_0^1 dx~\psi_{mn}^\dagger(x)\psi_{mn}(x) = 1.
\end{equation}

This state is a bound state of the null plane Hamiltonian, which means it must satisfy the equation
\begin{equation}
  \lim\limits_{N_c\rightarrow\infty}P^-\left|\psi, p\right> = \lim\limits_{N_c\rightarrow\infty}(K^*+V_M)\left|\psi, p\right> = \frac{\mu^2}{2p}\left|\psi, p\right>,
\end{equation}
where $\mu$ is the mass of the meson. Using the definitions and commutation relations above, we can apply the Hamiltonian to the state to get
\begin{align}
  \lim\limits_{N_c\rightarrow\infty}K^*\left|\psi, p\right> &= \frac{1}{2p}\int_0^1 dx~\left(\frac{M_m^2}{x}+\frac{M_n^2}{1-x}\right)\psi_{mn}(x)L_{mn}^\dagger(p,x)\left|0\right>, \\
  \lim\limits_{N_c\rightarrow\infty}V_M\left|\psi, p\right> &= \frac{1}{2p}\int_0^1 dx\left(-\xi^2\int_0^1 dy~\mathscr{P}\left[\frac{1}{(x-y)^2}\right]\psi_{mn}(y)\right)L_{mn}^\dagger(p,x)\left|0\right>,
\end{align}
where $M_n^2$ is the shifted mass squared for a quark of flavor $n$. Putting these two equations together, we get an equation for the meson mass of
\begin{equation}
  \left(\frac{M_m^2}{x}+\frac{M_n^2}{1-x}\right)\psi_{mn}(x)-\xi^2\int_0^1 dy~\mathscr{P}\left[\frac{1}{(x-y)^2}\right]\psi_{mn}(y)=\mu^2\psi_{mn}(x).
\end{equation}
This is precisely the wave equation that 't Hooft obtained in his 1974 paper. From here, we can numerically solve this equation to find both the wavefunctions and their corresponding mass values in the meson spectrum.

\section{Discrete Light-Cone Quantized Hamiltonian}
\label{sec:DLCQ}

\subsection{Overview of DLCQ}
While 1+1d QCD in the large-$N_c$ limit is exactly solvable, the same cannot necessarily be said of QCD with a finite number of colors. Thus we will need a way to solve the theory approximately in a way that makes it computationally tractable. Since there is no clear perturbative expansion that can be formulated to obtain an entanglement entropy, we will instead discretize the theory to render the Hilbert space finite. The bulk of the investigation into VS and SQ entanglement will be done using discrete light-cone quantization (DLCQ), using the same null plane Hamiltonian $P^-$ defined in Eq.~\eqref{Pminus} of Section \ref{sec:tHooft}. DLCQ is discussed in detail in \cite{Brodsky:1997de, Brodsky:2004tq, Heinzl:2000ht}, with specific applications to $1+1$ QCD given in \cite{Hornbostel:1988fb, Hornbostel:1988ne, Burkardt:1989wy}. In DLCQ, we discretize momentum space of quarks to half-integers with spacing $\delta p$. The Hamiltonian $P^-$ can be separated into smaller terms based on eigenstates of total momentum $P^+$. Since this momentum is strictly positive, the total momentum $P_{tot}^+=K~\delta p$ provides a natural cutoff for the constituent particles, so no separate large momentum cutoff is necessary. In addition, $\mathcal{M}^2=2P^+P^-$ has no explicit dependence on $\delta p$, but it does enter implicitly through $K$, so the continuum limit is approached as $K$ increases toward infinity. Notes about the specifics of the simulation architecture can be found in Appendix \ref{Algorithm}.

\subsection{The Hamiltonian}
Instead of working with the Hamiltonian directly, we will use the Poincar\'{e} invariant mass squared operator given by
\begin{equation}
  \mathcal{M}^2 = 2P^+P^- =  \mathcal{K}+\mathcal{V}_M+\mathcal{V}_B+\mathcal{V}_g+\mathcal{V}_s-\Omega.
\end{equation}
All of these terms are directly analogous to the terms of the continuum null plane Hamiltonian defined in Section \ref{sec:tHooft}. $\mathcal{K}$ is the free particle kinetic term, given by
\begin{equation}
  \mathcal{K} = n_p\sum_{n}\left(b^\dagger_{a,f,n}\frac{m_q^2}{n}b_{a,f,n}+d^\dagger_{a,f,n}\frac{m_q^2}{n}d_{a,f,n}\right).
\end{equation}
With $\xi^2=\frac{g^2N_c}{2\pi}$, the discretized potentials are
\begin{align}
\nonumber \mathcal{V}_M =& -\frac{\xi^2n_p}{N_c}\sum_{n_1,n_2,n_3,n_4}\mathscr{P}\left[\frac{1}{(n_2-n_3)^2}\right]\delta_{n_1+n_2,n_3+n_4} \\
  & *\left(b_{a,f,n_1}^\dagger d_{a,g,n_2}^\dagger d_{b,g,n_3} b_{b,f,n_4}-\frac{1}{N_c}b_{a,f,n_1}^\dagger d_{b,g,n_2}^\dagger d_{b,g,n_3} b_{a,f,n_4}\right), \\
\nonumber \mathcal{V}_B =&~ \frac{\xi^2 n_p}{2N_c}\sum_{n_1,n_2,n_3,n_4}\mathscr{P}\left[\frac{1}{(n_2-n_3)^2}\right]\delta_{n_1+n_2,n_3+n_4} \\
  & *\left(b_{a,f,n_1}^\dagger b_{b,g,n_2}^\dagger b_{a,g,n_3} b_{b,f,n_4}-\frac{1}{N_c}b_{a,f,n_1}^\dagger b_{b,g,n_2}^\dagger b_{b,g,n_3}b_{a,f,n_4}+(b\rightarrow d)\right), \\
\nonumber \mathcal{V}_g =&~ \frac{\xi^2n_p}{N_c}\sum_{n_1,n_2,n_3,n_4}\frac{1}{(n_1+n_2)^2}\delta_{n_1+n_2,n_3+n_4} \\
  & *\left(b_{a,f,n_1}^\dagger d_{b,f,n_2}^\dagger d_{b,g,n_3}b_{a,g,n_4}-\frac{1}{N_c}b_{a,f,n_1}^\dagger d_{a,f,n_2}^\dagger d_{b,g,n_3}b_{b,g,n_4}\right), \\
\nonumber \mathcal{V}_s =&~ \frac{\xi^2n_p}{N_c}\sum_{n_1,n_2,n_3,n_4}\frac{1}{(n_1+n_2)^2}\delta_{n_1+n_2+n_3,n_4}\left(\delta_{ad}\delta_{bc}-\frac{1}{N_c}\delta_{ab}\delta_{cd}\right) \\
  & *\left(b_{a,f,n_1}^\dagger d_{b,f,n_2}^\dagger (b_{c,g,k_3}^\dagger b_{d,g,n_4}-d_{d,g,k_3}^\dagger d_{c,g,n_4})+(h.c.)\right).
\end{align}
Since in our numerical work we are always using vector spaces that share the same total $+$-momentum, there is no practical difference between working with the Hamiltonian or working with $\mathcal{M}^2$. However, the single particle energy levels in $\mathcal{M}^2$ stay discrete when extrapolated to the continuum, so it is much easier to identify these states in $\mathcal{M}^2$ when varying the total number of momentum units $K$.

There is no term analogous to $V_2$ from Eq.~\eqref{V2} in our discrete version of the theory. This is because we choose to define the principle value for the discrete potential to give the same result as it would in the continuum, specifically so that
\begin{equation}
\sum_{n_3,n_4=1/2}^{\infty}\mathscr{P}\left[\frac{-1}{(n_2-n_3)^2}\right]\delta_{n_1+n_2,n_3+n_4}=\frac{1}{n_1}+\frac{1}{n_2},
\end{equation}
for positive half-integers $n_1$ and $n_2$. This leads to a value of $f_n$ as defined in Appendix \ref{PrincipalValue} of
\begin{equation}
f_n = \sum_{k=1}^{n-1/2}\frac{1}{k^2}+\frac{1}{n}.
\end{equation}
The extra factor of $\frac{1}{n_1}+\frac{1}{n_2}$ coming from this principle value cancels out the $V_2$ term in the Hamiltonian. This prescription is necessary in order to get a more accurate mass spectrum as the mass becomes small compared to the coupling strength $g$. With the more conventional choice of $\frac{\pi^2}{6}$ for the principle value, we would see some negative $m^2$ values emerge as the potential became stronger, which is a result of the zero modes that appear when the quarks are massless becoming states with negative $m^2$ values upon discretization. Our prescription for the principal value guarantees that the zero modes in the theory with massless quarks will continue to be zero modes in DLCQ, and all of the eigenstates of $\mathcal{M}^2$ do appear to have physically sound masses with this choice. Details about this choice of principle value are given in Appendix \ref{PrincipalValue}.

The diagonal parts of the Hamiltonian applied to a color singlet vector are given by
\begin{align}
\nonumber \mathcal{V}_0\left|\psi,\alpha\right> =&~ \frac{\xi^2n_p}{N_c}\sum_{I,J}\frac{\delta_{f_If_J}}{(n_I+n_J)^2}\left(\left|\psi,\alpha\right>-\left|\psi,\Delta_{IJ}(\alpha)\right>\right) \\
\nonumber & -\frac{\xi^2n_p}{N_c}\sum_{I,I'}\frac{\theta_\times(n_{I'}-n_I)\delta_{f_If_{I'}}}{(n_{I'}-n_I)^2}\left(\left|\psi,\alpha\right>-\frac{1}{N_c}\left|\psi,X^b_{II'}(\alpha)\right>\right) \\
\nonumber & -\frac{\xi^2 n_p}{N_c}\sum_{J,J'}\frac{\theta_\times(n_{J'}-n_J)\delta_{f_Jf_{J'}}}{(n_{J'}-n_J)^2}\left(\left|\psi,\alpha\right>-\frac{1}{N_c}\left|\psi,X^d_{JJ'}(\alpha)\right>\right) \\
  & +\xi^2n_p\left(1-\frac{1}{N^2_c}\right)(F_q+F_{\bar{q}})\left|\psi,\alpha\right>.
\end{align}
with $F_q$ and $F_{\bar{q}}$ defined in Appendix \ref{PrincipalValue} with the $f_n$ discussed above.

\section{Valence-Sea Entanglement}
\label{sec:VSE}

This section details the construction of the valence-sea entanglement for various hadrons. The initial state $\left|0\right>$ is taken to be the light-cone vacuum, while the VS entanglement operator is a product of multiple $Q_i$ operators from single-quark entanglement, one for each valence quark. All of our results are obtained using DLCQ, so all of our expressions listed here are given using operators in discretized 1+1d momentum space. The continuum analog can be obtained fairly straightforwardly by starting with the basic expression for the density matrix using the definitions for $Q_i$ and $\bar{Q}_i$ in Eqs.~(\ref{Qdef}, \ref{Qdef2}). Many of the analytic results derived in this section carry over to the continuum theory.

\subsection{VS Entanglement for Mesons}
The simplest example of a valence-sea density matrix is that of a meson where the quark and antiquark have different flavors. This serves as the 1+1d analog of the $\pi^+$ meson, where the valence particles have flavors $u$ for the quark and $d$ for the antiquark. The VS density matrix for the $\pi^+$ analog state with total momentum $K$ is given by 
\begin{align}
\nonumber \rho_{\pi^+} &= \mathcal{N}~\mathrm{Tr}_1\left[Q_u\bar{Q}_d\left|\pi^+,K\right>_1\left|0\right>_2\left<0\right|_2\left<\pi^+,K\right|_1Q_d\bar{Q}_u\right] \\
\label{pi+dens}
&= \mathcal{N}\sum_{a,b,c,d}\sum_{k_i}f_{k_1k_2k_3k_4,abcd}\left(b^\dagger_{k_1,u,a}d^\dagger_{k_2,d,b}\left|0\right>\left<0\right|d_{k_3,d,c}b_{k_4,u,d}\right),
\end{align}
with discrete momenta $k_i$, color indices $a,b,c,d$ for $SU(N_c)$ ranging from $1$ to $N_c$, and a normalization constant $\mathcal{N}$. The $Q$ operators in discrete momentum space are defined by 
\begin{align}
 Q_i &= \sum_{k,c}(b^\dagger_{k,i,c,2}b_{k,i,c,1}+d_{k,i,c,2}d^\dagger_{k,i,c,1}), \\
  \bar{Q}_i &= \sum_{k,c}(b_{k,i,c,2}b^\dagger_{k,i,c,1}+d^\dagger_{k,i,c,2}d_{k,i,c,1}),
\end{align}
and $f$ above is defined to be
\begin{equation}
  f_{k_1k_2k_3k_4,abcd} = \left<\pi^+,K\right|b^\dagger_{k_4,u,d}d^\dagger_{k_3,d,c}d_{k_2,d,a}b_{k_1,u,a}\left|\pi^+,K\right>.
\end{equation}
Since the meson state is a color singlet, only color singlet configurations of the creation and annihilation operators in $f$ will give nonzero contributions. The easiest way to get the color singlet configurations is to get the irreducible representations (irreps) of the creation operators and the annihilation operators separately, and then join the matching irreps in as many ways as possible to get the number of singlets. In the case of the $\pi^+$ meson, we have two pairs of operators of the forms $d b$ and $b^\dagger d^\dagger$, which both form into a singlet and an adjoint representation. This originates from the group theory of $SU(N_c)$, where the quark in the fundamental representation and the antiquark in the antifundamental representation combine so that $N_c\otimes\bar{N_c}=1\oplus(N_c^2-1)$, where the 1 on the right-hand side is the singlet and the $N_c^2-1$ is the adjoint representation. For $N_c=3$, this reduces to the more familiar form $3\otimes\bar{3}=1\oplus8$. The singlet is formed by matching the two color indices of the operators together and summing over them, while the adjoint is formed from every configuration of the color indices such that it commutes with the conjugate of the singlet operator. This gives us two singlets overall, one for the combination of the two singlet representations from the pairs of operators, and one for the singlet combination of the two adjoint representations. The matrix element $f$ can be decomposed into two terms corresponding to the singlet and adjoint unification terms
\begin{align}
  \Pi^S_{abcd} &= \frac{\delta_{ab}\delta_{cd}}{N_c}, \\
  \Pi^A_{abcd} &= \left(\delta_{ad}\delta_{bc}-\frac{1}{N_c}\delta_{ab}\delta_{cd}\right), \\
  f^{S,A}_{k_1k_2k_3k_4} &= \Pi^{S,A}_{abcd}f_{k_1k_2k_3k_4,abcd},
\end{align}
with an implicit sum over all color indices in the last equation. This means that $f$ can be written as
\begin{equation}
\label{ftenSA}
f_{k_1k_2k_3k_4,abcd} = \Pi^S_{abcd}f^S_{k_1k_2k_3k_4}+\frac{1}{N^2_c-1}\Pi^A_{abcd}f^A_{k_1k_2k_3k_4}.
\end{equation}
This implies that the VS density matrix can be separated into an average of two density matrices 
\begin{equation}
\rho_{\pi^+}=\omega^S(\Pi^S\otimes\rho_S)+\frac{1}{N^2_c-1}\omega^A(\Pi^A\otimes\rho_A),
\end{equation}
where $\rho_{S,A}$ are density matrices in momentum space, while $\omega^S$ and $\omega^A=1-\omega^S$ form a Bernoulli distribution, representing the chances of measuring the valence quarks to be in either a singlet or an adjoint representation.

In order to get the Von Neumann entropy of the VS density matrix in terms of $f^S$ and $f^A$, we need to know what matrix multiplication looks like in terms of the color and momentum tensor $f$. If we have two density matrices of the same form as that of the VS density matrix in Eq.~\eqref{pi+dens}, then $\rho_1.\rho_2$ becomes 
\begin{equation}
(f_1.f_2)_{k_1k_2k_3k_4,abcd}=\sum_{A,B}\sum_{q_1,q_2}(f_1)_{k_1k_2q_2q_1,abBA}(f_2)_{q_1q_2k_3k_4,ABcd}
\end{equation}
at the level of the tensors $f_1,f_2$. According to this definition of the product of tensors, the two color tensors $\Pi^S$ and $\Pi^A$ have the properties $\Pi^{S,A}.\Pi^{S,A}=\Pi^{S,A}$ and $\Pi^S.\Pi^A=0$, which makes then projection operators for two distinct vector subspaces in color space. For the remaining degrees of freedom in momentum space, we can define matrix multiplication and the trace for these distributions as 
\begin{align}
  (A.B)_{k_1k_2k_3k_4} &= \sum_{q_1,q_2}A_{k_1k_2q_2q_1}B_{q_1q_2k_3k_4}, \\
  \mathrm{Tr}[A] &= \sum_{k_1,k_2}A_{k_1k_2k_2k_1}.
\end{align}
With these definitions, we can define the matrix logarithm for $f^{S,A}$ and therefore the Von Neumann entropy of the density matrix in Eq.~\eqref{pi+dens}. The VS entanglement entropy for the $\pi^+$ meson can be written as
\begin{align}
  S_{\pi^+} &= \mathcal{N}\left(-\mathrm{Tr}\left[f^S.\log f^S\right]-\mathrm{Tr}\left[f^A.\log f^A\right]+ \mathrm{Tr}\left[f^A\right]\log(N^2_c-1)\right)+\log(\mathcal{N}^{-1}), \\
  \mathcal{N}^{-1} &= \mathrm{Tr}\left[f^S\right]+\mathrm{Tr}\left[f^A\right].
\end{align}
This can also be cast in the more illustrative form 
\begin{align}
  \omega^{S,A} &= \mathcal{N}~\mathrm{Tr}\left[f^{S,A}\right], \\
  S_{S,A} &= \frac{1}{\mathrm{Tr}\left[f^{S,A}\right]}\left(-\mathrm{Tr}\left[f^{S,A}.\log f^{S,A}\right]\right)+\log \left(\mathrm{Tr}\left[f^{S,A}\right]\right), \\
  S_{\pi^+} &= \omega^SS_S+\omega^AS_A+\omega^A\log(N^2_c-1)-\omega^S\log\omega^S-\omega^A\log\omega^A.
\end{align}
From this form we can see that the VS entanglement entropy of the $\pi^+$ meson can be separated into three contributions, stemming from the fact that our density matrix can be decomposed into
\begin{equation}
\rho_{\pi^+}=\omega^S(\Pi^S\otimes\rho_S)+\frac{1}{N^2_c-1}\omega^A(\Pi^A\otimes\rho_A),
\end{equation}
with $\Pi_S.\Pi_A=0$. The first contribution is the average of the entropies $S_{S,A}$ of $\rho_{S,A}$, corresponding to the momentum space degrees of freedom. The second part is the average of the entropies of $\Pi_S$ and $\frac{1}{N^2_c-1}\Pi^A$, which serve as the density matrices in color space. The entropy of $\Pi_S$ is zero since it only projects one state, while the entropy of $\frac{1}{N^2_c-1}\Pi^A$ is the $\log(N^2_c-1)$ we see in the full entropy expression. Finally, the last term is the entropy of the distribution formed by $\omega^S$ and $\omega^A$. This decomposition of entropy contributions also holds for a continuous momentum variable.

Thus far, we have found expressions for the VS density matrix and entanglement entropy in terms of a set of expectation values $f$ for a meson state vector under the assumption that the valence quarks in our meson had to have different flavors. If the valence quark and antiquark share the same flavor, we can define an entanglement entropy $S_{\mathrm{sing}}$ in terms of the corresponding $f_S$ and $f_A$ tensors in much the same way as we did above for the $\pi^+$. However, the singlet density matrix $\rho_{\mathrm{sing}}$ has an extra vacuum state contribution given by
\begin{align}
\nonumber \rho_{\mathrm{sing}} &= \mathcal{N}_{\mathrm{sing}}~\mathrm{Tr}_1\left[Q_f\bar{Q}_f\left|\mathrm{sing},K\right>_1\left|0\right>_2\left<0\right|_2\left<\mathrm{sing},K\right|_1Q_f\bar{Q}_f\right] \\
&= \mathcal{N}_{\mathrm{sing}}\left(\sum_{a,b,c,d}\sum_{k_i}f_{k_1k_2k_3k_4,abcd}\left(b^\dagger_{k_1,f,a}d^\dagger_{k_2,f,b}\left|0\right>\left<0\right|d_{k_3,f,c}b_{k_4,f,d}\right)+f_{\mathrm{vac}}\left|0\right>\left<0\right|\right), \\
f_{\mathrm{vac}} &= \sum_{a,b}\sum_{k_1,k_2}\left<\mathrm{sing},K\right|d^\dagger_{k_2,f,b}d_{k_2,f,b}d^\dagger_{k_1,f,a}d_{k_1,f,a}\left|\mathrm{sing},K\right>.
\end{align}
Thus the von Neumann entropy of this density matrix is somewhat different from $S_{\mathrm{sing}}$. They are related by
\begin{equation}
  -\mathrm{Tr}\left[\rho_{\mathrm{sing}}\log(\rho_{\mathrm{sing}})\right] = \mathcal{N}_{\mathrm{sing}}\left(\mathcal{N}^{-1}\left(S_{\mathrm{sing}}-\log(\mathcal{N}^{-1})\right)-f_{\mathrm{vac}}\log(f_{\mathrm{vac}})\right)+\log(\mathcal{N}^{-1}_{\mathrm{sing}}),
\end{equation}
where $\mathcal{N}_{\mathrm{sing}}=(\mathcal{N}^{-1}+f_{\mathrm{vac}})^{-1}$ is the normalization constant including the vacuum contribution, while $\mathcal{N}$ is the normalization without this contribution. In our work, we will define $S_{\mathrm{sing}}$ without the vacuum contribution, as it does not correspond to a valence quark configuration and it will only serve to obscure the conclusions we are able to draw from our data. In principle the vacuum contribution can be important when establishing the gauge invariance of the density matrix. However, since writing down the light-cone Hamiltonian we have been working in light-cone gauge exclusively, and the vast majority of our expressions would pick up Wilson lines in any other choice of gauge, complicating the interpretation of the entropy as a measure of quark entanglement. Thus the gauge invariance consideration is of lesser importance in this work, though it could become more important in a $3+1$ dimensional QCD calculation.

\subsection{VS Entanglement for Baryons}
The process of calculating the valence-sea entanglement for baryons is largely the same as it is for mesons. The main difference is that baryons have the number of valence quarks equal to the number of colors. This means that the density matrix becomes much more complicated as the number of colors increases. It is for this reason that we will only consider 2 and 3 color baryons in this work.

In $SU(2)$, the antifundamental representation is the same as the fundamental, so the VS density matrix for $SU(2)$ baryons will look very similar to the matrix for $SU(2)$ mesons. We have for a baryon $B_{mn}$ with valence flavors $m,n$ and total momentum $K$
\begin{align}
\nonumber \rho_{B_{mn}} &= \mathcal{N}~\mathrm{Tr}_1\left[Q_mQ_n\left|B_{mn},K\right>_1\left|0\right>_2\left<0\right|_2\left<B_{mn},K\right|_1\bar{Q}_n\bar{Q}_m\right] \\
\label{B2dens}
&= \mathcal{N}\sum_{a,b,c,d}\sum_{k_i}f_{k_1k_2k_3k_4,abcd}\left(b^\dagger_{k_1,m,a}b^\dagger_{k_2,n,b}\left|0\right>\left<0\right|b_{k_3,n,c}b_{k_4,m,d}\right).
\end{align}
All of the operators and indices are defined as they were in Eq.~\eqref{pi+dens} for mesons. $f$ above for baryons is defined to be
\begin{equation}
  f_{k_1k_2k_3k_4,abcd} = \left<B_{mn},K\right|b^\dagger_{k_4,m,d}b^\dagger_{k_3,n,c}b_{k_2,n,a}b_{k_1,m,a}\left|B_{mn},K\right>.
\end{equation}
We can use the fact that the baryon states are color singlets to constrain the color structure of $f$, just as we did for mesons. In fact, since the fundamental and antifundamental representations of $SU(2)$ are the same, the representation breakdown of the operators also gives us two singlet configurations of the four-quark operator. The group theory of $SU(2)$ for combining two fundamental representations tells us that $2\otimes2=1\oplus3$, where the 1 on the right-hand side is the singlet and the $N_c^2-1$ is the triplet. Thus the $b b$ and $b^\dagger b^\dagger$ operator pairs each form a singlet and a triplet, where the triplet is analogous to the adjoint representation in the meson case. The two singlets and triplets each form one overall color singlet. These two components of $f$ are given by 
\begin{align}
  \Pi^S_{abcd} &= \frac{\delta_{ad}\delta_{bc}-\delta_{ac}\delta_{bd}}{2}, \\
  \Pi^T_{abcd} &= \frac{\delta_{ad}\delta_{bc}+\delta_{ac}\delta_{bd}}{2}, \\
  f^{S,T}_{k_1k_2k_3k_4} &= \Pi^{S,T}_{abcd}f_{k_1k_2k_3k_4,abcd},
\end{align}
with an implicit sum over all color indices in the last equation. This means that $f$ can be written as
\begin{equation}
  f_{k_1k_2k_3k_4,abcd} = \Pi^S_{abcd}f^S_{k_1k_2k_3k_4}+\frac{1}{3}\Pi^T_{abcd}f^A_{k_1k_2k_3k_4}.
\end{equation}
Once again, we see that the VS density matrix can be separated into an average of two density matrices
\begin{equation}
\rho_{B_{mn}}=\omega^S(\Pi^S\otimes\rho_S)+\frac{1}{3}\omega^T(\Pi^T\otimes\rho_T),
\end{equation}
where $\rho_{S,T}$ are density matrices in momentum space, while $\omega^S$ and $\omega^T=1-\omega^S$ form a Bernoulli distribution, representing the chances of measuring the valence quarks to be in either a singlet or a triplet representation.

The multiplication of density matrices leads to a tensor product that is exactly the same as we found in the meson case. If we have two density matrices of the same form as that of the VS density matrix in Eq.~\eqref{B2dens}, then $\rho_1.\rho_2$ becomes
\begin{equation}
(f_1.f_2)_{k_1k_2k_3k_4,abcd}=\sum_{A,B}\sum_{q_1,q_2}(f_1)_{k_1k_2q_2q_1,abBA}(f_2)_{q_1q_2k_3k_4,ABcd}
\end{equation} 
at the level of the tensors $f_1,f_2$. According to this definition of the product of tensors, the two color tensors $\Pi^S$ and $\Pi^T$ have the properties $\Pi^{S,T}.\Pi^{S,T}=\Pi^{S,T}$ and $\Pi^S.\Pi^T=0$, which makes then projection operators for two distinct vector subspaces in color space. For the remaining degrees of freedom in momentum space, we can define matrix multiplication and the trace for these distributions as 
\begin{align}
  (A.B)_{k_1k_2k_3k_4} &= \sum_{q_1,q_2}A_{k_1k_2q_2q_1}B_{q_1q_2k_3k_4}, \\
  \mathrm{Tr}[A] &= \sum_{k_1,k_2}A_{k_1k_2k_2k_1}.
\end{align}
With this definition, the VS entanglement entropy for an $N_c=2$ baryon can be written as
\begin{align}
\nonumber S_{B_{mn}} &= \mathcal{N}\left(-\mathrm{Tr}\left[f^S.\log f^S\right]-\mathrm{Tr}\left[f^T.\log f^T\right]+\mathrm{Tr}\left[f^T\right]\log(3)\right)+\log(\mathcal{N}^{-1}), \\
  \mathcal{N}^{-1} &= \mathrm{Tr}\left[f^S\right]+\mathrm{Tr}\left[f^T\right].
\end{align}
This can also be cast in the more illustrative form 
\begin{align}
  \omega^{S,T} &= \mathcal{N}~\mathrm{Tr}\left[f^{S,T}\right], \\
  S_{S,T} &= \frac{1}{\mathrm{Tr}\left[f^{S,T}\right]}\left(-\mathrm{Tr}\left[f^{S,T}.\log f^{S,T}\right]\right)+\log \left(\mathrm{Tr}\left[f^{S,T}\right]\right), \\
  S_{B_{mn}} &= \omega^SS_S+\omega^TS_T+\omega^T\log(3)-\omega^S\log\omega^S-\omega^T\log\omega^T.
\end{align}
This form, just as in the meson case, can be separated into three contributions, stemming from the fact that our density matrix can be decomposed into
\begin{equation}
\rho_{B_{mn}}=\omega^S(\Pi^S\otimes\rho_S)+\frac{1}{3}\omega^T(\Pi^T\otimes\rho_T),
\end{equation}
with $\Pi_S.\Pi_T=0$. The first contribution is the average of the entropies $S_{S,T}$ of $\rho_{S,T}$, corresponding to the momentum space degrees of freedom. The second part is the average of the entropies of the color space representations, with factors of $\log(1)=0$ from the singlet and $\log(3)$ from the triplet. Finally, the last term is the entropy of the distribution formed by $\omega^S$ and $\omega^T$. This decomposition of entropy contributions also holds for a continuous momentum variable.

The case of $SU(3)$ baryons becomes more complicated as there is a third valence quark which will combine to form more than two irreducible representations. We have for some baryon $B$ with flavors $l,m,n$ and total momentum $K$
\begin{align}
\nonumber \rho_{B_{lmn}} &= \mathcal{N}~\mathrm{Tr}_1\left[Q_lQ_mQ_n\left|B_{lmn},K\right>_1\left|0\right>_2\left<0\right|_2\left<B_{lmn},K\right|_1\bar{Q}_n\bar{Q}_m\bar{Q}_l\right] \\
  &= \mathcal{N}\sum_{a,b,c,d,e,f}\sum_{k_i}f_{k_1k_2k_3k_4k_5k_6,abcdef}\left(b^\dagger_{k_1,l,a}b^\dagger_{k_2,m,b}b^\dagger_{k_3,n,c}\left|0\right>\left<0\right|b_{k_4,n,d}b_{k_5,m,e}b_{k_6,l,f}\right).
\end{align}
Again, the operators and indices are defined as they have been previously. $f$ above is defined to be
\begin{equation}
  f_{k_1k_2k_3k_4k_5k_6,abcdef} = \left<B_{lmn},K\right|b^\dagger_{k_6,l,f}b^\dagger_{k_5,m,e}b^\dagger_{k_4,n,d}b_{k_3,n,c}b_{k_2,m,b}b_{k_1,l,a}\left|B_{lmn},K\right>.
\end{equation}
Following the procedure we have outlined previously, the group theory of $SU(3)$ for combining three fundamental representations tells us that $3\otimes3\otimes3=1\oplus8\oplus8\oplus10$. Thus we find that the operator groupings $b b b$ and $b^\dagger b^\dagger b^\dagger$ form four different irreps of $SU(3)$: one antisymmetric singlet, two different 8-dimensional adjoint representations of mixed symmetry, and a 10-dimensional fully symmetric representation. Either of the two adjoints from the $b b b$ operator can form a color singlet with any of the two adjoints from the $b^\dagger b^\dagger b^\dagger$, giving us a total of four singlets that can be made from the mixed symmetry irreps. This gives us a total of six color singlets when the symmetric and antisymmetric irreps are included. The color tensors needed for these terms are given by 
\begin{align}
  \mathcal{S}_{abcdef} &= \frac{1}{6}(\delta_{af}(\delta_{be}\delta_{cd}+\delta_{bd}\delta_{ce})+\delta_{ae}(\delta_{bf}\delta_{cd}+\delta_{bd}\delta_{cf})+\delta_{ad}(\delta_{be}\delta_{cf}+\delta_{bf}\delta_{ce})), \\
  \mathcal{A}_{abcdef} &= \frac{1}{6}(\delta_{af}(\delta_{be}\delta_{cd}-\delta_{bd}\delta_{ce})-\delta_{ae}(\delta_{bf}\delta_{cd}-\delta_{bd}\delta_{cf})-\delta_{ad}(\delta_{be}\delta_{cf}-\delta_{bf}\delta_{ce})), \\
  M^{++}_{abcdef} &= \frac{1}{2}(\delta_{af}\delta_{be}+\delta_{ae}\delta_{bf})\delta_{cd}-\mathcal{S}_{abcdef}, \\
  M^{+-}_{abcdef} &= \frac{1}{2\sqrt{3}}((\delta_{bf}\delta_{ce}-\delta_{be}\delta_{cf})\delta_{ad}+(\delta_{af}\delta_{ce}-\delta_{ae}\delta_{cf})\delta_{bd}), \\
  M^{-+}_{abcdef} &= \frac{1}{2\sqrt{3}}((\delta_{ae}\delta_{bd}-\delta_{ad}\delta_{be})\delta_{cf}+(\delta_{af}\delta_{bd}-\delta_{ad}\delta_{bf})\delta_{ce}), \\
  M^{--}_{abcdef} &= \frac{1}{2}(\delta_{af}\delta_{be}-\delta_{ae}\delta_{bf})\delta_{cd}-\mathcal{A}_{abcdef}.
\end{align}
These tensors are analogous to the $\Pi$ color tensors form before. The $+$ and $-$ superscripts on the mixed symmetry tensors denote whether it is symmetric or antisymmetric in the first two color indices (for the first superscript) or the last two indices (for the second). The six components of $f$ are then given by
\begin{align}
  f^A_{k_1k_2k_3k_4k_5k_6} &= \mathcal{A}_{abcdef}f_{k_1k_2k_3k_4k_5k_6,abcdef}, \\
  f^S_{k_1k_2k_3k_4k_5k_6} &= \mathcal{S}_{abcdef}f_{k_1k_2k_3k_4k_5k_6,abcdef}, \\
  f^{st}_{k_1k_2k_3k_4k_5k_6} &= M^{st}_{abcdef}f_{k_1k_2k_3k_4k_5k_6,abcdef},
\end{align}
for all $s,t = \pm$ and  with implicit sums over the color indices. $f$ can then be written as
\begin{equation}
  f_{k_1k_2k_3k_4k_5k_6,abcdef} = \mathcal{A}_{abcdef}f^A_{k_1k_2k_3k_4k_5k_6}+\frac{1}{8}\sum_{s,t=\pm}M^{st}_{abcdef}f^{st}_{k_1k_2k_3k_4k_5k_6}+\frac{1}{10}\mathcal{S}_{abcdef}f^S_{k_1k_2k_3k_4k_5k_6}.
\end{equation}
Unlike in previous cases, these six tensors will not split the entanglement entropy into six disconnected contributions. To see this, we can look at how these color tensors stitch together when the density matrix is multiplied to itself, which we need to know in order to apply the replica trick to derive the entanglement entropy. Matrix multiplication of baryon density matrices translates to a product of $f$'s that is defined by
\begin{equation}
(f_1.f_2)_{k_1k_2k_3k_4k_5k_6,abcdef}=\sum_{A,B,C}\sum_{q_1,q_2,q_3}(f_1)_{k_1k_2k_3q_3q_2q_1,abcCBA}(f_2)_{q_1q_2q_3k_4k_5k_6,ABCdef}.
\end{equation}
The relevant nonzero products of color tensors for density matrix multiplication are
\begin{align}
  \mathcal{A}_{abcCBA}\mathcal{A}_{ABCdef} &= \mathcal{A}_{abcdef}, \\
  \mathcal{S}_{abcCBA}\mathcal{S}_{ABCdef} &= \mathcal{S}_{abcdef}, \\
  M^{st}_{abcCBA}M^{tu}_{ABCdef} &= M^{su}_{abcdef}.
\end{align}
for all $s,t,u = \pm$. These tensors split the density matrix into three contributions corresponding to the 3 irreducible representations of $S_3$. The density matrix thus splits into three parts given by
\begin{equation}
\rho_{B_{lmn}}=\omega^\mathcal{A}(\mathcal{A}\otimes\rho_\mathcal{A})+\frac{1}{8}\omega^M(M\otimes\rho_M)+\frac{1}{10}\omega^\mathcal{S}(\mathcal{S}\otimes\rho_\mathcal{S}).
\end{equation}
Note also the four mixed symmetry tensors $M^{st}_{abcdef}$ combine in a way that resembles matrix multiplication of a $2\times2$ matrix. Thus for the fully symmetric and antisymmetric contributions we can define matrix multiplication and the trace in momentum space much like before, where 
\begin{align}
  (A.B)_{k_1k_2k_3k_4k_5k_6} &= \sum_{q_1,q_2}A_{k_1k_2k_3q_3q_2q_1}B_{q_1q_2q_3k_4k_3k_4}, \\
  \mathrm{Tr}[A] &= \sum_{k_1,k_2,k_3}A_{k_1k_2k_3k_3k_2k_1}.
\end{align}
However, for the mixed symmetry contributions there is an extra step where they must be organized into a $2\times2$ matrix matrix such that
\begin{equation}
  f^M_{k_1k_2k_3k_4k_5k_6} = 
\left(\begin{matrix}
f^{--}_{k_1k_2k_3k_4k_5k_6} & f^{-+}_{k_1k_2k_3k_4k_5k_6} \\
f^{+-}_{k_1k_2k_3k_4k_5k_6} & f^{++}_{k_1k_2k_3k_4k_5k_6}
\end{matrix}\right),
\end{equation}
and from here matrix multiplication and tracing is defined by both uniting the momentum indices as in the previous equation along with the standard multiplication and trace of the $2\times2$ array shown here.
With these definitions, the VS entanglement entropy for an $N_c=3$ baryon can be written as
\begin{align}
\nonumber S_{B_{lmn}} =&~\mathcal{N}\left(-\mathrm{Tr}\left[f^\mathcal{A}.\log f^\mathcal{A}\right]-\mathrm{Tr}\left[f^M.\log f^M\right]-\mathrm{Tr}\left[f^\mathcal{S}.\log f^\mathcal{S}\right]+\mathrm{Tr}\left[f^M\right]\log(8)+\mathrm{Tr}\left[f^\mathcal{S}\right]\log(10)\right) \\ & +\log(\mathcal{N}^{-1}), \\
  \mathcal{N}^{-1} =&~\mathrm{Tr}\left[f^\mathcal{A}\right]+\mathrm{Tr}\left[f^M\right]+\mathrm{Tr}\left[f^\mathcal{S}\right].
\end{align}
This can also be cast in the more illustrative form 
\begin{align}
  \omega^{\mathcal{A},M,\mathcal{S}} &= \mathcal{N}~\mathrm{Tr}\left[f^{\mathcal{A},M,\mathcal{S}}\right], \\
  S_{\mathcal{A},M,\mathcal{S}} &= \frac{1}{\mathrm{Tr}\left[f^{\mathcal{A},M,\mathcal{S}}\right]}\left(-\mathrm{Tr}\left[f^{\mathcal{A},M,\mathcal{S}}.\log f^{\mathcal{A},M,\mathcal{S}}\right]\right)+\log\left(\mathrm{Tr}\left[f^{\mathcal{A},M,\mathcal{S}}\right]\right), \\
  S_{B_{lmn}} &= \omega^\mathcal{A}S_\mathcal{A}+\omega^MS_M+\omega^\mathcal{S}S_\mathcal{S}+\omega^M\log(8)+\omega^\mathcal{S}\log(10)-\omega^\mathcal{A}\log\omega^\mathcal{A}-\omega^M\log\omega^M-\omega^\mathcal{S}\log\omega^\mathcal{S}.
\end{align}
Once again, we see that the VS entanglement entropy splits into averages over momentum space entropies, color space entropies, and the entropy of the $\omega$ distribution, which also holds in the continuum.

\section{Data and Results}
\label{sec:plots}

In this section we calculate the valence-sea entanglement of an assortment of hadrons given a variety of parameters such as total momentum, number of colors, and the coupling constant. To do this, we begin by constructing the discrete light-cone quantized  Hamiltonian for $1+1$d QCD and finding the eigenstates with the lowest energies. Then we form the appropriate $f$ tensors described in Section \ref{sec:VSE} by applying the eigenstates we obtained to a set of state-independent four- or six-quark operators. Finally, we arrange the $f$ tensors into matrices in momentum space, diagonalize them, and calculate $\mathrm{Tr}[f]$ and $-\mathrm{Tr}[f.\log f]$ for each color singlet. For the figures in this section, we plot the entropy against either the number of colors, total hadron momentum, or the coupling strength. The quark masses are degenerate unless otherwise specified. For the coupling strength, since the Hamiltonian separates like $2P^+H=m^2\mathcal{O}_{kin}+g^2\mathcal{O}_{pot}$ and both $m$ and $g$ have units of mass, the eigenstates only depend upon the ratio $g/m$. It is more convenient to define the variables $\xi^2=\frac{g^2N_c}{2\pi}$ and $\lambda=\frac{\xi^2}{m^2+\xi^2}$. We use $\xi^2$ because it remains constant in the large-$N_c$ limit and it removes a recurring factor of $1/2\pi$ in the calculations. The variable $\lambda$ is preferable as it ranges from $0$ to $1$, where $\lambda=0$ corresponds to a theory of free quarks while $\lambda=1$ gives us massless QCD, making it more amenable to plotting than $g/m$ which ranges from $0$ to $\infty$.

\begin{figure}
	\centering
	\includegraphics[width=0.6\textwidth]{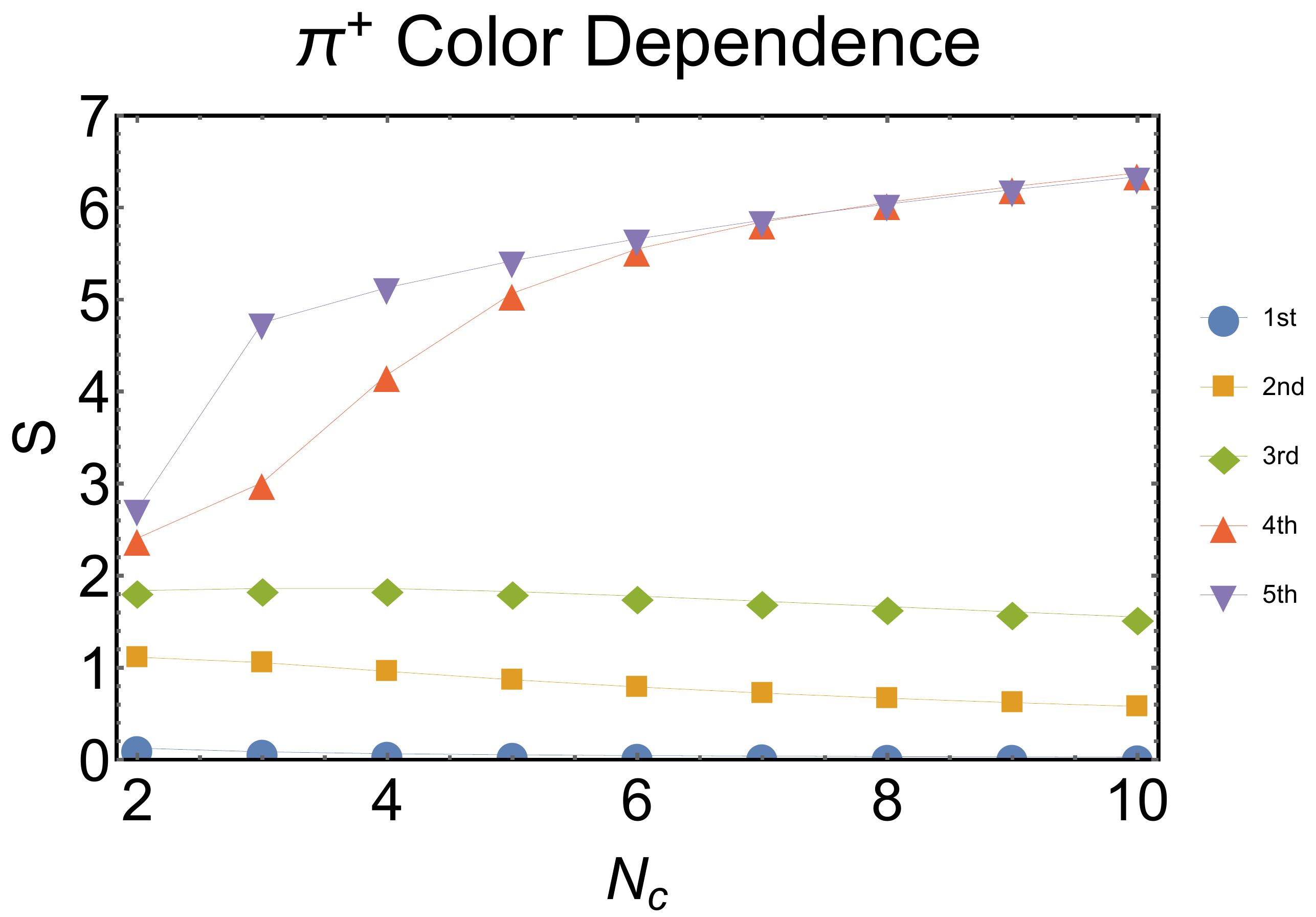}
	\includegraphics[width=0.6\textwidth]{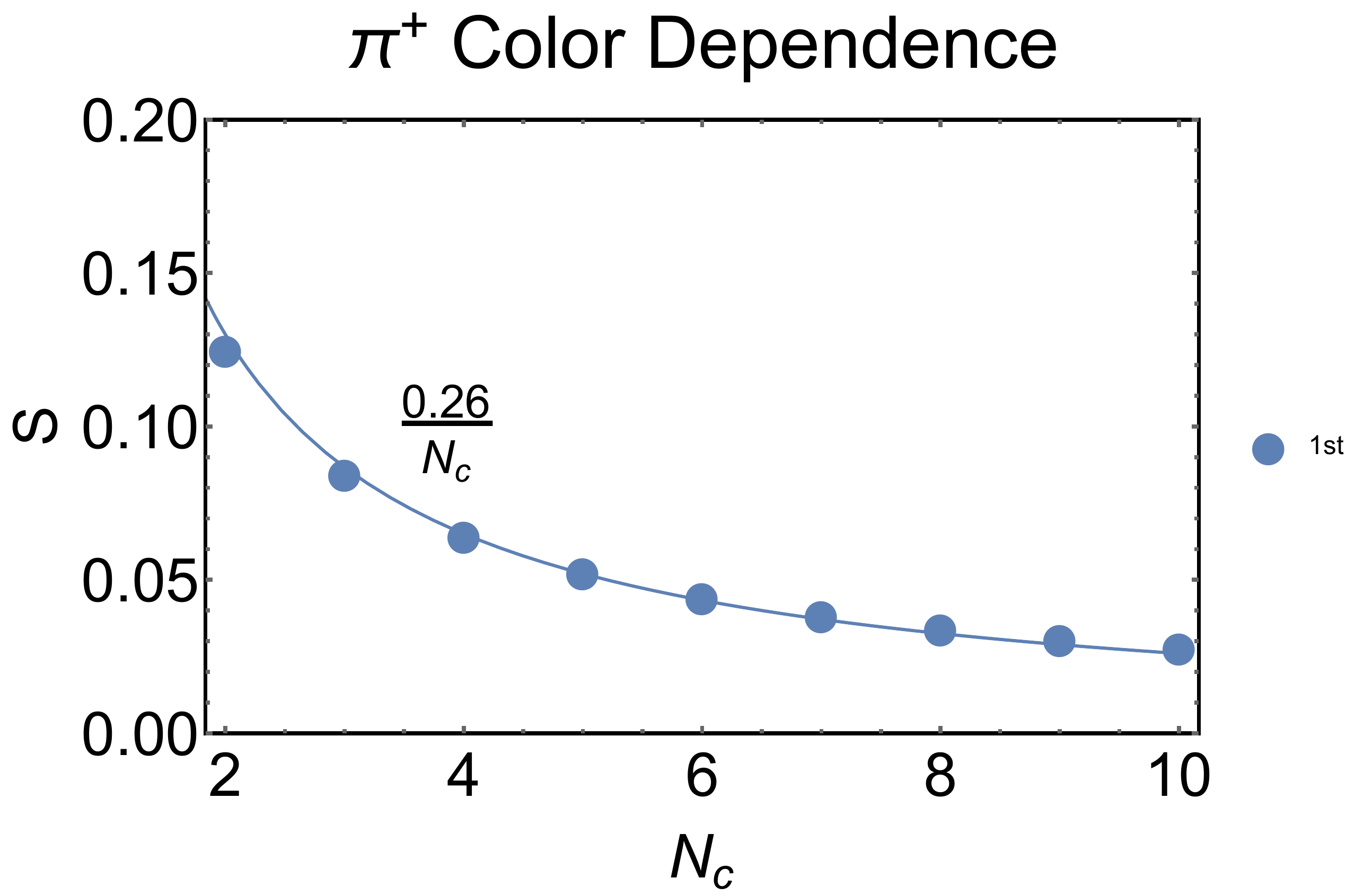}
	\caption{Entanglement entropy for the 5 lowest energy $\pi^+$ states with $N_f=2$ that have $K_{tot}=8$ units of momentum, with $m^2=\frac{g^2N_c}{2\pi}$ and $g^2\sim\frac{1}{N_c}$. The lines connecting data points in the first plot are merely for visual effect and do not communicate additional data. The second plot shows the entanglement entropy of the ground state meson with a $1/N_c$ fit. The theoretical maximum entropy is $S_{max}=2~\mathrm{Log}_2(6N_c)$.}
	\label{pi+color}
\end{figure}

\begin{figure}
	\centering
	\includegraphics[width=0.6\textwidth]{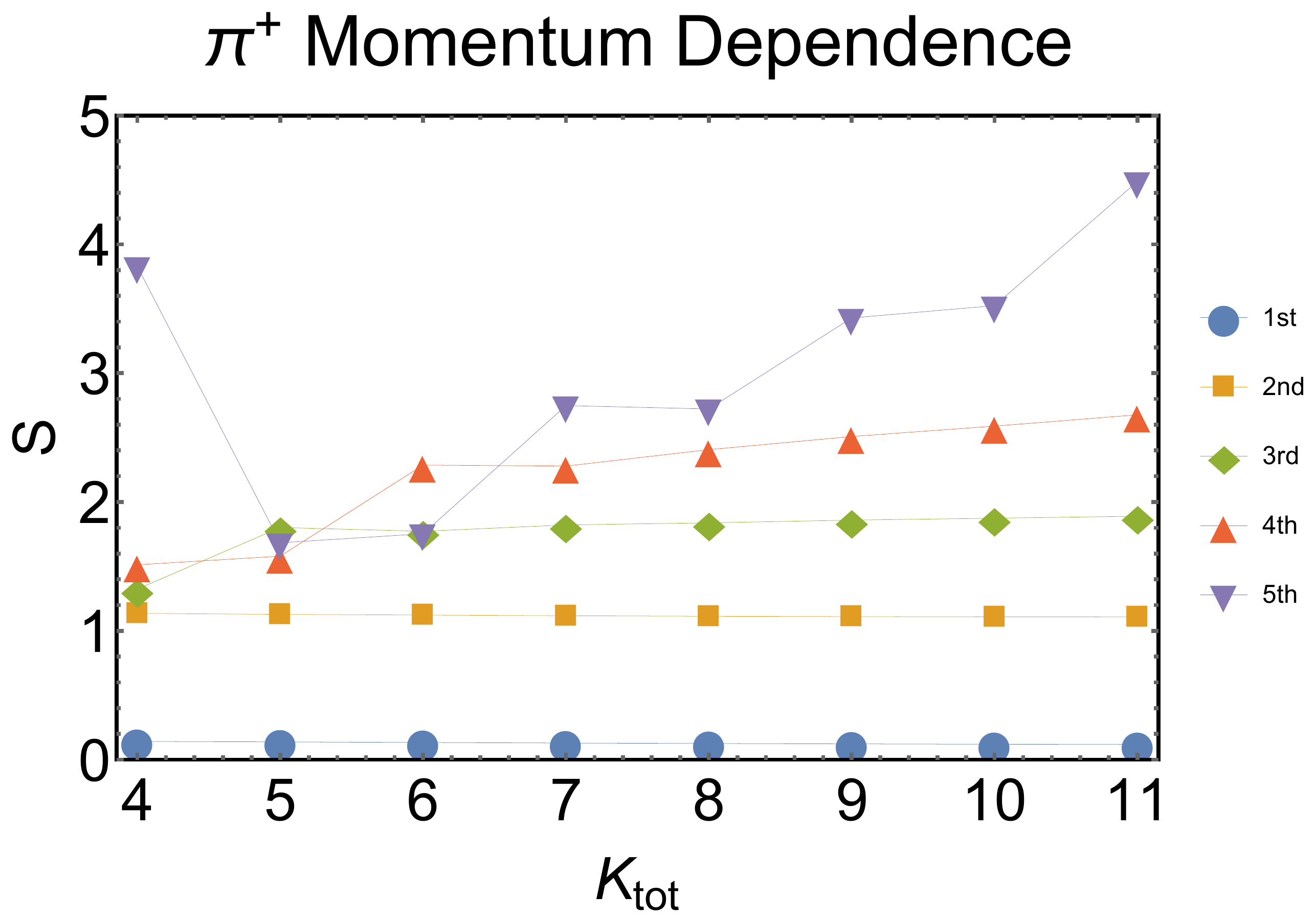}
	\caption{Entanglement entropy for the 5 lowest energy $\pi^+$ states with $N_f=2$, $N_c=2$, and $m^2=\frac{g^2}{\pi}$.  The theoretical maximum entropy is $S_{max}=\mathrm{Log}_2(2K_{tot}(K_{tot}+1))$. The lines connecting these data points are merely for visual effect and do not communicate additional data.}
	\label{pi+mom}
\end{figure}

\begin{figure}
	\centering
	\includegraphics[width=0.6\textwidth]{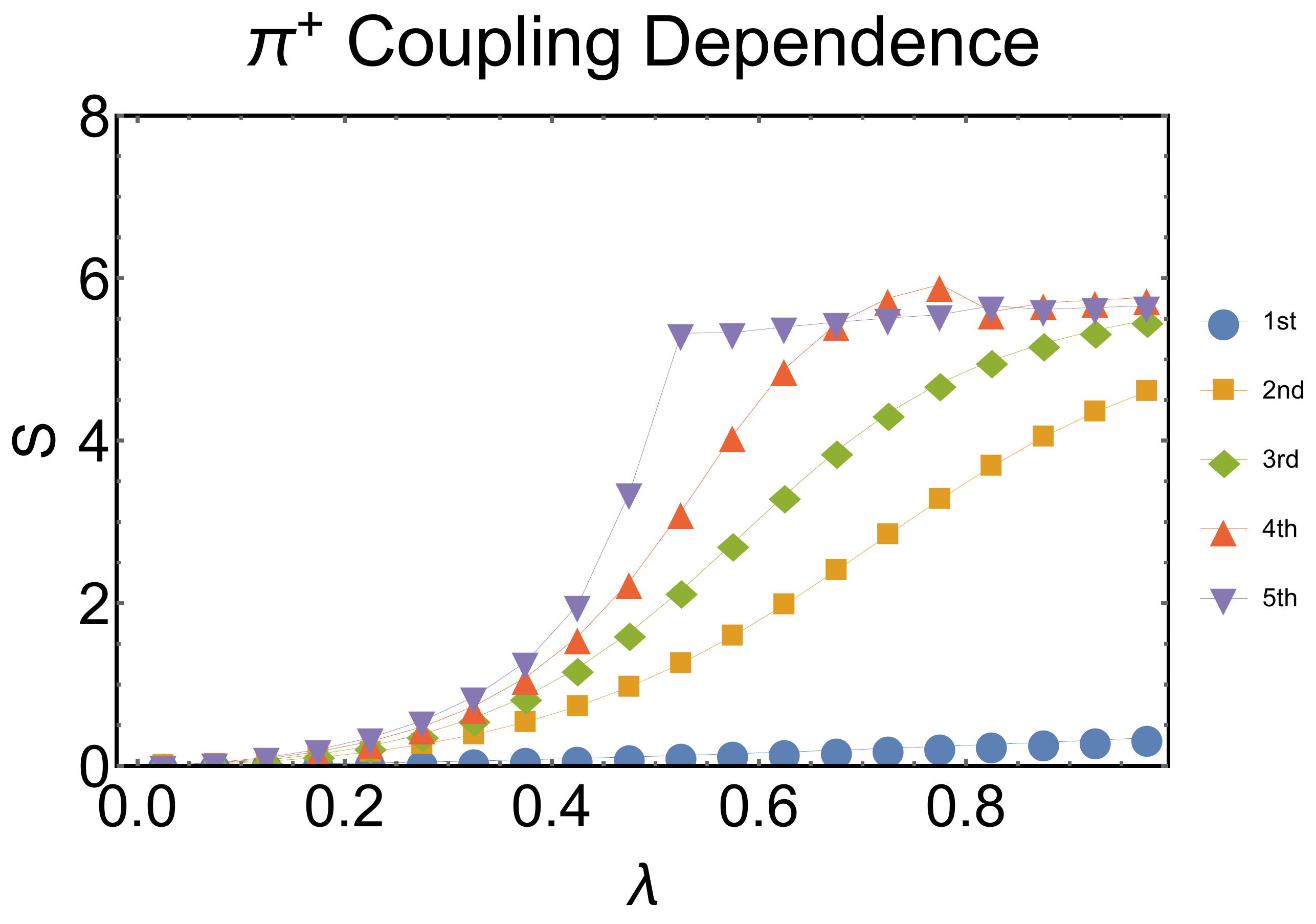}
	\caption{Entanglement entropy for the 5 lowest energy $\pi^+$ states with $N_f=2$, $N_c=2$, $K_{tot}=11$, and $\lambda=\frac{g^2}{\pi m^2+g^2}$. The theoretical maximum entropy is $S_{max}=\mathrm{Log}_2(2*11*12)\approx8.04$. The lines connecting these data points are merely for visual effect and do not communicate additional data.}
	\label{pi+coup}
\end{figure}

\begin{figure}
	\centering
	\includegraphics[width=0.6\textwidth]{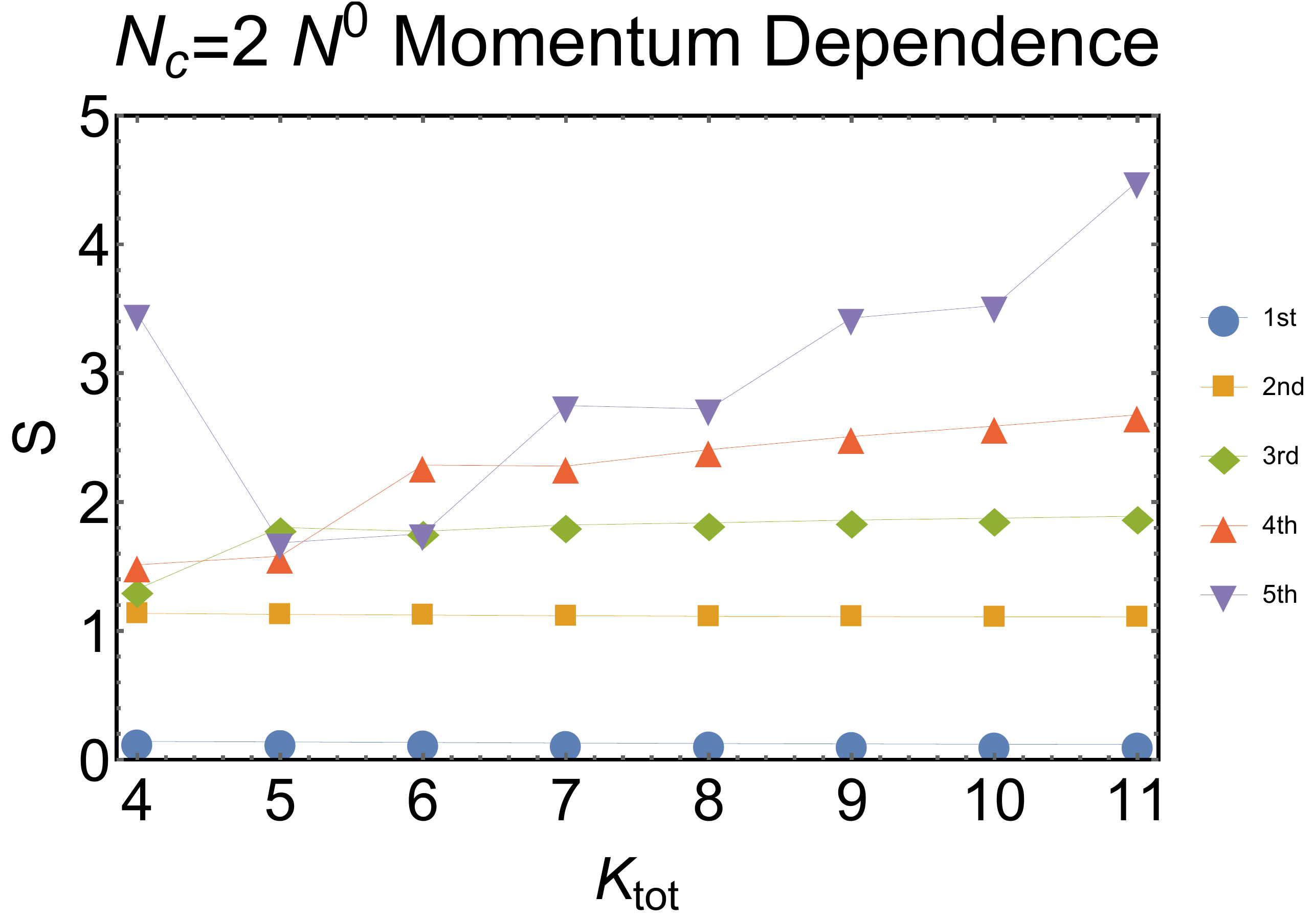}
	\includegraphics[width=0.6\textwidth]{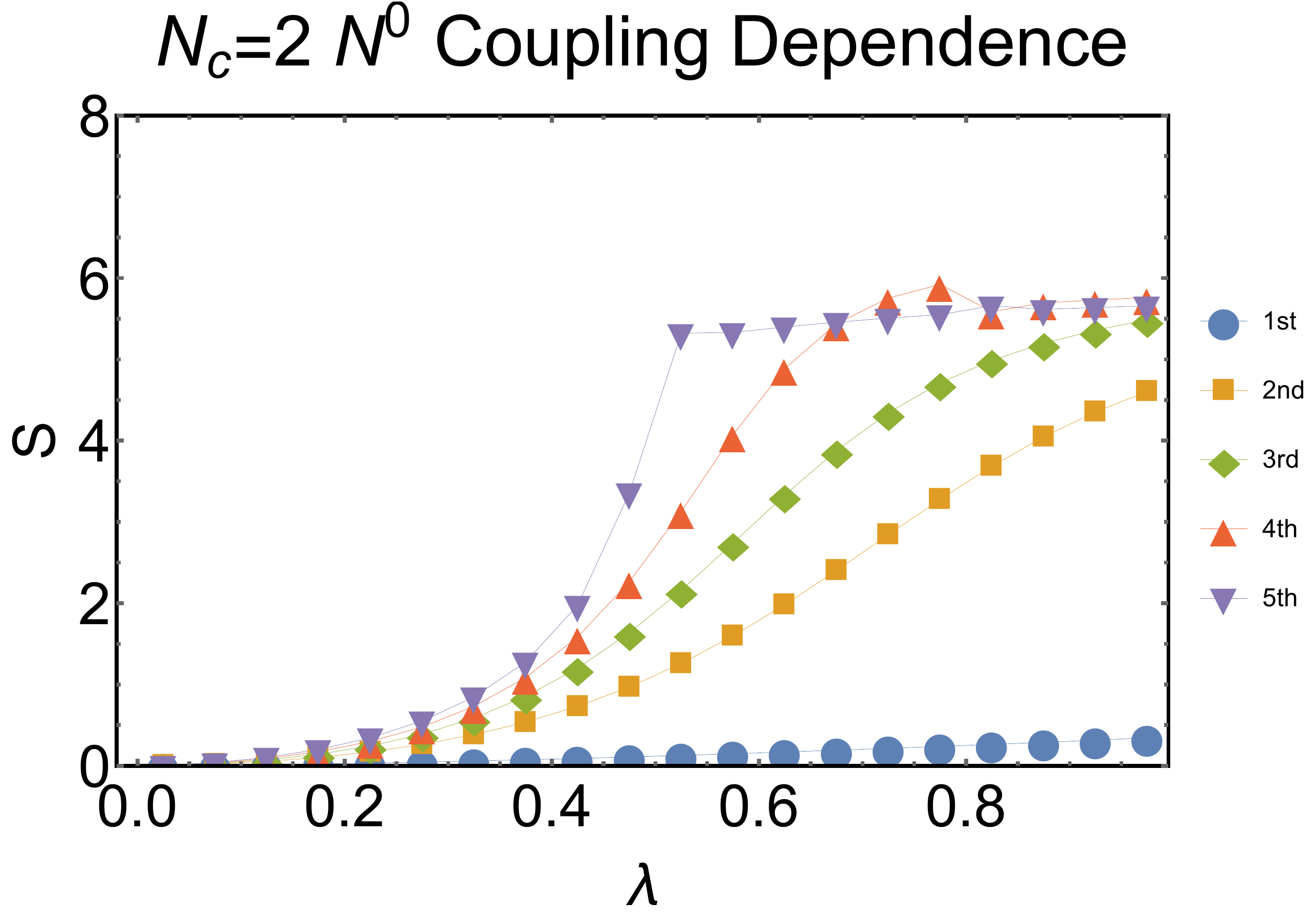}
	\caption{Entanglement entropy for the 5 lowest energy 2-color baryon states with $N_f=2$ where the constituent quarks have different flavors, somewhat analogous to the neutron $N^0$. In the first plot, $m^2=\frac{g^2}{\pi}$. In the second plot, $K_{tot}=11$ and $\lambda=\frac{g^2}{\pi m^2+g^2}$. Note the extreme similarity to the results for the $\pi^+$ above. The lines connecting these data points are merely for visual effect and do not communicate additional data.}
	\label{N0c2}
\end{figure}

The following results show some general patterns in the VS entanglement entropy of low energy states that seem to hold for mesons and baryons simultaneously. First, we see that in Figs.~\ref{pi+mom}, \ref{N0c2}, and \ref{N0c3} these lower lying energy states tend to a specific value even at fairly modest values of the total momentum. This convergence becomes weaker as we increase the relative strength of the potential $\lambda$. In the plots against the relative coupling strength in Figs.~ \ref{pi+coup}, \ref{N0c2}, and \ref{Ndelc3}, we see that the ground state hadron always has low entropy, while the next few states steadily increase with $\lambda$, with some of them seeming to plateau near a particular value of $S$ near the middle of its possible range of values. The low entropy of the ground state is due to the state being dominated by the minimum particle state that has no quark sea at all, which persists even for very small quark masses \cite{Brodsky:1997de}. This is true to a lesser extent for the first few excited states, though they seem to have much larger contributions from Fock states with sea quarks as $\lambda$ increases. If the contributions get too large, there will be a level crossing with a multi-hadron state at some specific value of $\lambda$, which causes the behavior of $S$ as a function of $\lambda$ to suddenly change. This level crossing behavior is what causes the plateauing effect in our plots. These level crossings also make it difficult for comparisons to large-$N_c$ hadrons to be made, since that requires a clean method for distinguishing single and multi-hadron states from each other.

\subsection{Meson Results}
This section has plots of the VS entropy of the $\pi^+$ meson analog against number of colors in Fig.~\ref{pi+color}, total momentum in Fig.~\ref{pi+mom}, and the relative coupling strength in Fig.~\ref{pi+coup}. The results for the $\pi^0$ analog are similar, as are the results when a third degenerate quark flavor is added. In Fig.~\ref{pi+color}, the first three states look like they may eventually taper off to zero at large $N_c$, but the other two are increasing with $N_c$. This is because large-$N_c$ states with multiple mesons can have nonzero VS entropy due to entanglement. In fact, using the expression for the entropy in terms of $f_S$ and $f_A$ above, we can see that there is a term proportional to $\mathrm{Tr}\left[f^A\right]\log(N^2_c-1)$. The Fermi degeneracy creates entanglement between the mesons that feeds into $f_A$, so the VS entropy for multi-meson states at large $N_c$ will actually tend to infinity. However, $f_S$ remains zero for all large $N_c$ states, and seems to decrease asymptotically as $N_c$ increases for all 5 states. 

In general, the meson spectra seem to fall into two regimes depending on the size of $g/m$. If $g/m$ is small (roughly less than $2$), then the theory is in the large mass regime which contains both the free theory and the 't Hooft model. This regime is characterized by single meson states that are dominated by two-particle Fock states in momentum space. This makes it easy to distinguish the single meson states from the multiple meson states in DLCQ. When $g/m$ is large, the theory is in the strong potential regime which contains the Schwinger model \cite{Schwinger:1962tp,Coleman:1975pw} when we have a $U(1)$ gauge field. In this regime, the meson states are localized in position space and tend to have local color currents that are small. In addition, the presence of massless fermions leads to zero modes in the spectrum, meaning the quark-antiquark pairs can be generated fairly easily, so particle number in momentum space cannot be used to identify bound states in this regime. The transition between these two regimes can be seen in the plots of the VS entropy against coupling $\lambda$ where the excited state entropies seem to stop increasing beyond certain large values of $\lambda$.

The ground state mesons have a particularly low VS entropy compared to all other states. This is because it bears a close resemblance to its large-$N_c$ counterpart, and this resemblance fades for higher excited states. The large-$N_c$ connection is also what causes the Fock states with no sea contribution to dominate the Fock state expansion of the ground state. The resemblance to the large-$N_c$ counterpart also suggests that its VS entropy could be captured by a $1/N_c$ expansion, and indeed we see in Fig.~\ref{pi+mom} that it has a good fit to a $1/N_c$ curve. The value of $0.26$ for the fit was chosen because it minimizes the sum of the squares of the differences between the data points and the fit curve. If this perturbative behavior persists into $3+1$ dimensions, this could open the door for a perturbative calculation of VS entropy for ground state hadrons in real QCD.

\begin{figure}
	\centering
	\includegraphics[width=0.6\textwidth]{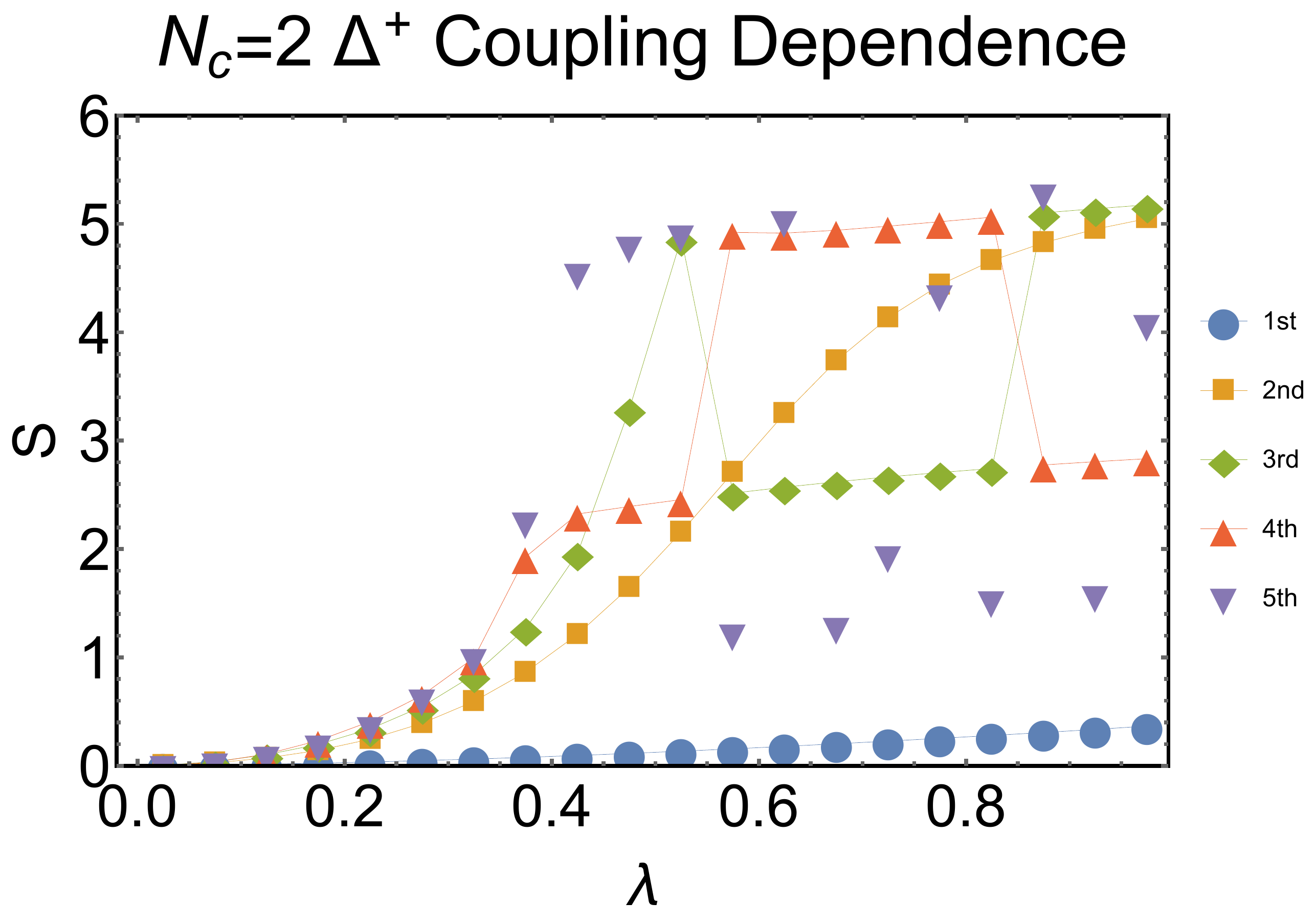}
	\includegraphics[width=0.6\textwidth]{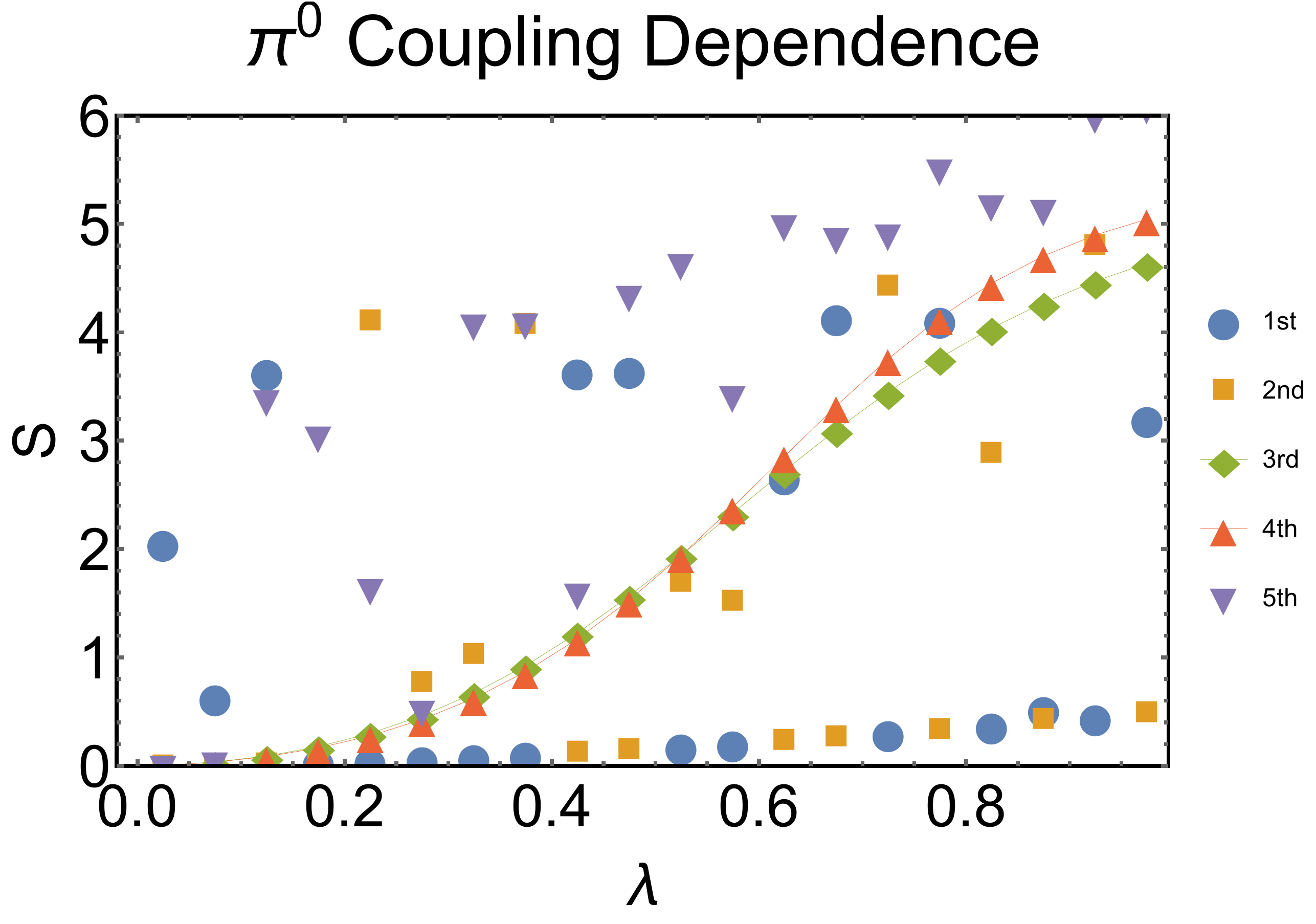}
	\caption{The first plot shows the entanglement entropy for the 5 lowest energy 2-color baryon states where the constituent quarks have the same flavor, somewhat analogous to the delta $\Delta^+$. The second plot shows the entanglement entropy for the 5 lowest energy $\pi^0$ states. In both plots, $N_f=N_c=2$, $K_{tot}=11$, and $\lambda=\frac{g^2}{\pi m^2+g^2}$. The theoretical maximum entropy for the $\Delta^+$ is $S_{max}\approx6.98$, while for $\pi^0$ it is $S_{max}\approx8.04$. The lines connecting these data points are merely for visual effect and do not communicate additional data.}
	\label{delpidir}
\end{figure}

\begin{figure}
	\centering
	\includegraphics[width=0.6\textwidth]{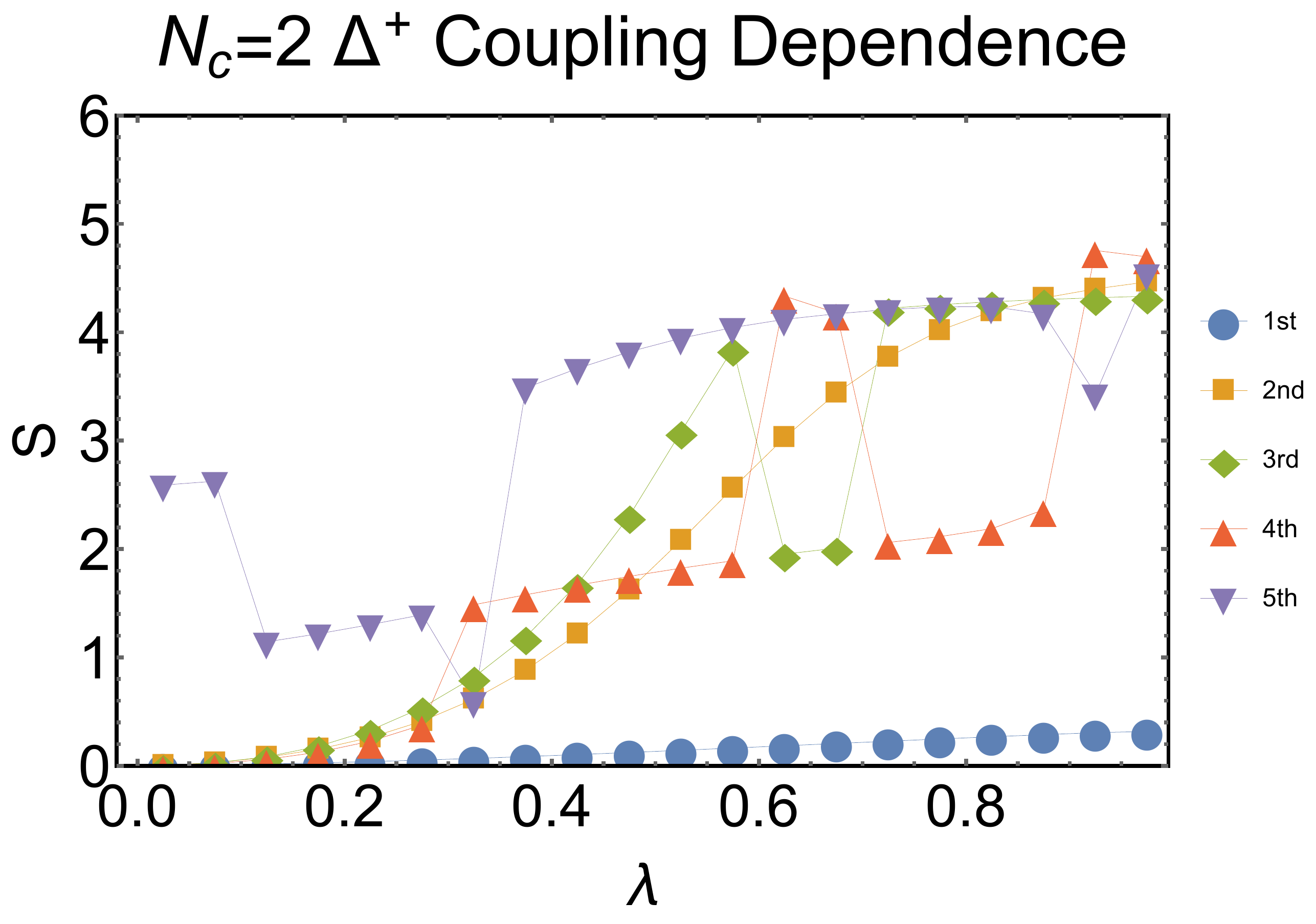}
	\includegraphics[width=0.6\textwidth]{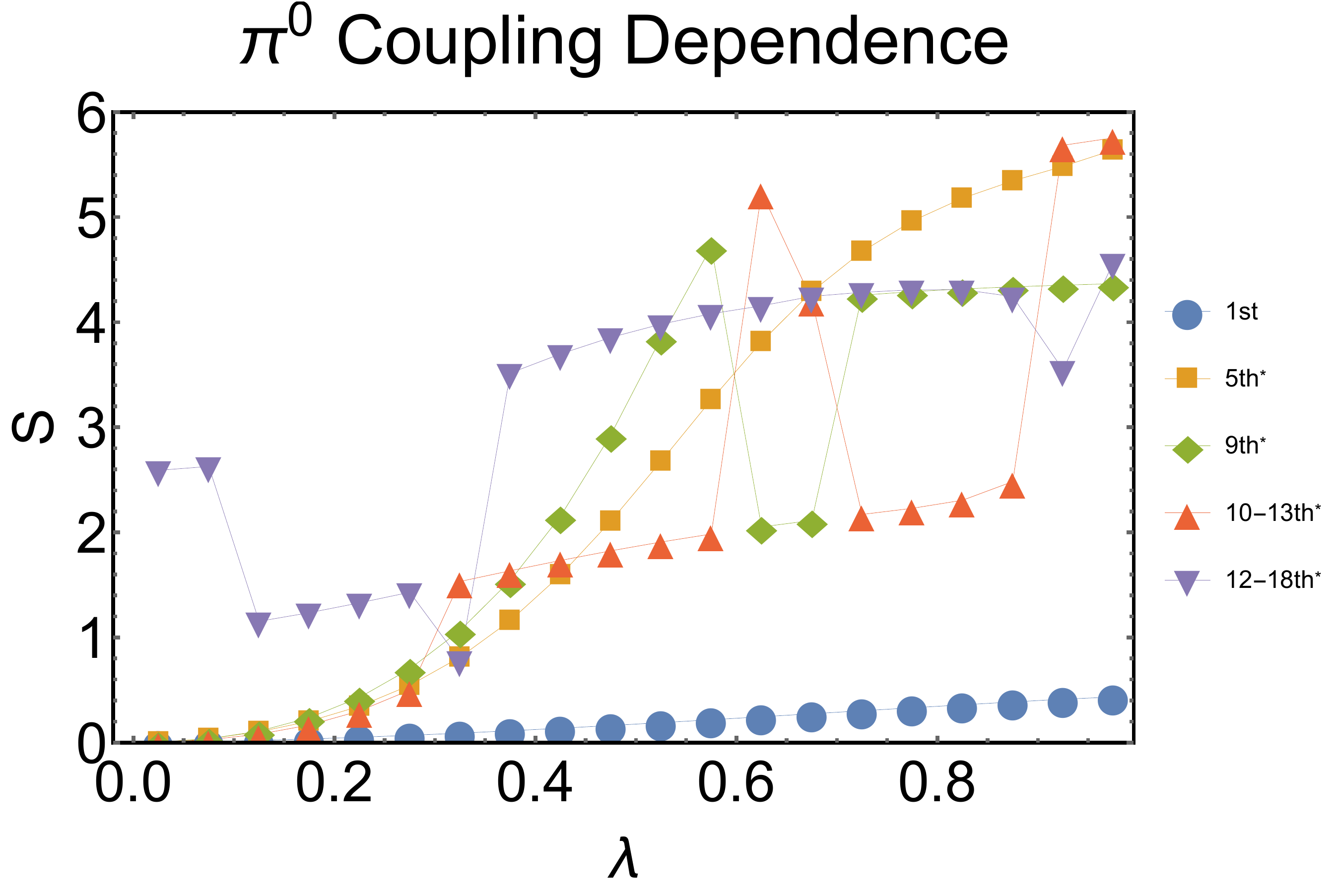}
	\caption{In these plots, there is a small difference in the quark masses given by $m_2^2-m_1^2=0.01(\bar{m}^2+g^2/\pi)$, where $\bar{m}^2=(m_1^2+m_2^2)/2$. The first plot shows the entanglement entropy for the 5 lowest energy 2-color $\Delta^+$-analogue states. The second plot shows the entanglement entropy for the 5 $\pi^0$ states whose eigenvalues match those of the $\Delta^+$ shown above. The asterisk indicates states whose relative ordering in the full eigenvalue spectrum changes depending on the coupling; the number or range of numbers listed indicates the usual position of the state over most of the given values of $\lambda$. In both plots, $N_f=N_c=2$, $K_{tot}=8$, and $\lambda=\frac{g^2}{\pi\bar{m}^2+g^2}$. The theoretical maximum entropy for the $\Delta^+$ is $S_{max}\approx6.09$, while for $\pi^0$ it is $S_{max}\approx7.17$. The lines connecting these data points are merely for visual effect and do not communicate additional data.}
	\label{delpieig}
\end{figure}

\subsection{Baryon Results}
The first figure in this section, Fig.~\ref{N0c2}, has plots of the VS entropy of the 2-color $N^0$ baryon analog against total momentum and the relative coupling strength. A surprising result is that these plots are identical to Figs.~\ref{pi+mom} and \ref{pi+coup} for the pion. It turns out that when $N_c=2$, the Hamiltonians for the baryon and meson channels where the valence particles have different flavors are not only of identical size, but have an identical eigenvalue spectrum. In other words, these Hamiltonians differ only by a unitary transformation, one that the VS entropy is apparently insensitive to. In the case that the flavors of the two particles are the same (as in Figs.~\ref{delpidir} and \ref{delpieig}), the meson Hamiltonian is over twice as large as the baryon Hamiltonian, but there is a subsection of the meson's mass spectrum that matches the full spectrum of the baryon. A comparison of states with matching eigenvalues is given in Fig.~\ref{delpieig}.

\begin{figure}
	\centering
	\includegraphics[width=0.6\textwidth]{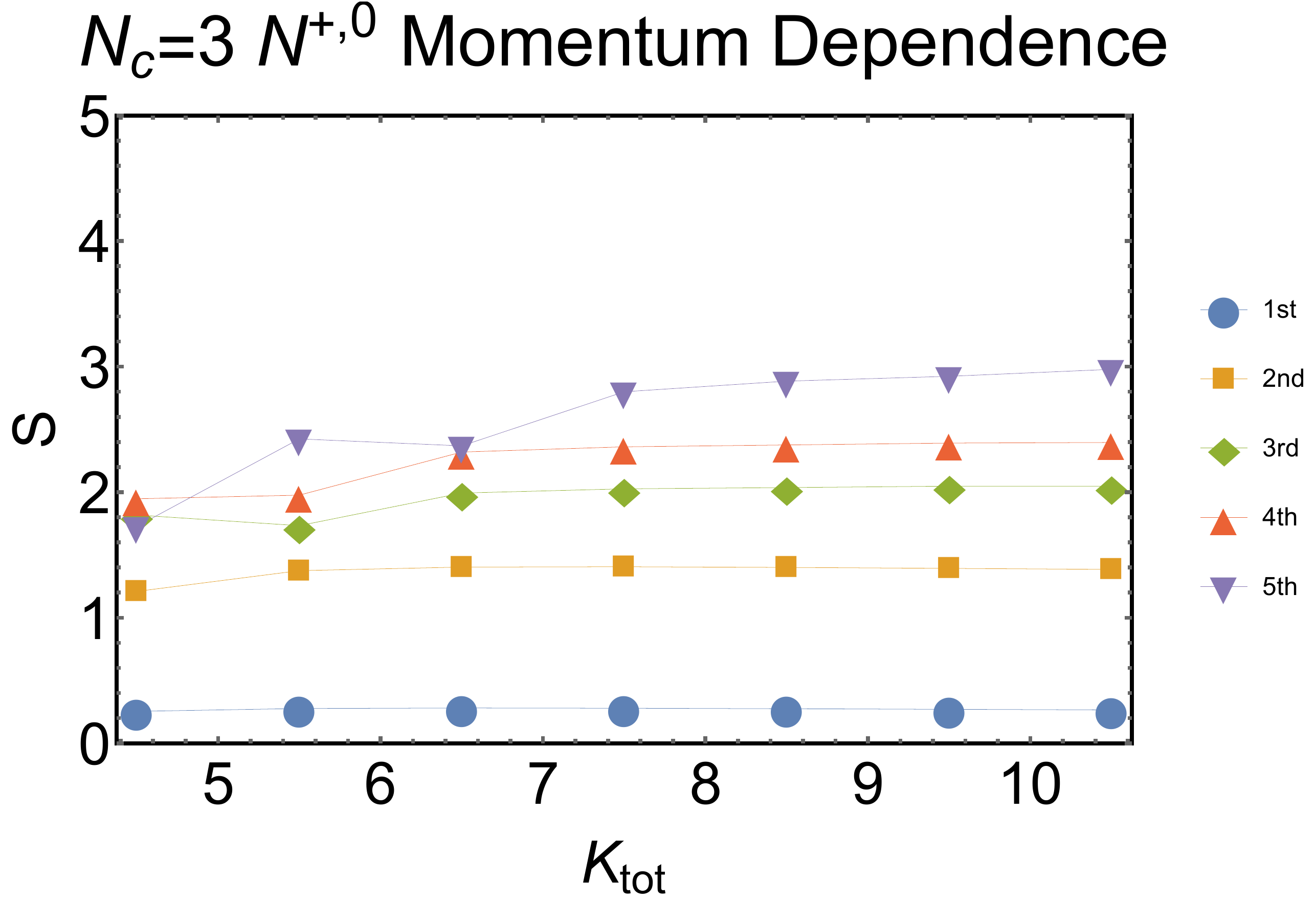}
	\caption{Entanglement entropy for the 5 lowest energy 3-color baryon states with $N_f=2$ where one constituent quark has a different flavor from the other two, analogous to the nucleon $N^{+,0}$ in QCD. The quark masses are degenerate, and $m^2=\frac{3g^2}{2\pi}$. The lines connecting these data points are merely for visual effect and do not communicate additional data.}
	\label{N0c3}
\end{figure}

\begin{figure}
	\centering
	\includegraphics[width=0.6\textwidth]{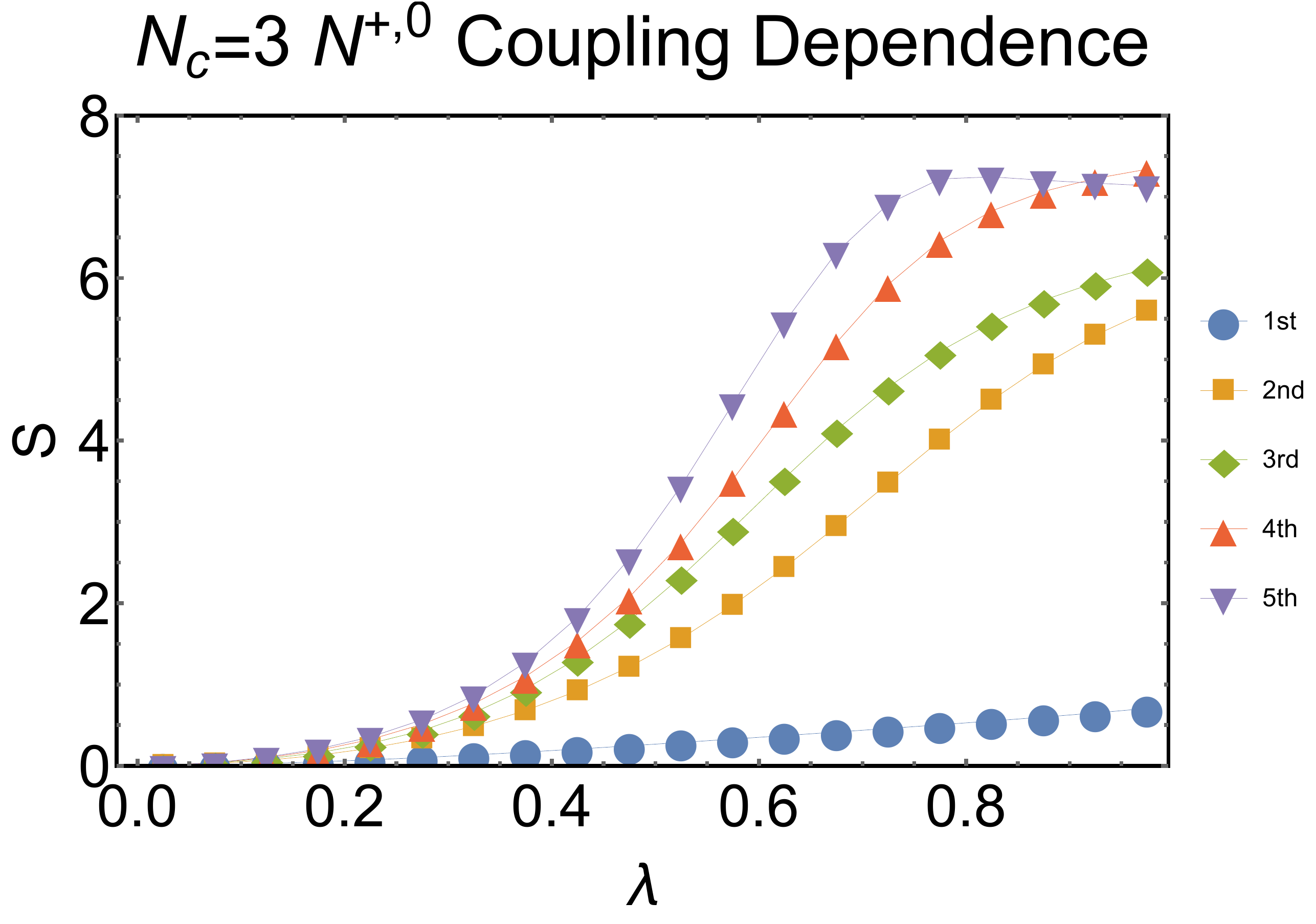}
	\includegraphics[width=0.6\textwidth]{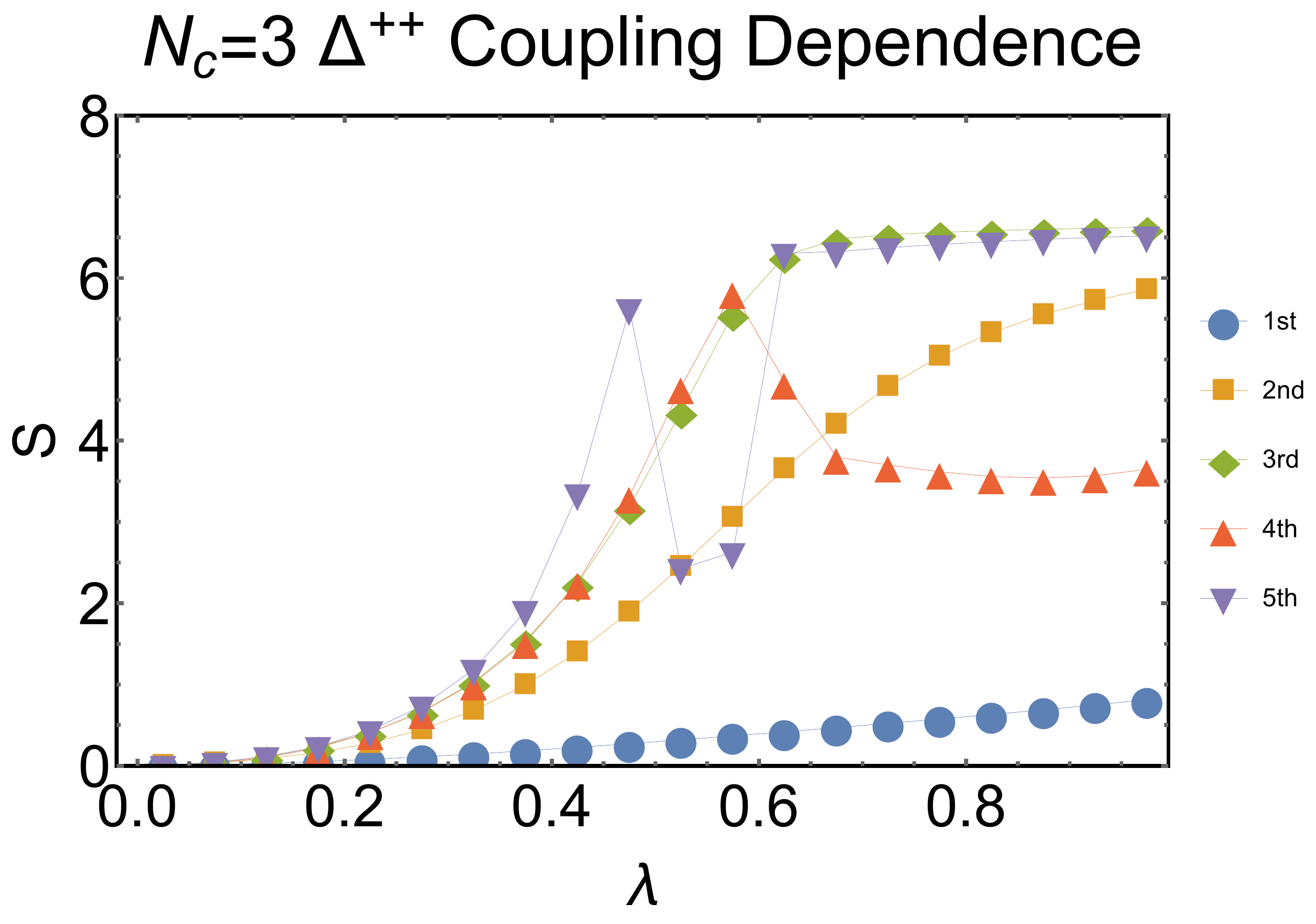}
	\caption{These plots show the entanglement entropy for the 5 lowest energy 3-color baryon states where $N_f=2$, $N_c=3$, $K_{tot}=\frac{21}{2}$, the quark masses are degenerate, and $\lambda=\frac{3g^2}{2\pi m^2+3g^2}$. The first plot shows the entanglement entropy for states where one constituent quark has a different flavor from the other two, analogous to the nucleon $N^{+,0}$ in QCD. The second plot shows states where the constituent quarks all have the same flavor, analogous to the delta $\Delta^{++}$ in QCD. The theoretical maximum entropy for the $\Delta^{++}$ is $S_{max}\approx9.75$, while for $N^{+,0}$ it is $S_{max}\approx11.47$. The lines connecting these data points are merely for visual effect and do not communicate additional data.}
	\label{Ndelc3}
\end{figure}

\begin{figure}
	\centering
	\includegraphics[width=0.6\textwidth]{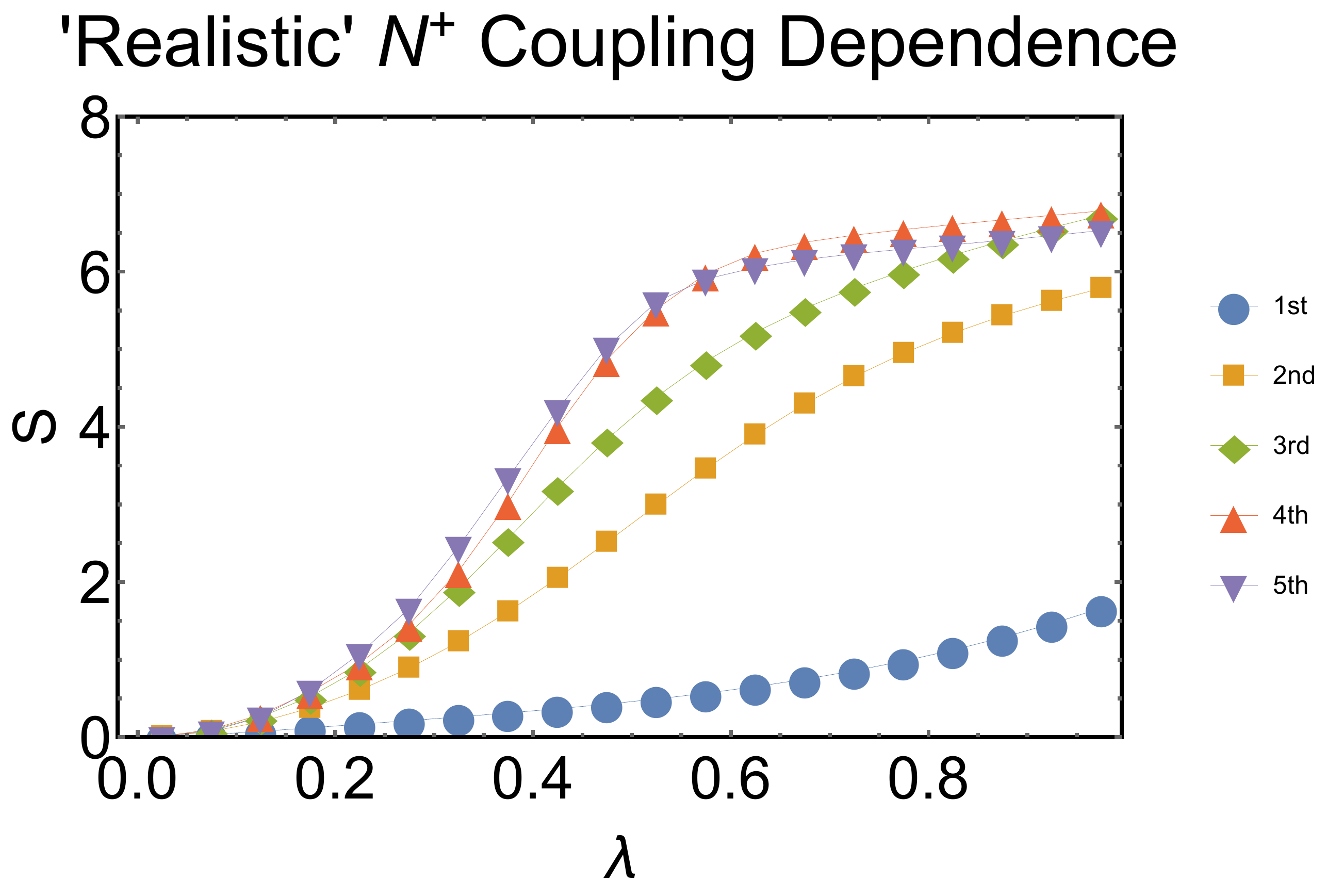}
	\caption{This plot shows the entanglement entropy for the 5 lowest energy 3-color baryon states analogous to the proton $N^+$ in QCD, where $N_f=3$, $N_c=3$, and $K_{tot}=\frac{17}{2}$. It is "realistic" in the sense that the quark masses are defined using approximately physical values relative to the average of the squares of the valence quark masses $\bar{m}^2=\frac{2m_u^2+m_d^2}{3}$. The values used for the masses are $\frac{m_u^2}{\bar{m}^2}=0.45$, $\frac{m_d^2}{\bar{m}^2}=2.1$, $\frac{m_s^2}{\bar{m}^2}=834$, and $\lambda=\frac{3g^2}{2\pi \bar{m}^2+3g^2}$. The theoretical maximum entropy is $S_{max}\approx10.58$. The lines connecting these data points are merely for visual effect and do not communicate additional data.}
	\label{Nreal}
\end{figure}

Fig.~\ref{N0c3} shows a plot of the VS entropy of the nucleon analog against total momentum, while Fig.~\ref{Ndelc3} shows the coupling dependence of both the nucleon analog and the $\Delta^{++}$ analog. The results for other baryons are similar, as are the results when a third degenerate quark flavor is added. Fig.~\ref{Nreal} shows the coupling dependence of the nucleon analog for realistic quark masses. Removing the heavy strange quarks from the system barely affects the VS entropy. All of these 3-color baryon entropy plots exhibit similar qualitative features to those of the mesons and 2-color baryons, such as the very low VS entropy of the ground state, the dominance of the singlet contribution $f_S$ for these low lying energy states, and the separation of theories into two broad regimes of large mass and strong coupling. These similarities persist despite these being fermions with 3 valence quarks instead of bosons with only two, and there is no symmetry that can relate them to the mesonic states like with the 2-color case. There is also an additional $SU(N_c)$ singlet contribution with these baryons corresponding to the mixed symmetry irrep of $S_3$, but this addition does not seem to greatly affect the general patterns of the VS entropy of low energy states. This suggests that these states also maintain features of their large-$N_c$ counterparts, despite the large-$N_c$ baryon wavefunction being largely inaccessible without making a sweeping assumption about its general form. Since the connection to large-$N_c$ physics seems to hold in $3+1$ dimensions for $N_c=3$, it may even be possible to calculate the VS entropy of real nucleons using a large-$N_c$ expansion, assuming that the approximate $1/N_c$ behavior of the meson ground state entropy also holds for baryons.

\section{Conclusion}
\label{sec:conclu}

In this paper, we have defined the first rigorous measure of valence-sea entanglement in QCD, despite the lack of a clear distinction between valence quarks and sea quarks, and analyzed the VS entanglement entropy in 1+1d QCD. The framework used to define VS entanglement can potentially be used to define other types of entanglement in situations when a clear bipartition of the Hilbert space does not exist. It would most likely be useful for defining entanglement between constituent fields in the bound states of other QFTs, but there could in principle be more clever applications of the method. This also naturally leads to a definition of single quark (or antiquark) entanglement in a hadron using the same framework, in which the elements of the corresponding density matrix turn out to be parton distribution functions. This provides an interpretation of PDFs as measures of quark entanglement, which may highlight some connection between quark entanglement and the operator product expansion of the DIS cross section.

To get the VS entanglement, we also described much of the framework needed to do the calculation of the entanglement entropy in 1+1d QCD. We have derived the terms in the Hamiltonian formalism of 1+1d light front QCD, recovered the spectrum of the 't Hooft model by taking the large-$N_c$ limit in said formalism, and applied discrete light-cone quantization to render the Hamiltonian as a finite matrix. The expression for the VS entanglement in these hadrons was found to separate into different contributions coming from the different $SU(N_c)$ irreps of the valence quarks, as well as separating additively into momentum space and color space terms. Appendix \ref{Algorithm} also details a method for deriving DLCQ operators in a basis of color singlet states that is manifestly orthonormal and complete from the beginning. To our knowledge, this has not been done in the literature as of yet, since the papers that have done explicit QCD calculations in 1+1 dimensions mention rendering the Hamiltonian in an incomplete basis as an intermediate step \cite{Hornbostel:1988fb,Hornbostel:1988ne,Burkardt:1989wy}.

Our numerical results, which provide the first quantification of VS entanglement from hadronic state vectors in QCD, show that the VS entanglement is unusually low for low energy eigenstates of the QCD Hamiltonian. For mesons, this results from the fact that these states at finite $N_c$ have almost no sea quarks and closely resemble the corresponding eigenstates in the large-$N_c$ limit. For baryons, the connection to large-$N_c$ cannot be made easily, but the states seem to have next to no sea quarks as well. If we treat VS entanglement as a measure of the applicability of the parton model as discussed in the introduction, then this would imply that the parton model is only applicable to the lowest mass hadrons in 1+1d QCD. This is of course consistent with our observations of ground state nucleons in real QCD, but it remains to be seen if the model breaks down for highly excited hadrons. The low values of the VS entropy also suggest that the large-$N_c$ expansion may fail for higher excited states, perhaps even suggesting a connection between the applicability of the parton model and the large-$N_c$ expansion. This may not be the case in real QCD, as several papers have shown that large-$N_c$ results for excited baryons in 3+1 dimensions generally agree with experiment for $N_c=3$ \cite{Pirjol:1997bp,Pirjol:1997sr,Cohen:2003jk}.

We also found for mesons in their ground states that the VS entanglement entropy approximately follows a $1/N_c$ curve, indicating that the entropy for these specific states can be calculated perturbatively in a next to leading order $1/N_c$ expansion. Given that the ground state baryons also have very low VS entropy, they may also be perturbative in the same way. Given that the large-$N_c$ expansion works surprisingly well in real QCD for ground state hadrons, it may be reasonable to expect the VS entropy in real QCD to be calculable in a $1/N_c$ expansion as well. The elements of the VS density matrix resemble PDFs, which are scale dependent in real QCD, so the corresponding VS entanglement entropy ought to be scale dependent as well. Also, it would likely be small at both large scales where the coupling is weak and small scales where the hadrons resemble large-$N_c$ states, but this may not hold true in between. Having access to this quantity could therefore lead to some very interesting insights into the transition between quark and hadron degrees of freedom, as it would likely be scale dependent and function as an order parameter for this transition. However, the presence of extra spin degrees of freedom and transverse gauge fields compared to 1+1d QCD makes it difficult to make solid claims about what VS entanglement would look like in 3+1d QCD.

\begin{figure}
	\centering
	\includegraphics[width=0.6\textwidth]{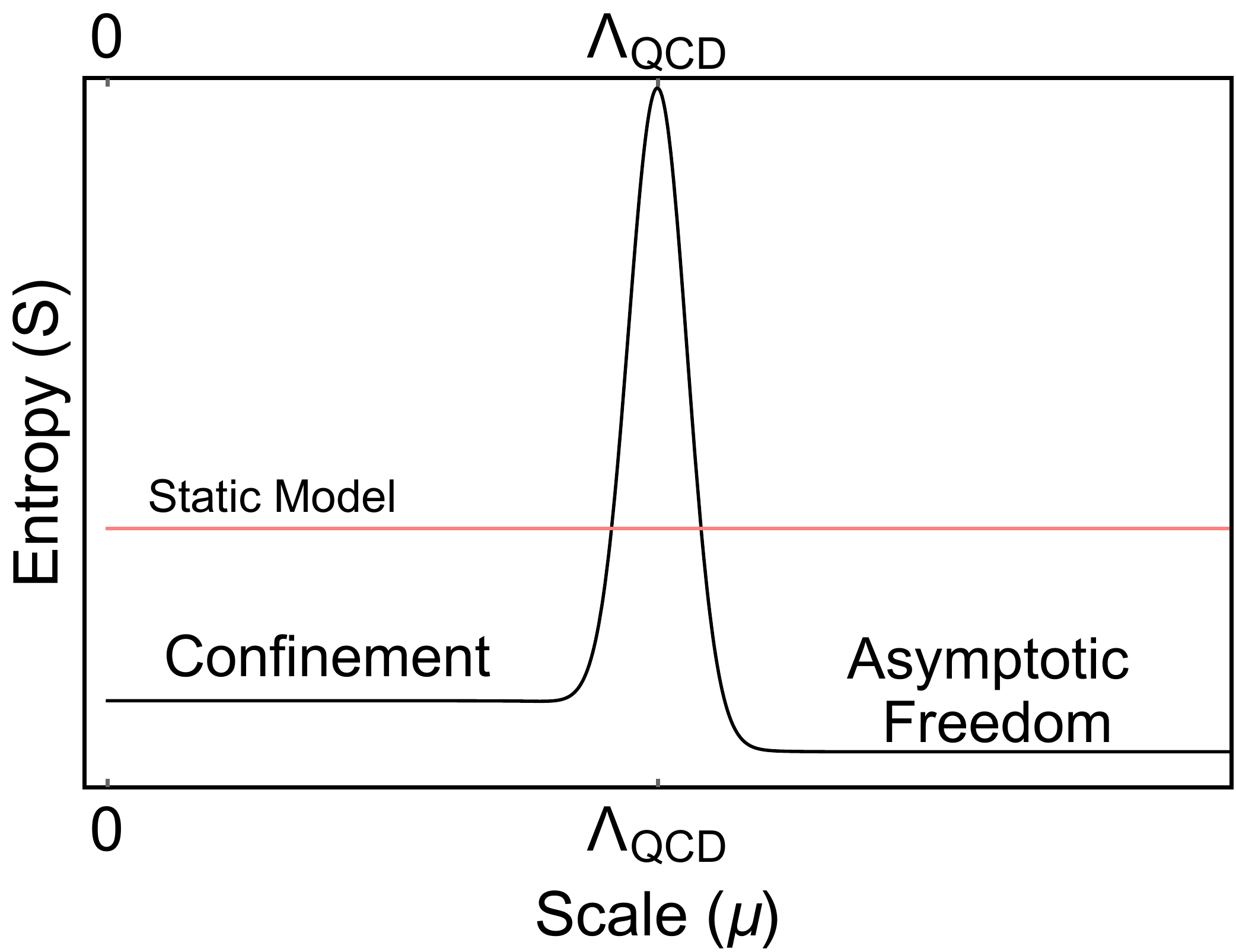}
	\caption{This sketch illustrates what the VS entropy of a ground state hadron in real QCD is expected to look like. The red line labeled "Static Model" shows what results a model built from only static properties would give, such as in \cite{Beane:2019loz}. At high energy scales where the quarks are asymptotically free, the entropy should be close to zero. Our results suggest that it will also be low in the confining phase. This implies that the region near $\Lambda_{QCD}$ must have high entropy, or else the static model entropy would be much lower.}
	\label{SpecEnt}
\end{figure}

In spite of these differences, the results of our previous work \cite{Beane:2019loz} along with some arguments about the scale dependence of the VS entanglement entropy can give us a rough expectation of what to expect in real QCD, which is sketched out in Fig.~\ref{SpecEnt}. Due to asymptotic freedom, the quark interactions are weak at very short length scales, so we can expect low VS entanglement at scales well beyond $\Lambda_{QCD}$. The results of this work suggest that the VS entanglement is also low for ground state hadrons at scales below $\Lambda_{QCD}$ where quarks are confined. The question remains as to how entangled the quarks are near $\Lambda_{QCD}$, but our previous work, which investigated entanglement between valence and sea quarks due to chiral symmetry breaking in real QCD, gave results consistent with a large overall VS entropy. We say "overall" here because the VS entropy was obtained by fitting parameters in a chiral symmetry focused model to low-energy nuclear coupling constants, which can be thought of from an effective field theory point of view as containing some integrated form of the information about the physics at length scales below that of the interaction. Since we have ruled out VS entanglement at scales beyond $\Lambda_{QCD}$, the only possible source for that entanglement must therefore be at scales near $\Lambda_{QCD}$, which also tracks with chiral symmetry breaking being the source of entanglement in our previous work. Thus we expect that the VS entropy in real QCD should act as an order parameter for the transition between quark and hadron degrees of freedom, as it should remain low except near the scale at which hadrons begin to form and chiral symmetry is spontaneously broken.

\section*{Acknowledgments}
I would like to thank the Nuclear Theory Group at the University of Washington for providing me this research opportunity, Silas Beane for advising on this project, and Gerald Miller for helpful discussions. The preprint number for this document is NT@UW-22-10. This work was supported by the U.S. Department of Energy grant DE-FG02-97ER-41014.

\clearpage

\phantomsection
\printbibliography

\appendix

\section{Algorithm Details and Young's Orthogonal Basis} \label{Algorithm}
In this appendix, we will explain some of the details of the algorithm used to generate the DLCQ Hamiltonian, as well as the other algorithms needed to obtain the entanglement entropies. A suitable computational basis in DLCQ is a Fock state basis comprised of the finite set of states with a given total momentum and flavor content, where states are labeled by the momenta, flavors, and color charges of each quark in a state. Doing calculations in this basis is fairly straightforward, but since  physical quantities are always calculated using color singlet states using this basis would be incredibly inefficient due to the large number of color charged states that ultimately do not contribute to any quantities of interest. We would rather restrict our computational basis to the subset of color singlet Fock states to improve the efficiency of the calculation.

A convenient way to construct color singlets from a Fock state with $n_q$ quarks and $n_{\bar{q}}$ antiquarks is to organize them both into the $n_q$-dimensional and $n_{\bar{q}}$-dimensional irreducible representations (irreps) of $SU(N_c)$, and then unite the irreps for the quarks and antiquarks that are dual to each other, which is guaranteed to produce exactly one singlet per quark-antiquark irrep combination. The irreps of $SU(N_c)$ can be organized and parametrized using Young tableaux, and the number of copies of a given irrep that is formed from $n$ quarks is equal to the number of ways that the tableau for that irrep can be formed from $n$ boxes, which is in turn equal to the dimension of the irrep of the symmetric group $S_n$ corresponding to the same tableau. Furthermore, operations in color space can be implemented through permutations that swap the color indices of the creation operators within the color singlet states. These permutations can be written as matrices that can be derived from the tableaux, in a basis known as Young's orthogonal basis. Thus we can exploit the structural similarities between the irreps of $SU(N_c)$ and $S_n$, which are encoded into Young tableaux, to implement operations in color space via matrix operations in the symmetric group. The vast majority of this appendix will be devoted to describing how this is done.

The implementation of the Hamiltonian and density matrices can be separated into three broad categories. The kinetic term of the Hamiltonian is a two-quark operator that conserves momentum and flavor, so it can be implemented by simply adding a factor of $n_p \frac{m_q^2}{k}$ for each quark and antiquark in the state to the relevant diagonal element of the Hamiltonian. The potential in the Hamiltonian as well as the density matrices for mesons and 2-color baryons are composed of four-quark operators, which can be broken down into a product of two-quark color singlet operators that may not conserve momentum or flavor. Finally, we also have six-quark operators needed for the 3-color baryon density matrices, but these can also be broken down into two-quark color singlet operators. Thus we can get everything we need by focusing on the actions of general two-quark color singlet operators on our Fock state basis. These come in four different types: operators of the form $b_a^\dagger b_a$, $d_a^\dagger d_a$, $d_a b_a$, and $b_a^\dagger d_a^\dagger$, where $a$ is the color index that is implicitly being summed over. 

The action of singlet operators of the form $b_a^\dagger b_a$ or $d_a^\dagger d_a$ can be implemented by changing the momentum and flavor indices of the Fock state to reflect the removal and addition of one particle, followed by the application of a permutation matrix in color space. The matrix stems from having to restore the normal ordering of the creation operators, which necessitates a permutation of the color indices. Since the action of these operators in color space amounts to a simple permutation matrix, they do not mix states in different irreps. Some simplifications can be made along the diagonal terms of the Hamiltonian matrix since they map states onto themselves. In this case only one permutation matrix is applied instead of two since the best a four-quark operator made only of $b_a^\dagger b_a$ or $d_a^\dagger d_a$ can do on the diagonal is swap two creation operators. Details on the construction of the permutation matrices are given in Appendix \ref{app:mats}. For color singlet operators of the form $d_a b_a$, a more involved color matrix is applied to reflect that the $SU(N_c)$ irreps get truncated by the operator. Details on the construction of that matrix are given in Appendix \ref{app:dbtype}. Singlet operators of the form $b_a^\dagger d_a^\dagger$ can be obtained from Hermitian conjugation of $d_a b_a$.

\subsection{Permutation Matrices in Young's Orthogonal Basis}
\label{app:mats}

Young's orthogonal basis is a vector basis for the symmetric group in which the permutation matrices are rendered orthogonal. The orthogonality is of vital importance as it corresponds to the unitarity of the operators that swap quark and antiquark creation operators. The systematic method for deriving these permutation matrices from Young tableaux is detailed here. First, for a given Young tableau, we write out all the ways that the boxes of a tableau can be numerically labeled from 1 to $n$, where $n$ is the number of boxes in the tableau. Each numbered tableau represents one of the color singlet basis vectors in the irrep represented by the tableau. The labeling is ordered such that the upper-left most box is always labeled $1$, each following numerical label must be to the right of and below other labeled boxes, unless it is in the same column or row as the first box. This is essentially counting and labeling the distinct ways in which a given tableau can be assembled from $n$ individual boxes, such that we are always assembling valid tableau of sizes ranging from 1 to $n$ along the way. The number of distinct ways to form a tableau is the same as the dimension of the $S_n$ irrep represented by the tableau.

Now that we have our basis states organized and labeled, we can use this to derive the orthogonal permutation matrices. To get the permutation matrices that swap elements $i$ and $i+1$, where $i$ ranges from 1 to $n-1$, we can get the nonzero elements from the positions of boxes $i$ and $i+1$ in the numbered tableaux. The diagonal elements are given by $\pm1/d$, where $d$ is the length under the taxicab metric between boxes $i$ and $i+1$, and the sign is positive(negative) when $i+1$ is to the right(left) of and/or above(below) $i$. The off-diagonal elements are only non-zero when the two numbered tableau involved nearly identical except that boxes $i$ and $i+1$ are swapped. The value of these elements are given by $\sqrt{(d^2-1)/d}$, using the value $d$ corresponding to the labeled tableau involved. The permutations that swap elements $i$ and $i+1$ form a multiplicative basis for all other permutations, so all of the other permutation matrices can be formed from products of the above matrices. In practice, the only permutations that are needed are ones that either swap two different elements and leave the rest alone or ones that cycle the order of a string of adjacent elements forward or back by one step.

\begin{figure}
	\centering
	\includegraphics[width=1.0\textwidth]{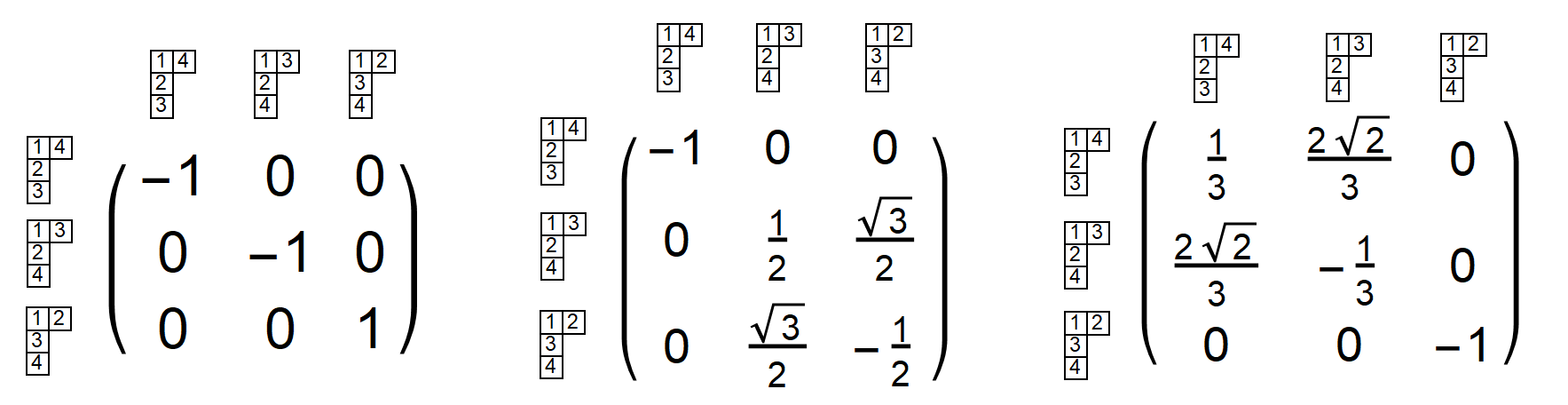}
	\caption{The three permutation matrices $T_{12}$, $T_{23}$, and $T_{34}$ for the antifundamental irrep of $S_4$ written in Young's orthogonal basis. The numbered Young tableaux correspond to the basis elements of the vector space and are required to derive these matrices as described in Appendix \ref{Algorithm}.}
	\label{Tset}
\end{figure}

Examples of this process are given in Fig.~\ref{Tset}. For $T_{12}$ on the left, all of the diagonal elements are $\pm1$ since boxes 1 and 2 in the numbered tableaux representing the basis states are always adjacent to each other. The first two are negative since box 2 is always found under box 1, while the third is positive because its corresponding tableau has box 2 to the right of box one. All of the off diagonal elements involving this state are equal to 0.

For $T_{23}$ in the middle, we need to look at the relative positions of boxes 2 and 3. In the first tableau, box 3 is directly below box 2, so $d=1$ and the sign of the diagonal element is negative, so the diagonal element is $-1$ and all off-diagonal elements involving this state are 0. The remaining two tableaux are related by a swap of boxes 2 and 3, so they will have the same value of $d$, with one diagonal element positive and one negative, with off-diagonal elements between them given by $\sqrt{(d^2-1)/d}$. For these tableaux, boxes 2 and 3 are two boxes apart since they are separated by box 1, so $d=2$, and since the first of the two tableaux has box 3 above and to the right of box 2, the diagonal elements are $+1/2$ followed by $-1/2$, with off diagonal elements given by $\sqrt{(2^2-1)/2}=\sqrt{3/2}$.

For $T_{34}$ on the right, boxes 3 and 4 are relevant. The first two tableaux are related by a swap of boxes 3 and 4, so their off-diagonal elements will be non-zero. There are two boxes between boxes 3 and 4, so $d=3$ and therefore the off-diagonal elements are given by $\sqrt{(3^2-1)/3}=2\sqrt{2/3}$. The first tableau has box 4 above and to the right of box 3, so the diagonal elements are $+1/3$ followed by $-1/3$. Finally, the last tableau has boxes 3 and 4 adjacent to each other with box 4 below box 3, so the last remaining nonzero element is the third diagonal, which is given by $-1$.

\subsection{Transformation from standard basis}
To further convince the reader that Young's orthogonal basis is a valid choice of basis for color singlet Fock states, we can take a look at how this basis is related to a more naive but more intuitive choice of defining the color singlet Fock states. If we assume that we have a number of quarks $n$ and an equal number of antiquarks, we can define all color singlet states of this type using the form
\begin{equation}
\left|n,\mathcal{P}_n\right> = \delta_{a_1...a_n;\mathcal{P}_n[b_1...b_n]}~b^\dagger_{1,a_1}...b^\dagger_{n,a_n}d^\dagger_{1,b_1}...d^\dagger_{n,b_n}\left|0\right>,
\end{equation}
where $\delta_{a_1...a_n;b_1...b_n}=\delta_{a_1b_1}\delta_{a_2b_2}...\delta_{a_nb_n}$ and $\mathcal{P}_n$ is some permutation of $n$ elements. We also assume that each quark and antiquark operator is distinguishable from all others by quantum numbers other than color. This set of states is unnormalized, not orthogonal, and if the number of colors is less than $n$, it is overcomplete. However, this basis does have the advantage that an operator $\mathcal{T}_b(\mathcal{P}_n)$ which permute the quark creation operators by a permutation $\mathcal{P}_n$ has the effect
\begin{equation}
\mathcal{T}_b(\mathcal{P}_n)\left|n,\mathcal{P}_n'\right> = (-1)^{n-\ell(\mathcal{P}_n)}\left|n,\mathcal{P}_n.\mathcal{P}_n'\right>,
\end{equation}
where $\ell(\mathcal{P}_n)$ gives the number of disjoint cycles in the permutation $\mathcal{P}_n$, including 1-cycles corresponding to particles that do not get moved. Likewise, the similar operator $\mathcal{T}_d(\mathcal{P}_n)$ that permutes the antiquark operators gives
\begin{equation}
\mathcal{T}_d(\mathcal{P}_n)\left|n,\mathcal{P}_n'\right> = (-1)^{n-\ell(\mathcal{P}_n)}\left|n,\mathcal{P}_n'.\mathcal{P}_n^{-1}\right>.
\end{equation}
The overlap of two states in this basis is given by
\begin{equation}
\left<n,\mathcal{P}_n\right|\left|n,\mathcal{P}_n'\right> = \delta_{a_1...a_n;(\mathcal{P}_n^{-1}.\mathcal{P}_n')[a_1...a_n]} = N_c^{\ell(\mathcal{P}_n^{-1}.\mathcal{P}_n')}.
\end{equation}

In contrast, states corresponding to Young's orthogonal basis form a complete orthonormal basis. They are labeled by the state $\left|\lambda;u_\lambda,v_\lambda\right>$, where $\lambda$ refers one of the irreducible representations of the symmetric group $S_n$ and both $u_\lambda$ and $v_\lambda$ are vector space indices for the irrep $\lambda$. The effects of the permutation operators are 
\begin{equation}
\mathcal{T}_b(\mathcal{P}_n)\left|\lambda;u_\lambda,v_\lambda\right> = (-1)^{n-\ell(\mathcal{P}_n)}(T_{\mathcal{P}_n})_{u_\lambda'u_\lambda}\left|\lambda;u_\lambda',v_\lambda\right>, \quad \mathcal{T}_d(\mathcal{P}_n)\left|\lambda;u_\lambda,v_\lambda\right> = (-1)^{n-\ell(\mathcal{P}_n)}(T_{\mathcal{P}_n})_{v_\lambda'v_\lambda}\left|\lambda;u_\lambda,v_\lambda'\right>,
\end{equation}
where $T_{\mathcal{P}_n}$ is the permutation matrix for irrep $\lambda$ written in Young's orthogonal basis, which was discussed earlier in Appendix \ref{app:mats}. The Fock states can be written in terms of this basis by
\begin{equation}
\left|n,\mathcal{P}_n\right> = \sum_{\lambda}\sum_{u_\lambda,v_\lambda=1}^{d_\lambda} \sqrt{\mathcal{D}_\lambda}~(T_{\mathcal{P}_n})_{u_\lambda v_\lambda}\left|\lambda;u_\lambda,v_\lambda\right>,
\end{equation}
where $\lambda$ is summed over the irreducible representations of the symmetric group $S_n$, $d_\lambda$ is the dimension of the irrep $\lambda$, $\mathcal{D}_\lambda$ is the dimension of the irrep $\lambda$, $u_\lambda$ and $v_\lambda$ are vector indices for the irrep $\lambda$, and $T_{\mathcal{P}_n}$ is the permutation matrix in Young's orthogonal basis discussed above corresponding to the irrep $\lambda$.

To see how this works, consider the actions of $\mathcal{T}_b(\mathcal{P}_n)$ and $\mathcal{T}_d(\mathcal{P}_n)$ on the right side of the equation. Permuting the quarks gives
\begin{align}
\nonumber (-1)^{n-\ell(\mathcal{P}_n)}\mathcal{T}_b(\mathcal{P}_n)\left|n,\mathcal{P}_n'\right> &= \sum_{\lambda}\sum_{u_\lambda,v_\lambda=1}^{d_\lambda} \sqrt{\mathcal{D}_\lambda}~(T_{\mathcal{P}_n'})_{u_\lambda v_\lambda}(T_{\mathcal{P}_n})_{u_\lambda'u_\lambda}\left|\lambda;u_\lambda',v_\lambda\right>\\
  &= \sum_{\lambda}\sum_{u_\lambda,v_\lambda=1}^{d_\lambda} \sqrt{\mathcal{D}_\lambda}~(T_{(\mathcal{P}_n.\mathcal{P}_n')})_{u_\lambda v_\lambda}\left|\lambda;u_\lambda,v_\lambda\right> = \left|n,\mathcal{P}_n.\mathcal{P}_n'\right>,
\end{align}
so this matches. The antiquark permutation gives
\begin{align}
\nonumber (-1)^{n-\ell(\mathcal{P}_n)}\mathcal{T}_d(\mathcal{P}_n)\left|n,\mathcal{P}_n'\right> &= \sum_{\lambda}\sum_{u_\lambda,v_\lambda=1}^{d_\lambda} \sqrt{\mathcal{D}_\lambda}~(T_{\mathcal{P}_n'})_{u_\lambda v_\lambda}(T_{\mathcal{P}_n})_{v_\lambda'v_\lambda}\left|\lambda;u_\lambda,v_\lambda'\right>\\
  &= \sum_{\lambda}\sum_{u_\lambda,v_\lambda=1}^{d_\lambda} \sqrt{\mathcal{D}_\lambda}~(T_{(\mathcal{P}_n'.\mathcal{P}_n^{-1})})_{u_\lambda v_\lambda}\left|\lambda;u_\lambda,v_\lambda\right> = \left|n,\mathcal{P}_n'.\mathcal{P}_n^{-1}\right>,
\end{align}
so this also matches. Finally, the overlap is given by
\begin{align}
\nonumber \left<n,\mathcal{P}_n\right|\left|n,\mathcal{P}_n'\right> &= \sum_{\lambda}\sum_{u_\lambda,v_\lambda=1}^{d_\lambda} \mathcal{D}_\lambda~(T_{\mathcal{P}_n})_{u_\lambda v_\lambda}(T_{\mathcal{P}_n'})_{u_\lambda v_\lambda} \\
  &= \sum_{\lambda}\mathcal{D}_\lambda~\mathrm{Tr}\left[T_{(\mathcal{P}_n^{-1}.\mathcal{P}_n')}\right] = \sum_{\lambda}\mathcal{D}_\lambda~\chi_\lambda(\mu(\mathcal{P}_n^{-1}.\mathcal{P}_n')),
\end{align}
where $\chi_\lambda(\mu)$ is the character associated with the irrep $\lambda$ and the conjugacy class $\mu$ and $\mu(\mathcal{P}_n))$ is the conjugacy class of the partition $\mathcal{P}_n$.

To prove that $\sum_{\lambda}\mathcal{D}_\lambda~\chi_\lambda(\mu(\mathcal{P}_n))=N_c^{\ell(\mathcal{P}_n)}$, we will need to introduce the Frobenius formula \cite{hewitt2012elements,harris1991representation} for calculating the values of characters. We start with the polynomial given by
\begin{align}
  P_n(\vec{x},\mu) &= A_n(\vec{x})\prod_{k=1}^{\ell(\mu)}\left(\sum_{i=1}^{N_c}x_i^{\mu_k}\right) = \sum_{\alpha_1,...,\alpha_{N_c}=1}^{N_c+n-1}c_{\alpha_1...\alpha_{N_c}}(\mu)\prod_{i=1}^{N_c}x_i^{\alpha_i}, \\
  A_n(\vec{x}) &= \prod_{i=1}^{N_c}\prod_{j=i+1}^{N_c}(x_i-x_j) = \sum_{a_1,...,a_{N_c}=1}^{N_c+n-1}\epsilon_{a_1...a_{N_c}}\prod_{i=1}^{N_c}x_i^{N_c-a_i},
\end{align}
where $\vec{x}$ is a $N_c$-element vector of variables, $\mu_k$ is the $k$th element of the integer partition of $n$ corresponding to the conjugacy class $\mu$, $\ell(\mu)$ is the length of this partition, and $\epsilon_{a_1...a_{N_c}}$ is the Levi-Civita tensor with $\epsilon_{1...N_c}=+1$. The Frobenius formula states that the polynomial $P_n(\vec{x},\mu)$ is a generator of the characters in the sense that
\begin{equation}
c_{\alpha_1>\alpha_2>...>\alpha_{N_c}}(\mu) = \chi_\lambda(\mu), \quad \lambda_i = \alpha_i-n+i,
\end{equation}
where $\lambda_i$ is the length of the $i$th row of the tableau associated with the irrep $\lambda$, so that the second equation above gives a connection between the $\alpha_i$'s and the irrep $\lambda$ of the character. From the formulas of $P_n(\vec{x},\mu)$ and $A_n(\vec{x})$ above, we can see that both polynomials are antisymmetric under any exchange of two of the $x_i$ variables, so we can go a step further and say that
\begin{equation}
c_{\alpha_1...\alpha_{N_c}}(\mu) = -\chi_\lambda(\mu)\epsilon_{\alpha_1...\alpha_{N_c}}.
\end{equation}

To connect this back to our original expression $\sum_{\lambda}\mathcal{D}_\lambda~\chi_\lambda(\mu)$, we will need an explicit expression for $\mathcal{D}_\lambda$. It can be written as
\begin{equation}
\mathcal{D}_\lambda = \prod_{i=1}^{N_c}\frac{(N_c+\lambda_i-i)!}{(N_c-i)!}\frac{d_\lambda}{n!},
\end{equation}
where $d_\lambda=\chi_\lambda(\mathbb{1})$ is the dimension of the irrep $\lambda$, equivalent to the character of the irrep under the conjugacy class of the identity matrix. We can then show that our original expression is equivalent up to a constant to the quantity
\begin{align}
\nonumber P_n(\vec{\nabla},\mathbb{1})P_n(\vec{x},\mu) &= \left(\sum_{\beta_1,...,\beta_{N_c}=1}^{N_c+n-1}c_{\beta_1...\beta_{N_c}}(\mathbb{1})\prod_{j=1}^{N_c}\nabla_j^{\beta_j}\right) \sum_{\alpha_1,...,\alpha_{N_c}=1}^{N_c+n-1}c_{\alpha_1...\alpha_{N_c}}(\mu)\prod_{i=1}^{N_c}x_i^{\alpha_i} \\
\nonumber &= \sum_{\alpha_1,...,\alpha_{N_c}=1}^{N_c+n-1}c_{\alpha_1...\alpha_{N_c}}(\mathbb{1}) c_{\alpha_1...\alpha_{N_c}}(\mu)\left(\prod_{i=1}^{N_c}(\alpha_i)!\right) \\
\nonumber &= \sum_{\alpha_1,...,\alpha_{N_c}=1}^{N_c+n-1}d_\lambda~\chi_\lambda(\mu)(\epsilon_{\alpha_1...\alpha_{N_c}})^2\left(\prod_{i=1}^{N_c}(N_c+\lambda_i-i)!\right) \\
  &= n!(N_c)!\left(\prod_{i=1}^{N_c}(N_c-i)!\right)\left(\sum_{\lambda}\mathcal{D}_\lambda~\chi_\lambda(\mu)\right).
\end{align}

On the other hand, if we use the other definition of $P_n$ for this expression, we get
\begin{equation}
P_n(\vec{\nabla},\mathbb{1})P_n(\vec{x},\mu) = A_n(\vec{\nabla})\left(\sum_{i=1}^{N_c}\nabla_i\right)^n\left(A_n(\vec{x})\prod_{k=1}^{\ell(\mu)}\left(\sum_{i=1}^{N_c}x_i^{\mu_k}\right)\right).
\end{equation}
Since $A_n(\vec{x})$ is a product of differences between pairs of variables and $\left(\sum_{k=1}^{N_c}\nabla_k\right)(x_i-x_j)=0$ for any pairing of indices $i$ and $j$, we have that $\left(\sum_{i=1}^{N_c}\nabla_i\right)A_n(\vec{x})=0$. Also, if we expand the product $\prod_{k=1}^{\ell(\mu)}\left(\sum_{i=1}^{N_c}x_i^{\mu_k}\right)$, each term in the sum will have exactly $n$ factors of $x$ variables, so the $n$th derivative of this $\left(\sum_{k=1}^{N_c}\nabla_k\right)^n\prod_{k=1}^{\ell(\mu)}\left(\sum_{i=1}^{N_c}x_i^{\mu_k}\right)$ is equal to $n!$ times the number of terms in the polynomial. Thus we have
\begin{equation}
P_n(\vec{\nabla},\mathbb{1})P_n(\vec{x},\mu) = \left(A_n(\vec{\nabla})A_n(\vec{x})\right)n!N_c^{\ell(\mu)}.
\end{equation}
The remaining polynomial and derivative term gives
\begin{align}
\nonumber A_n(\vec{\nabla})A_n(\vec{x}) &= \sum_{b_1,...,b_{N_c}=1}^{N_c+n-1}\epsilon_{b_1...b_{N_c}}\prod_{j=1}^{N_c}\nabla_j^{N_c-b_j}\left(\sum_{a_1,...,a_{N_c}=1}^{N_c+n-1}\epsilon_{a_1...a_{N_c}}\prod_{i=1}^{N_c}x_i^{N_c-a_i}\right) \\
  &= \sum_{a_1,...,a_{N_c}=1}^{N_c+n-1}\left(\epsilon_{a_1...a_{N_c}}\right)^2\prod_{i=1}^{N_c}(N_c-a_i)! = (N_c)!\prod_{i=1}^{N_c}(N_c-i)!.
\end{align}
Now, with our two different expressions for $P_n(\vec{\nabla},\mathbb{1})P_n(\vec{x},\mu)$ we obtain the relation
\begin{equation}
\frac{1}{n!(N_c)!\prod_{i=1}^{N_c}(N_c-i)!}P_n(\vec{\nabla},\mathbb{1})P_n(\vec{x},\mu) = \sum_{\lambda}\mathcal{D}_\lambda~\chi_\lambda(\mu) = N_c^{\ell(\mu)}.
\end{equation}
Thus we can finally show that the overlap of the Fock states is given by
\begin{equation}
  \left<n,\mathcal{P}_n\right|\left|n,\mathcal{P}_n'\right> = \sum_{\lambda}\mathcal{D}_\lambda~\chi_\lambda(\mu(\mathcal{P}_n^{-1}.\mathcal{P}_n')) = N_c^{\ell(\mu(\mathcal{P}_n^{-1}.\mathcal{P}_n'))},
\end{equation}
using the assumption that it can be written in terms of the Young's orthogonal basis states.

We can also show what the Young's orthogonal basis states are in terms of the Fock states. To derive this, we will need to find a closed form expression for the quantity
\begin{equation}
S_{u_\lambda u_\Lambda,v_\lambda v_\Lambda } = \sum_{\mathcal{P}_n}(T_{\mathcal{P}_n})_{u_\lambda v_\lambda}(T_{\mathcal{P}_n})_{u_\Lambda v_\Lambda},
\end{equation}
where $T_{\mathcal{P}_n}$ is the permutation matrix in Young's orthogonal basis for a permutation $\mathcal{P}_n$ and an irrep $\lambda$ or $\Lambda$. Since we are summing over all permutations and the $T_{\mathcal{P}_n}$ are orthogonal, we can show that
\begin{align}
\nonumber (T_{\mathcal{P}_n})_{u_\lambda u'_\lambda}S_{u'_\lambda u_\Lambda,v_\lambda v_\Lambda} &= \sum_{\mathcal{P}_n'}(T_{\mathcal{P}_n.\mathcal{P}_n'})_{u_\lambda v_\lambda}(T_{\mathcal{P}_n'})_{u_\Lambda v_\Lambda} = \sum_{\mathcal{P}_n''}(T_{\mathcal{P}_n''})_{u_\lambda v_\lambda}(T_{\mathcal{P}_n^{-1}.\mathcal{P}_n''})_{u_\Lambda v_\Lambda} \\ &= S_{u_\lambda u'_\Lambda,v_\lambda v_\Lambda}(T_{\mathcal{P}_n})_{u'_\Lambda u_\Lambda}.
\end{align}
Likewise, we also have
\begin{equation}
(T_{\mathcal{P}_n})_{v_\lambda v'_\lambda}S_{u_\lambda u_\Lambda,v'_\lambda v_\Lambda} = S_{u_\lambda u_\Lambda,v_\lambda v'_\Lambda}(T_{\mathcal{P}_n})_{v'_\Lambda v_\Lambda}.
\end{equation}
If we view $T_{\mathcal{P}_n}$ as a matrix not just for a specific irrep but as a block diagonal matrix over all irreps, then these formulas tell us that $S$ commutes with all of the $T_{\mathcal{P}_n}$'s as a matrix over the $u$ and $v$ indices separately. The only way that this is possible is if $S$ has the form
\begin{equation}
S_{u_\lambda u_\Lambda,v_\lambda v_\Lambda} = \alpha_\lambda~\delta_{\lambda\Lambda}~\delta_{u_\lambda u_\Lambda} \delta_{v_\lambda v_\Lambda},
\end{equation}
for some yet to be determined constant $\alpha_\lambda$. In other words, the tensor $S$ is zero unless the two irreps are the same, in which case it is proportional to the outer product of the identity matrix in that irrep  with itself.

To find this constant, we can take the sum over the diagonal elements of $S$ using its definition and by the form above. On one hand, the sum over the diagonal elements of $S$ is just $\alpha_\lambda d_\lambda^2$, where $d_\lambda$ is the dimension of the irrep $\lambda$. On the other hand, the original definition of $S$ gives us a sum that looks like
\begin{equation}
\sum_{u_\lambda,v_\lambda}\sum_{\mathcal{P}_n}(T_{\mathcal{P}_n})_{u_\lambda v_\lambda}(T_{\mathcal{P}_n})_{u_\lambda v_\lambda}=\sum_{u_\lambda}\sum_{\mathcal{P}_n}1=n!~d_\lambda,
\end{equation}
where $n$ corresponds to the symmetric group $S_n$ and we used the fact that the matrix $T_{\mathcal{P}_n}$ is orthogonal. Thus we have that
\begin{equation}
S_{u_\lambda u_\Lambda,v_\lambda v_\Lambda } = \sum_{\mathcal{P}_n}(T_{\mathcal{P}_n})_{u_\lambda v_\lambda}(T_{\mathcal{P}_n})_{u_\Lambda v_\Lambda} = \frac{n!}{d_\lambda}~\delta_{\lambda\Lambda}~\delta_{u_\lambda u_\Lambda} \delta_{v_\lambda v_\Lambda}.
\label{Stensor}
\end{equation}
From here, it is relatively easy to achieve the original goal of writing a Young's orthogonal basis state in terms of Fock states. We have that
\begin{equation}
\sum_{\mathcal{P}_n}(T_{\mathcal{P}_n})_{u_\lambda v_\lambda}\left|n,\mathcal{P}_n\right> = \sum_{\Lambda}\sum_{u_\Lambda,v_\Lambda} \sqrt{\mathcal{D}_\Lambda}~\left(\sum_{\mathcal{P}_n}(T_{\mathcal{P}_n})_{u_\lambda v_\lambda}(T_{\mathcal{P}_n})_{u_\Lambda v_\Lambda}\right)\left|\Lambda;u_\Lambda,v_\Lambda\right> = \sqrt{\mathcal{D}_\lambda}~\frac{n!}{d_\lambda}\left|\lambda;u_\lambda,v_\lambda\right>.
\end{equation}
Thus the Young's orthogonal basis state can be written as
\begin{equation}
\left|\lambda;u_\lambda,v_\lambda\right> = \frac{d_\lambda}{n!\sqrt{\mathcal{D}_\lambda}}\sum_{\mathcal{P}_n}(T_{\mathcal{P}_n})_{u_\lambda v_\lambda}\left|n,\mathcal{P}_n\right>.
\end{equation}

So far, we have discussed Fock states with an equal number of quarks and antiquarks, which are directly relevant to forming an orthonormal basis for mesons. This basis conversion can also be used for baryons as well, by attaching a Levi-Civita tensor to one of the quarks and treating it as a fully antisymmetrized collection of $N_c-1$ antiquarks.

\subsection{Action of $d_cb_c$ type operators in Young's orthogonal basis}
\label{app:dbtype}

The goal of this subsection is to illustrate how to find the matrix element $\left<\lambda,u_\lambda v_\lambda\right|d_{j,c}b_{i,c}\left|\Lambda,u_\Lambda v_\Lambda\right>$, where $\left|\Lambda,u_\Lambda v_\Lambda\right>$ is a Young's orthogonal basis state composed of $n$ distinct quarks and antiquarks for an irrep $\Lambda$ in $S_n$, $\left|\lambda,u_\lambda v_\lambda\right>$ is a Young's orthogonal basis state composed of the same quarks and antiquarks except for the $i$th quark and the $j$th antiquark for an irrep $\lambda$ in $S_{n-1}$, and $d_{j,c}b_{i,c}$ annihilates quark $i$ and antiquark $j$ with an implicit sum over the color index $c$. To simplify things, we can use permutation matrices to move the $i$th quark and the $j$th antiquark to their respective $n$th positions in the Fock state so that the annihilation operators always manipulate the $n$th color indices. This means that as far as the color structures are concerned the matrix element is numerically equivalent to
\begin{equation}
\left<\lambda,u_\lambda v_\lambda\right|d_{j,c}b_{i,c}\left|\Lambda,u_\Lambda v_\Lambda\right> = (-1)^{i+j}\left(\prod_{k=i}^{n-1}T_{k,k+1}\right)_{u_\Lambda u'_\Lambda}\left(\prod_{l=j}^{n-1}T_{l,l+1}\right)_{v_\Lambda v'_\Lambda}\left<\lambda,u_\lambda v_\lambda\right|d_{n,c}b_{n,c}\left|\Lambda,u_\Lambda' v_\Lambda'\right>,
\end{equation}
where $T_{k,k+1}$ is the permutation matrix in the $\Lambda$ irrep of $S_n$ that swaps the color indices of operators $k$ and $k+1$. This is helpful because it gives us the explicit $i$ and $j$ dependence of the matrix element relative to the choice where $i=j=n$.

To find the remaining matrix element, it will be most convenient to find the action of $d_{n,c}b_{n,c}$ on Fock states, and then change the basis to Young's orthogonal basis. The permutation $\mathcal{P}_n$ in $S_n$ can either be written as $p_{n-1}\otimes\mathbf{1}$, meaning that it is a permutation $p_{n-1}$ in $S_{n-1}$ that does not move element $n$, or if $\mathcal{P}_n(i)=n$ it can be written in the form $\mathcal{P}_n=T_{in}.(p_{n-1}\otimes\mathbf{1})$, where $T_{in}$ swaps the elements $i$ and $n$. In the former case the operator simply annihilates the $n$th quark and antiquark and so
\begin{equation}
d_{n,c}b_{n,c}\left|n,p_{n-1}\otimes\mathbf{1}\right> = (-1)^{n-1}N_c\left|n,p_{n-1}\right>,
\end{equation}
where the factor of $N_c$ comes from the sum over the color index $c$. For the latter case, when the $n$ quark and antiquark operators get annihilated, their color indices get matched together so that the resulting permutation moves $i\rightarrow n\rightarrow p_{n-1}(i)$. Thus we get
\begin{equation}
d_{n,c}b_{n,c}\left|n,T_{in}.(p_{n-1}\otimes\mathbf{1})\right> = (-1)^{n-1}\left|n,p_{n-1}\right>.
\end{equation}
A Young's orthogonal basis matrix acted upon by the operator becomes
\begin{align}
\nonumber d_{n,c}b_{n,c}\left|\Lambda,u_\Lambda v_\Lambda\right> =&~ \frac{d_\Lambda}{n!\sqrt{\mathcal{D}_\Lambda}}\sum_{\mathcal{P}_n}(T_{\mathcal{P}_n})_{u_\Lambda v_\Lambda}d_{n,c}b_{n,c}\left|n,\mathcal{P}_n\right> \\
\nonumber =&~ \frac{(-1)^{n-1}d_\Lambda}{n!\sqrt{\mathcal{D}_\Lambda}}\left(N_c\sum_{p_{n-1}}(T_{p_{n-1}\otimes\mathbf{1}})_{u_\Lambda v_\Lambda}+\sum_{i}\sum_{p_{n-1}}(T_{in}.T_{p_{n-1}\otimes\mathbf{1}})_{u_\Lambda v_\Lambda}\right)\left|n-1,p_{n-1}\right> \\
\nonumber =&~ (-1)^{n-1}\frac{d_\Lambda}{n!}\sum_{\lambda}\sqrt{\frac{\mathcal{D}_\lambda}{\mathcal{D}_\Lambda}}\sum_{u_\Lambda'}\sum_{u_\lambda,v_\lambda}\left(N_c\delta_{u_\Lambda u_\Lambda'}+\sum_{i=1}^{n-1}(T_{in})_{u_\Lambda u_\Lambda'}\right) \\
  & * \left(\sum_{p_{n-1}}(T_{p_{n-1}\otimes\mathbf{1}})_{u_\Lambda' v_\Lambda}(T_{p_{n-1}})_{u_\lambda v_\lambda}\right)\left|\lambda;u_\lambda,v_\lambda\right>.
\end{align}
Thus the remaining matrix element is
\begin{equation}
\left<\lambda,u_\lambda v_\lambda\right|d_{n,c}b_{n,c}\left|\Lambda,u_\Lambda v_\Lambda\right> = (-1)^{n-1}\frac{d_\Lambda}{n!}\sqrt{\frac{\mathcal{D}_\lambda}{\mathcal{D}_\Lambda}}\sum_{u_\Lambda'}\mathcal{R}_{u_\Lambda u_\Lambda'}\mathcal{S}_{u_\lambda u_\Lambda', v_\lambda v_\Lambda},
\end{equation}
with
\begin{equation}
\mathcal{S}_{u_\lambda u_\Lambda, v_\lambda v_\Lambda} = \sum_{p_{n-1}}(T_{p_{n-1}})_{u_\lambda v_\lambda}(T_{p_{n-1}\otimes\mathbf{1}})_{u_\Lambda v_\Lambda}
\end{equation}
and
\begin{equation}
\mathcal{R}_{u_\Lambda u_\Lambda'} = \left(N_c\delta_{u_\Lambda u_\Lambda'}+\sum_{i=1}^{n-1}(T_{in})_{u_\Lambda u_\Lambda'}\right).
\end{equation}

For the first tensor $\mathcal{S}$, we can see that it is very similar to the tensor $S$ given in Eq.~\eqref{Stensor}. This new tensor has the similar property
\begin{equation}
(T_{p})_{u_\lambda u'_\lambda}\mathcal{S}_{u_\lambda' u_\Lambda, v_\lambda v_\Lambda} = \mathcal{S}_{u_\lambda u_\Lambda', v_\lambda v_\Lambda}(T_{p\otimes\mathbf{1}})_{u'_\Lambda u_\Lambda}
\end{equation}
for a permutation $p$ in $S_{n-1}$, with a similar equation for the $v$ indices. As with the tensor $S$, these relations imply that $\mathcal{S}$ is a direct product of two matrices, so that
\begin{equation}
\mathcal{S}_{u_\lambda u_\Lambda, v_\lambda v_\Lambda} = \beta_{\lambda\Lambda}(M_{\lambda\Lambda})_{u_\Lambda u_\lambda}(M_{\lambda\Lambda})_{v_\Lambda v_\lambda}
\end{equation}
for some matrix $M_{\lambda\Lambda}$ with the property
\begin{equation}
T_{p\otimes\mathbf{1}}.M_{\lambda\Lambda} = M_{\lambda\Lambda}.T_{p}.
\end{equation}
This implies that the matrix $M_{\lambda\Lambda}^T.M_{\lambda\Lambda}$ is proportional to the identity in the smaller basis. Since we define $M_{\lambda\Lambda}$ through $\mathcal{S}$ using an undetermined constant $\beta_{\lambda\Lambda}$, we can choose $M_{\lambda\Lambda}$ to be defined such that $M_{\lambda\Lambda}^T.M_{\lambda\Lambda}=\mathbb{1}$ exactly. This also implies that the rows of the matrix $M_{\lambda\Lambda}$ are orthonormal vectors, which means that the matrix $M_{\lambda\Lambda}.M_{\lambda\Lambda}^T$ is a projection operator in the larger basis.

\begin{figure}
	\centering
	\includegraphics[width=0.75\textwidth]{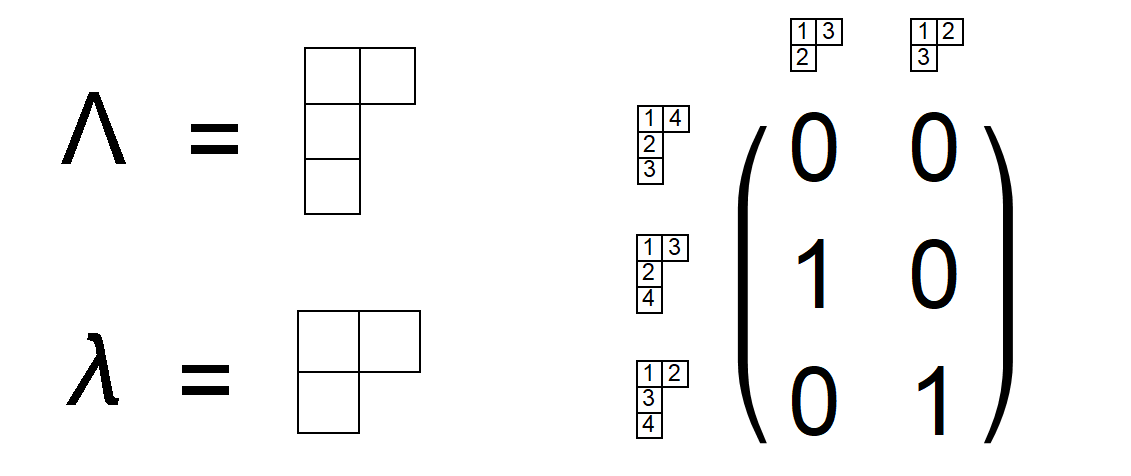}
	\caption{The matrix $M_{\lambda\Lambda}$ where $\Lambda$ is the antifundamental irrep of $S_4$ and $\lambda$ is the mixed symmetry irrep of $S_3$, all written in Young's orthogonal basis. The numbered Young tableaux correspond to the basis elements of the vector space and are required to derive this matrix as described in Appendix \ref{Algorithm}. Note that the only nonzero elements occur when the numbered tableau on the left matches the numbered tableau on top with box $4$ removed.}
	\label{Mset}
\end{figure}

Looking at the way that Young's orthogonal matrices are constructed, we can see that the matrix $M_{\lambda\Lambda}$ is actually a section of the identity matrix in the larger space, where the ones appear where the numbered tableau on the smaller basis is the same as the numbered tableau in the larger basis with box $n$ removed, as seen in Fig.~\ref{Mset}.

To find $\beta_{\lambda\Lambda}$, we can once again take traces of the tensor $\mathcal{S}$ to get
\begin{equation}
\sum_{u_\lambda, u_\Lambda}\mathcal{S}_{u_\lambda u_\Lambda, u_\lambda u_\Lambda} = \beta_{\lambda\Lambda}~d_\lambda = \sum_{\mu}d_\mu~\chi_\lambda(\mu)\chi_\Lambda(\nu),
\end{equation}
where $d_\lambda$ is the dimension of the smaller irrep $\lambda$, $\mu$ is one of the conjugacy classes of the permutations in $S_{n-1}$, $\nu$ is the conjugacy class in $S_n$ such that if $\mu$ corresponds to a permutation $p$, then $\nu$ corresponds to $p\otimes\mathbf{1}$, the characters $\chi_\lambda(\mu)$ and $\chi_\Lambda(\nu)$ originate from traces of permutation matrices, and $d_\mu$ is the dimension of the conjugacy class $\mu$, equivalent to the number of matrices in the conjugacy class $\mu$. The character $\chi_\Lambda(\nu)$ can be written in terms of the characters $\chi_\lambda(\mu)$ by using the Frobenius formula. The polynomial corresponding to the conjugacy class $\nu$ can be written in terms of the smaller ones for $\mu$ by
\begin{equation}
F_n(\mathbf{x},\nu) = F_{n-1}(\mathbf{x},\mu)\left(\sum_{i=1}^{n}x_i\right).
\end{equation}
This implies that $\chi_\Lambda(\nu)=\sum_{i=1}^{n}\chi_{\lambda_i}(\mu)$, where $\lambda_i$ is the irrep obtained from taking the tableau associated with $\Lambda$ and removing one box from row $i$. The character will be equal to zero if $\lambda_i$ does not correspond to a valid tableau. Finally, we note that $\chi_\lambda(\mu)$ is orthogonal as a vector in the irreps  $\lambda$, and if that vector is normalized it will also be orthonormal in the conjugacy classes $\mu$. The normalization factor is equal to $(n-1)!/d_\mu$ for a class $\mu$ in $S_{n-1}$. Thus we can put this all together to get 
\begin{equation}
\beta_{\lambda\Lambda} = \frac{1}{d_\lambda}\sum_{\mu}\sum_{i=1}^nd_\mu~\chi_\lambda(\mu)\chi_{\lambda_i}(\mu)= \frac{(n-1)!}{d_\lambda}\sum_{i=1}^n\delta_{\lambda_i\lambda}.
\end{equation}
This tells us that the tensor $\mathcal{S}$ is zero unless the tableau for the irrep $\Lambda$ is the same as $\lambda$ except for one extra box. We now have that, when the tensor $\mathcal{S}$ is nonzero, it is equal to
\begin{equation}
\mathcal{S}_{u_\lambda u_\Lambda, v_\lambda v_\Lambda} = \sum_{p_{n-1}}(T_{p_{n-1}})_{u_\lambda v_\lambda}(T_{p_{n-1}\otimes\mathbf{1}})_{u_\Lambda v_\Lambda} = \frac{(n-1)!}{d_\lambda}(M_{\lambda\Lambda})_{u_\Lambda u_\lambda}(M_{\lambda\Lambda})_{v_\Lambda v_\lambda}.
\end{equation}

The other tensor $\mathcal{R}$ is a matrix that can be shown to commute with any permutation of the form $p\otimes\mathbf{1}$ since that permutation will only rearrange the indices $i$ in the sum over matrices that defines $\mathcal{R}$. This implies that the matrix $M_{\lambda\Lambda}^T.\mathcal{R}.M_{\lambda\Lambda}$ is proportional to the identity since it commutes with every permutation matrix in $S_{n-1}$. Also, it can be shown using the definition of $\mathcal{S}$ as a sum over permutations that $(\mathcal{R}.M_{\lambda\Lambda})\otimes M_{\lambda\Lambda}=M_{\lambda\Lambda}\otimes(\mathcal{R}.M_{\lambda\Lambda})$, which further indicates that $\mathcal{R}.M_{\lambda\Lambda}$ is proportional to $M_{\lambda\Lambda}$. It turns out that the proportionality constant is given by
\begin{equation}
\mathcal{R}.M_{\lambda\Lambda} = (N_c+\Lambda_a-a)M_{\lambda\Lambda},
\end{equation}
where $a$ is the row of the tableau for $\Lambda$ that is different from the tableau for $\lambda$ and $\Lambda_a$ is the $a$th value of the permutation corresponding to $\Lambda$. This factor is equal to the $N_c$ dependent factor coming from the box in $\Lambda$ that is not in $\lambda$ when determining the dimension of the $SU(N_c)$ irrep corresponding to $\Lambda$.

Putting all of this together and using the formulas for the dimensions $d_\lambda$ and $\mathcal{D}_\lambda$ to simplify things, we find that
\begin{align}
\nonumber \left<\lambda,u_\lambda v_\lambda\right|d_{n,c}b_{n,c}\left|\Lambda,u_\Lambda v_\Lambda\right> &= (-1)^{n-1}\sqrt{\frac{\mathcal{D}_\lambda}{\mathcal{D}_\Lambda}}\frac{d_\Lambda}{n~d_\lambda}(N_c+\Lambda_a-a)(M_{\lambda\Lambda})_{u_\lambda u_\Lambda}(M_{\lambda\Lambda})_{v_\lambda v_\Lambda} \\
  &= (-1)^{n-1}\sqrt{\frac{\mathcal{D}_\Lambda}{\mathcal{D}_\lambda}}(M_{\lambda\Lambda})_{u_\Lambda u_\lambda}(M_{\lambda\Lambda})_{v_\Lambda v_\lambda},
\end{align}
where we have used that $\frac{\mathcal{D}_\Lambda}{\mathcal{D}_\lambda}=\frac{d_\Lambda}{d_\lambda}\frac{1}{n}(N_c+\Lambda_a-a)$. Thus the full matrix element for any operator $d_{j,c}b_{i,c}$ is given by
\begin{equation}
\left<\lambda,u_\lambda v_\lambda\right|d_{i,c}b_{j,c}\left|\Lambda,u_\Lambda v_\Lambda\right> = (-1)^{n-1+i+j}\sqrt{\frac{\mathcal{D}_\Lambda}{\mathcal{D}_\lambda}}\left(\left(\prod_{k=i}^{n-1}T_{k,k+1}\right).M_{\lambda\Lambda}\right)_{u_\Lambda u_\lambda}\left(\left(\prod_{l=j}^{n-1}T_{l,l+1}\right).M_{\lambda\Lambda}\right)_{v_\Lambda v_\lambda}.
\end{equation}

\subsection{Near-degeneracy and scale factors in basis states}
So far, we have exclusively worked with Fock states that have completely distinguishable quarks and antiquarks. In practice, though, many of the basis states will have nearly degenerate quarks in them, meaning that there are multiple quarks that share all quantum numbers except for color charge. In those cases, the basis has to be truncated into one where the nearly degenerate quarks are antisymmetrized. We can do this by working in the full, non-degenerate basis and then truncate the result using the appropriate change of basis matrices for the states involved in the matrix element. Care must be taken when taking this approach with four-quark and six-quark operators, as the intermediate color singlet states formed by applying part of the full operator may have a higher degree of near-degeneracy that either of the two basis states in the matrix element. If this is not avoided or accounted for, then the result will have extra contributions from states that should not exist. This can be avoided by either finding an alternative way of calculating the matrix elements, as can be done for certain diagonal elements, or by carefully ordering the two-quark color singlet operators so that intermediate states alway have less antisymmetry than the basis states corresponding to the matrix element.

Like the permutation matrices, these projection matrices can be built from the labeled tableau. The procedure for doing this is to first note which quarks and antiquarks need to have their color indices antisymmetrized. Then, we group up all of the labeled tableau that are identical up to permutations of the numerical labels that are antisymmetrized. If a labeled tableau from the full basis does not match a tableau from the reduced basis up to antisymmetrized permutations, then the matrix element between those two tableau must be zero. Finally, we calculate the nonzero elements of the projection matrix using the distances $d$ as defined above between all pairs of boxes that are antisymmetrized. The non-zero elements are found by taking a factor of $1/\sqrt{n!}$ coming from the change in normalization, where $n$ is the number of near-degenerate quarks, and then multiplying a factor of $\mp\sqrt{1\mp1/d}$ for every pair within each group of antisymmetrized boxes, where $\pm d$ is defined as it is above for a pair of boxes.

\begin{figure}
	\centering
	\includegraphics[width=1.0\textwidth]{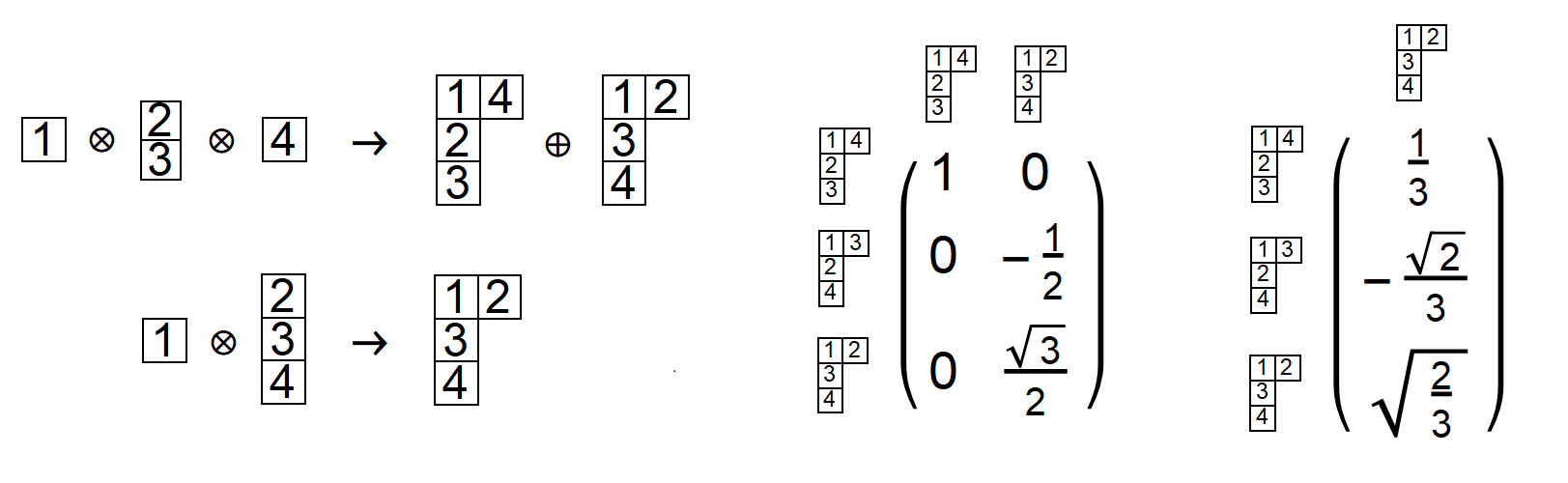}
	\caption{Two change of basis matrices for the antifundamental irrep of $S_4$ written in Young's orthogonal basis. The first matrix changes from Young's orthogonal basis to a basis where particles $2$ and $3$ are antisymmetrized, as illustrated by the first direct product. The second matrix changes from Young's orthogonal basis to a basis where particles $2$,$3$, and $4$ are antisymmetrized, as illustrated by the second direct product. The numbered Young tableaux correspond to the basis elements of the vector space and are required to derive these matrices as described in Appendix \ref{Algorithm}.}
	\label{Bset}
\end{figure}

Examples of these projection matrices are given in Fig.~\ref{Bset}. The first example shows the matrix needed when quarks 2 and 3 are antisymmetrized, with the two relevant 4-box tableaux derived in the first of the direct products shown on the left. The second example shows the matrix needed when particles 2, 3, and 4 are all antisymmetrized, with the single relevant 4-box tableau derived in the other direct product shown on the left.

For the first matrix on the left, we first need to group each of the numbered tableau in the full basis with one of the tableau in the reduced basis to see which elements must be zero. The first tableau in both bases are identical, and no other tableau match them up to permutations of boxes 2 and 3, so the first column must be all zeros except for the first one. The remaining two tableaux in the full basis match the second tableau in the reduced basis up to a permutation of boxes 2 and 3, so only the first element in the second column is guaranteed to be zero. For the first tableau, we have between boxes 2 and 3 a value $\pm d=-1$, so the value of the first element in the matrix is $\mp\sqrt{(1\mp1/d)/n!}=+\sqrt{(1+1)/2!}=1$. The values of $d$ between boxes 2 and 3 for the remaining two tableaux are $\pm d=2$ and $\pm d=-2$, so the remaining two elements of the matrix are $-\sqrt{(1-1/2)/2!}=-1/2$ and $+\sqrt{(1+1/2)/2!}=\sqrt{3}/2$.

For the second matrix on the right, there is only one state in the reduced basis, and all three tableaux in the full basis match it up to permutations in boxes 2, 3, and 4, so all elements of the matrix can be nonzero. To find these elements, we need the values of $d$ between boxes 2 and 3, between 2 and 4, and between 3 and 4 for each of the tableaux in the full basis. For the first element and tableau, we have $\pm d_{23}=-1,\pm d_{24}=2,\pm d_{34}=3$, so the value of the element is given by
\begin{equation}
(\mp)_{23}(\mp)_{24}(\mp)_{34}\sqrt{(1\mp1/d_{23})(1\mp1/d_{24})(1\mp1/d_{34})/n!}=+\sqrt{(1+1)(1-1/2)(1-1/3)/3!}=1/3.
\end{equation}
For the second element, we have $\pm d_{23}=2,\pm d_{24}=-1,\pm d_{34}=-3$, so the matrix element is equal to $-\sqrt{(1-1/2)(1+1)(1+1/3)/3!}=-\sqrt{2}/3$. Finally, the last one has $\pm d_{23}=-2,\pm d_{24}=-3,\pm d_{34}=-1$, which becomes $+\sqrt{(1+1/2)(1+1/3)(1+1)/3!}=\sqrt{2/3}$.

Finally, there will also be a scale factor needed to account for changes in normalization and degeneracy factors. Each creation operator added to a state incurs a factor given by the square root of the near-degeneracy of the added state {\it after} the operator is applied, while for any annihilation operator we get the square root of the near-degeneracy {\it before} the state is removed. This is because the near-degeneracy of states needs to be normalized by an extra factor of $1/\sqrt{g!}$, where $g$ is the number of near-degenerate copies of a specific type of quark or antiquark, so when a new state with the same quantum numbers is added we get a state that is $\sqrt{g+1}$ times bigger than its normalized counterpart. For annihilation operators, this means that the norm is initially off by $1/\sqrt{g}$, but the annihilation operator will also generate $g$ copies of this state since it fails to anticommute with each copy of the quark, so we get an over all factor of $\sqrt{g}$.

\section{Principal Values in DLCQ} \label{PrincipalValue}
This appendix elaborates on our choice of principle value for the momentum space potentials in 1+1d QCD. For a more general definition of the principle value given by
\begin{equation}
  \left(\mathscr{P}\left[\frac{-1}{(n_2-n_3)^2}\right]\delta_{n_1+n_2,n_3+n_4}\right)_{n_1=n_4,n_2=n_3}=f_{n_1}+f_{n_2}.
\end{equation}
The parts of the potential that require a principle value are given by
\begin{align}
\nonumber V_B =&~ -\frac{\xi^2 n_p}{2N_c}\sum_{n_1,n_2,n_3,n_4}\mathscr{P}\left[\frac{-1}{(n_2-n_3)^2}\right]\delta_{n_1+n_2,n_3+n_4} \\
  & *\left(b_{a,f,n_1}^\dagger b_{b,g,n_2}^\dagger b_{a,g,n_3} b_{b,f,n_4}-\frac{1}{N_c}b_{a,f,n_1}^\dagger b_{b,g,n_2}^\dagger b_{b,g,n_3}b_{a,f,n_4}+(b\rightarrow d)\right), \\
\nonumber V_M =&~ \frac{\xi^2n_p}{N_c}\sum_{n_1,n_2,n_3,n_4}\mathscr{P}\left[\frac{-1}{(n_2-n_3)^2}\right]\delta_{n_1+n_2,n_3+n_4} \\
  & *\left(b_{a,f,n_1}^\dagger d_{a,g,n_2}^\dagger d_{b,g,n_3} b_{b,f,n_4}-\frac{1}{N_c}b_{a,f,n_1}^\dagger d_{b,g,n_2}^\dagger d_{b,g,n_3} b_{a,f,n_4}\right).
\end{align}
When the incoming and outgoing momenta are the same, then we have
\begin{align}
\nonumber (V_B)_0 =&~ -\frac{\xi^2 n_p}{2N_c}\sum_{n_1,n_2}(f_{n_1}+f_{n_2}) \\
  & *\left(b_{a,f,n_1}^\dagger b_{b,g,n_2}^\dagger b_{a,g,n_2} b_{b,f,n_1}-\frac{1}{N_c}b_{a,f,n_1}^\dagger b_{b,g,n_2}^\dagger b_{b,g,n_2}b_{a,f,n_1}+(b\rightarrow d)\right), \\
\nonumber (V_M)_0 =&~ \frac{\xi^2n_p}{N_c}\sum_{n_1,n_2}(f_{n_1}+f_{n_2}) \\
  & *\left(b_{a,f,n_1}^\dagger d_{a,g,n_2}^\dagger d_{b,g,n_2} b_{b,f,n_1}-\frac{1}{N_c}b_{a,f,n_1}^\dagger d_{b,g,n_2}^\dagger d_{b,g,n_2} b_{a,f,n_1}\right).
\end{align}
With $F_q = \sum_{n}(f_n)b_{a,f,n}^\dagger b_{a,f,n}$ and $F_{\bar{q}} = \sum_{n}(f_n)d_{a,f,n}^\dagger d_{a,f,n}$, these potentials become
\begin{align}
  (V_B+V_M)_0 =&~ \frac{\xi^2n_p}{N^2_c}\left(N_c U_0+(N_q-N_{\bar{q}})(F_q-F_{\bar{q}})-F_{q}-F_{\bar{q}}\right), \\
\nonumber U_0 =& \sum_{n_1,n_2}(f_{n_1}+f_{n_2})\left(b_{a,f,n_1}^\dagger d_{a,g,n_2}^\dagger d_{b,g,n_2} b_{b,f,n_1}\right) \\
  & -\sum_{n_1,n_2}f_{n_1}\left(b_{a,f,n_1}^\dagger b_{b,g,n_2}^\dagger b_{a,g,n_2} b_{b,f,n_1}+(b\rightarrow d)\right)
\end{align}

The state vector for a set of quarks and antiquarks $\psi$ in a color singlet $\alpha$ is given by
\begin{equation}
  \left|\psi,\alpha\right> = T^\alpha_{\{a_i\},\{a_j\}}\prod_i\left(b^\dagger_{a_i,f_i,n_i}\right)\prod_j\left(d^\dagger_{a_j,f_j,n_j}\right)\left|0\right>,
\end{equation}
where $\left|0\right>$ is the light-cone vacuum. The operator $U_0$ acting on this state gives
\begin{align}
\nonumber U_0\left|\psi,\alpha\right> =& \left(\sum_{I}f_{n_I}\left(\sum_{J}T^{\Delta_{IJ}(\alpha)}_{\{a_i\},\{a_j\}}-\sum_{I'\ne I}T^{X^b_{II'}(\alpha)}_{\{a_i\},\{a_j\}}\right)+\sum_{J}f_{n_J}\left(\sum_{I}T^{\Delta_{IJ}(\alpha)}_{\{a_i\},\{a_j\}}-\sum_{J'\ne J}T^{X^d_{JJ'}(\alpha)}_{\{a_i\},\{a_j\}}\right)\right) \\
  & *\prod_i\left(b^\dagger_{a_i,f_i,n_i}\right)\prod_j\left(d^\dagger_{a_j,f_j,n_j}\right)\left|0\right>,
\end{align}
where
\begin{align}
T^{\Delta_{IJ}(\alpha)}_{\{a_i\},\{a_j\}} &= \sum_A\left(T^\alpha_{\{a_i\},\{a_j\}}\right)_{a_I=a_J=A}\delta_{a_I,a_J}, \\
T^{X^b_{II'}(\alpha)}_{\{a_i\},\{a_j\}} &= \left(T^\alpha_{\{a_i\},\{a_j\}}\right)_{a_I\leftrightarrow a_{I'}}, \\
T^{X^d_{JJ'}(\alpha)}_{\{a_i\},\{a_j\}} &= \left(T^\alpha_{\{a_i\},\{a_j\}}\right)_{a_J\leftrightarrow a_{J'}}.
\end{align}
Since $T^\alpha$ corresponds to an $SU(N_c)$ color singlet, the $\Delta_{IJ}$ piece summed over the antiquark indices can be written as
\begin{align}
\nonumber \sum_{J}T^{\Delta_{IJ}(\alpha)}_{\{a_i\},\{a_j\}} =& \sum_{J}\sum_{b_I,b_J}\left(T^\alpha_{\{a_i\},\{a_j\}}\right)_{a_I=b_I,a_J=b_J}\delta_{b_I,b_J}\delta_{a_I,a_J} \\
\nonumber =& \sum_{J}\sum_{b_I,b_J}\left(T^\alpha_{\{a_i\},\{a_j\}}\right)_{a_I=b_I,a_J=b_J}(2(T^A)_{a_I,b_I} (T^A)_{a_J,b_J}+\frac{1}{N_c}\delta_{a_I,b_I}\delta_{a_J,b_J}) \\
\nonumber =& \sum_{I'\ne I}\sum_{b_I,b_I'}\left(T^\alpha_{\{a_i\},\{a_j\}}\right)_{a_I=b_I,a_I'=b_I'}(2(T^A)_{a_I,b_I} (T^A)_{a_I',b_I'}) \\
\nonumber & +\sum_{b_I,c_I}(2(T^A)_{a_I,b_I} (T^A)_{b_I,c_I})\left(T^\alpha_{\{a_i\},\{a_j\}}\right)_{a_I=c_I} +\frac{1}{N_c}N_{\bar{q}}T^\alpha_{\{a_i\},\{a_j\}} \\
\nonumber =& \sum_{I'\ne I}\left(T^\alpha_{\{a_i\},\{a_j\}}\right)_{a_I\leftrightarrow a_I'}+\left(N_c-\frac{1}{N_c}(N_q-N_{\bar{q}})\right)T^\alpha_{\{a_i\},\{a_j\}} \\
  =& \sum_{I'\ne I}T^{X^b_{II'}(\alpha)}_{\{a_i\},\{a_j\}}+(N_c-B)T^\alpha_{\{a_i\},\{a_j\}}.
\end{align}
Similarly, the $\Delta_{IJ}$ piece summed over the quark indices can be written as
\begin{equation}
  \sum_{I}T^{\Delta_{IJ}(\alpha)}_{\{a_i\},\{a_j\}} = \sum_{J'\ne J}T^{X^d_{JJ'}(\alpha)}_{\{a_i\},\{a_j\}}+(N_c+B)T^\alpha_{\{a_i\},\{a_j\}}.
\end{equation}
The expressions in parenthesis above can then be written as
\begin{align}
  \left(\sum_{J}T^{\Delta_{IJ}(\alpha)}_{\{a_i\},\{a_j\}}-\sum_{I'\ne I}T^{X^b_{II'}(\alpha)}_{\{a_i\},\{a_j\}}\right) &= (N_c-B)T^\alpha_{\{a_i\},\{a_j\}}, \\
  \left(\sum_{I}T^{\Delta_{IJ}(\alpha)}_{\{a_i\},\{a_j\}}-\sum_{J'\ne J}T^{X^d_{JJ'}(\alpha)}_{\{a_i\},\{a_j\}}\right) &= (N_c+B)T^\alpha_{\{a_i\},\{a_j\}}.
\end{align}

Therefore, $U_0$ acting on a color singlet vector is given by
\begin{align}
\nonumber U_0\left|\psi,\alpha\right> &= \left(\sum_{I}f_{n_I}(N_c-B)+\sum_{J}f_{n_J}(N_c+B)\right)\left|\psi,\alpha\right> \\
  &= \left((F_q+F_{\bar{q}})N_c-(F_q-F_{\bar{q}})B\right)\left|\psi,\alpha\right>,
\end{align}
and the forward scattering part of the potentials are given by
\begin{align}
\nonumber (V_B+V_M)_0\left|\psi,\alpha\right> &= \frac{\xi^2n_p}{N^2_c}\left((F_q+F_{\bar{q}})N^2_c-(F_q-F_{\bar{q}})N_c B+N_c B(F_q-F_{\bar{q}})-F_{q}-F_{\bar{q}}\right)\left|\psi,\alpha\right> \\
  &= \xi^2n_p\left(1-\frac{1}{N^2_c}\right)(F_q+F_{\bar{q}})\left|\psi,\alpha\right>.
\end{align}
Thus even with a fairly general expression for the principle value the contribution to the Hamiltonian will be fairly simple, and this is what allows the $V_2$ term to be subsumed into the principle value of the momentum space potential.

\end{document}